\documentclass[12pt]{article}

\newcommand{\la}{\langle}
\newcommand{\ra}{\rangle}
\newcommand{\beq}{\begin{equation}}
\newcommand{\eeq}{\end{equation}}
\newcommand{\bea}{\begin{eqnarray}}
\newcommand{\eea}{\end{eqnarray}}

\def\NN		{\mathcal{N}}
\def\nn 		{\nonumber}
\def\Tr 		{\mathop{\rm Tr}\nolimits}
\def\tr		{\mathop{\rm tr}\nolimits}
\def\norma	{\mathscr{N}}
\def\normap	{\mathscr{N}_{p_1p_2p_3p_4}}

\def\OO	 	{\mathcal{O}}
\def\OOL		{\mathscr{O}}
\def\GG	 	{\mathpzc{G}}

\def\AA		{\mathcal{A}}
\def\HH   	 	{\mathcal{H}}
\def\HHp	        {\mathcal{H}_{p_1p_2p_3p_4}}
\def\CB		{\mathcal{B}}
\def\FFp	 	{\mathcal{F}_{p_1p_2p_3p_4}}

\def\MM 	 	{\mathcal{M}}
\def\LL		{\mathbb{L}}
\def\Hdyna 	{\mathcal{H}^{\rm dyna} }
\def\dm 		{d\mathcal{C}}
\def\Li 		{\mathrm{Li}}

\def\Pref 		{\mathcal{P}}
\def\Prefp		{\mathcal{P}_{p_1p_2p_3p_4}}
\def\Intri 		{\mathcal{I}}
\def\Dbar		{\overline{D}}
\def\SCW  	{\mathbb{S}}
\def\bF		{{\bf F}}
\def\bP		{{\bf P}}

\def\x 		{x_1}
\def\xb		{x_2}
\def\y 		{y_1}
\def\yb  		{y_2}

\def\tw		{t}
\def\ula{{\underline{\smash \lambda}}}
\def\cO {{\mathcal O}}
\def\cN {{\mathcal N}}

\def\xx{\mathsf{x}}
\def\muu{\mu}

\def \be  {\begin{equation}}
\def \ee  {\end{equation}}
\def \ba  {\begin{eqnarray}}
\def \ea  {\end{eqnarray}}

\def \phield {\phi}
\def \Nphield { } 
\def \n {q}
\def \symbol {\mathscr{M} }
\def \nameuno { \widehat{\symbol} (t|l) }
\def \namedue { \widehat{\AA} (t|l) }

\usepackage{sectsty}
\sectionfont{\fontsize{14}{19}\selectfont}
\subsectionfont{\fontsize{12}{15}\selectfont}

\usepackage{
	epsfig,
	amssymb,
	psfrag,cite
}

\usepackage{mathrsfs}
\DeclareMathAlphabet{\mathpzc}{OT1}{pzc}{m}{it}

\usepackage{amsmath}
\usepackage{setspace}

\usepackage[mathscr]{euscript}

\usepackage{color}   
\usepackage{hyperref}
\hypersetup{
    colorlinks=true, 
    linktoc=all,     
    linkcolor=black, 
    citecolor=black
}

\usepackage{tikz}
\usepackage{float}

\begin{document}

\thispagestyle{empty}

\null\vskip-43pt \hfill
\begin{minipage}[t]{30mm}
	DCPT-17/19
\end{minipage}

\null\vskip-12pt \hfill  \\
\null\vskip-12pt \hfill   \\

\vskip2.2truecm
\begin{center}
\vskip 0.2truecm {\Large\bf
{\Large Unmixing Supergravity}
}\\
\vskip 1truecm
{\bf F.~Aprile${}^{1,2}$, J.~M. Drummond${}^{2}$, P.~Heslop${}^{3}$, H.~Paul${}^{2}$ \\
}

\vskip 0.4truecm
 
{\it
${}^{1}$ Mathematical Sciences and STAG Research Centre, \\
University of Southampton, Highfield, SO17 1BJ,\\
\vskip .2truecm }
\vskip .2truecm
{\it
${}^{2}$ School of Physics and Astronomy and STAG Research Centre, \\
University of Southampton,
 Highfield,  SO17 1BJ,\\
\vskip .2truecm                        }
\vskip .2truecm
{\it
${}^{3}$ Mathematics Department, Durham University, \\
Science Laboratories, South Rd, Durham DH1 3LE \vskip .2truecm                        }
\end{center}

\vskip 1truecm 
\centerline{\bf Abstract} 

We examine the double-trace spectrum of $\mathcal{N} = 4$ super Yang-Mills theory in the supergravity limit. At large $N$ double-trace operators exhibit degeneracy. By considering free-field and tree-level supergravity contributions to four-point functions of half-BPS operators we resolve the degeneracy for a large family of double-trace operators. The mixing problem reveals a surprisingly simple structure which allows us to obtain their three-point functions at leading order in the large $N$ expansion as well as their leading anomalous dimensions. 

\medskip

 \noindent

\newpage
\setcounter{page}{1}\setcounter{footnote}{0}

\section{Introduction}

The behaviour of conformal field theories in the limit of large central charge has been a subject of great interest in recent years. One motivation for this interest is the AdS/CFT correspondence which relates gravitational theories in AdS to a CFT on the boundary \cite{1,2,3}. While much recent work has focussed on understanding general constraints on possible holographic theories \cite{0907.0151,Fitzpatrick:2010zm,Fitzpatrick:2011ia,1212.3616,1212.4103,1410.4717,Beem:2016wfs,1612.03891,Alday:2017gde}, it is also of interest to explore explicit examples to understand the details of the spectrum and interactions as these can sometimes reveal unexpected features. The archetypal holographic example is the correspondence between four-dimensional $\mathcal{N}=4$ super Yang-Mills theory and type IIB superstring theory on ${AdS}_5 \times {S}^5$.

The central quantities of interest under such a correspondence are the correlation functions of gauge-invariant local operators. In the case of $\mathcal{N}=4$ super Yang-Mills theory, such correlation functions are dependent on the gauge coupling $g$ and the choice of gauge group, which we take to be $SU(N)$. The limit of large central charge corresponds to the large $N$ limit and, when taken with the 't Hooft coupling $\lambda = g^2 N$ fixed and large, should lead to a regime of the theory where the massive string excitations decouple and which can be described by IIB supergravity on  ${AdS}_5 \times {S}_5$.

The massless string modes include the graviton and its superpartners. These fields can propagate in the ${AdS}_5$ directions, while the ${S}^5$ factor leads to a tower of Kaluza-Klein modes all carrying representations of $SU(4)$. The graviton multiplet corresponds to the energy-momentum multiplet of $\mathcal{N}=4$ super Yang-Mills theory and it is the simplest example of a half-BPS multiplet. There is an infinite tower of related half-BPS operators, corresponding to the associated Kaluza-Klein modes.
In terms of Yang-Mills fields the superconformal primary operators of these half-BPS multiplets take the form of a single trace over scalar fields $\phi^I$ which transform in the vector representation of $SO(6)$,
\be
\mathcal{O}_p(x,y) = y^{I_1} \ldots y^{I_p} \tr \bigl(\phield_{I_1}(x) \ldots \phield_{I_p} (x)\bigr)\,.
\ee
Here $y^I$ is an auxiliary null $SO(6)$-vector, $y^2=0$. The label $p$ denotes the fact that the primary sits in the $SU(4)$ representation $[0,p,0]$, with the case $p=2$ corresponding to the energy-momentum multiplet. The fact that the operators $\mathcal{O}_p$ are half-BPS means that they always possess their classical integer scaling dimensions. Their two-point and three-point functions also receive no quantum corrections and take their free field theory forms.

Here we will draw on general CFT techniques, in particular the operator product expansion (OPE), as well as explicit results for the tree-level supergravity contribution to correlation functions of half-BPS operators. A very compact solution for the most general half-BPS four-point function $\langle \mathcal{O}_{p_1} \mathcal{O}_{p_2} \mathcal{O}_{p_3} \mathcal{O}_{p_4} \rangle$ at tree-level in the supergravity limit was presented in \cite{Rastelli:2016nze}. The formula is given in Mellin space, and is deduced from general analytic principles applied to the Mellin representation, rather than a direct supergravity calculation. These properties are based on the existence of the OPE and in particular the presence of exchanged double-trace operators as well as other properties such as crossing symmetry. The resulting formula agrees with the cases available in the literature obtained from representations in terms of Witten diagrams and other techniques \cite{Liu:1998ty,hep-th/9903196,Arutyunov:1999fb,Arutyunov:2000py,Arutyunov:2002fh,Arutyunov:2003ae,Dolan:2006ec,Berdichevsky:2007xd,Uruchurtu:2008kp,Uruchurtu:2011wh}. Further analysis examining the consistency of the result of \cite{Rastelli:2016nze} with supergravity have been performed in \cite{Arutyunov:2017dti}.

Here we systematically analyse the OPE of a particular class of four-point functions at large $N$, using methods developed in many papers on the OPE of conformal and superconformal theories \cite{Dolan:2000ut,Eden:2001ec,Dolan:2001tt,Dolan:2002zh,Dolan:2004iy,Dolan:2006ec}. In the OPE of these correlators we expect both protected operators and unprotected ones. The only unprotected operators which we expect to be present in the  spectrum in the supergravity limit are multi-trace operators made from products of derivatives of the operators $\mathcal{O}_p$. This is because we expect all other long operators to correspond to string excitations which have acquired large mass in the supergravity limit. Furthermore, of the long multi-trace operators, we expect only the double-trace operators of the schematic form
\be
\mathcal{O}_p \Box^n \partial^l \mathcal{O}_q\,,
\label{doubletrace}
\ee
to appear in the OPE at leading order in $1/N^2$. Higher multi-trace operators should also appear, but only at higher orders in $1/N^2$.
Operators of the form (\ref{doubletrace}) have classical dimension $p+q+2n+l$ and spin $l$. More often we will refer to the twist which is the difference of the dimension and the spin (hence equal to $p+q+2n$ in the above case) instead of the dimension itself. In the strict large $N$ limit the dimension will be fixed to its classical value, regardless of the value of the Yang-Mills coupling. In a large $N$ expansion the operators (\ref{doubletrace}) will only acquire anomalous dimensions at order $1/N^2$.

In the first instance we will consider the $SU(4)$ singlet double-trace operators for which we need $p=q$ in (\ref{doubletrace}). In that case the only quantum numbers which distinguish them are the twist and the spin. It is therefore clear that the set of $(t-1)$ operators
\be
\{ \mathcal{O}_2 \Box^{t-2} \partial^l \mathcal{O}_2, \mathcal{O}_3 \Box^{t-3} \partial^l \mathcal{O}_3, \ldots , \mathcal{O}_t \Box^{0} \partial^l \mathcal{O}_t \}
\label{degenops}
\ee
are degenerate in the strict large $N$ limit since they all have twist $2t$ and spin $l$. Including the anomalous dimensions at order $1/N^2$ will lift the degeneracy however.

We label the $(t-1)$ degenerate operators with fixed $t$ and $l$ by $K_{t,l,i}$ for $i=1,\ldots,t-1$.
In order to resolve the degeneracy among the operators (\ref{degenops}) we consider four-point correlators of the form $\langle \mathcal{O}_p \mathcal{O}_p \mathcal{O}_q \mathcal{O}_q \rangle$ for $2 \leq p \leq q \leq t$. To perform our OPE analysis we need two pieces of information about each correlator. Firstly we need the leading large $N$ result which comes from disconnected contributions and can be obtained from free field theory. Secondly we need the first $1/N^2$ suppressed connected contribution, coming from the formula of \cite{Rastelli:2016nze}. With these two pieces of information we find that we have enough information to resolve the degeneracy of the sector of unprotected double trace operators. This yields the leading order three-point functions $\langle \mathcal{O}_p \mathcal{O}_p K_{t,l,i} \rangle$ for each of the operators $K_{t,l,i}$ as well as the $O(1/N^2)$ contribution to their anomalous dimensions. 

Above we discussed the singlet channel but we are able to generalise the analysis to consider long double-trace operators in the $[n,0,n]$ representation for any $n$. In this case we have $(t-n-1)$ degenerate operators of twist $2t$ and spin $l$ schematically given by
\be
\{ \mathcal{O}_{n+2} \Box^{t-n-2} \partial^l \mathcal{O}_{n+2}, \mathcal{O}_{n+3} \Box^{t-n-3} \partial^l \mathcal{O}_{n+3}, \ldots , \mathcal{O}_t \Box^{0} \partial^l \mathcal{O}_t \}
\label{degenopsn0n}
\ee
Again the information required to resolve the degeneracy can be obtained just by considering correlators of the form $\langle \mathcal{O}_p \mathcal{O}_p \mathcal{O}_q \mathcal{O}_q \rangle$ for $2+n \leq p \leq q \leq t$.

Even though the formula of \cite{Rastelli:2016nze} for the Mellin representation of the correlation functions is very simple, there is no guarantee that solution of the mixing problem will be simple. However, we find a surprisingly simple structure appearing in both the anomalous dimensions and the OPE coefficients. To exhibit the simplicity here we quote the formula for the anomalous dimensions of the $(t-n-1)$ double-trace operators in the $[n,0,n]$ representation with classical twist $2t$ and spin $l$. We write the full dimension as
\be
\Delta_{t,l,i}^{[n,0,n]} = 2t + l +\frac{2}{N^2} \eta^{[n,0,n]}_{t,l,i} + \ldots
\ee
where the ellipsis denotes terms of higher order in $1/N$. The quantity $\eta^{[n,0,n]}_{t,l,i}$ is given by
\be
\eta^{[n,0,n]}_{t,l,i}= -\frac{2 (t-1 - n) t (t+1)(t+2 + n) (t+l-n)(t+l+1)(t+l+2)(t+l+3+n) }{  (l+2i+n-1)_{6} } \,,
\label{etan0n}
\ee
and $i=1,\ldots,t-n-1$ is the extra label needed to distinguish the different operators which become degenerate at infinite $N$. In (\ref{etan0n}) we have used the Pochhammer symbol $(x)_r = x (x+1) \ldots(x+r-1)$ to compactify the expression.

The three-point functions $\langle \mathcal{O}_p \mathcal{O}_p K_{t,l,i} \rangle$ also exhibit a very nice structure with respect to their spin dependence. For fixed $t$ these naturally form a mixing matrix with $l$-dependent entries. We find that the $l$ dependence has a universal structure which can be precisely fixed by imposing orthogonality of the (normalised) matrix. Thus, having used the explicit data to identify this pattern, the three-point functions can then actually be determined with no more reference to the explicit correlation functions. We find this universal structure quite remarkable and suggests a further underlying structure yet to be identified.

Very recently the two papers \cite{Alday:2017xua,Aprile:2017bgs} appeared using the idea of resolving the degeneracy among the singlet double-trace operators to make statements about quantum corrections to the classical supergravity results. In \cite{Alday:2017xua}, the method of large spin perturbation theory (see \cite{Alday:2016njk}) was applied to derive formulas for the $O(1/N^4)$ corrections to the anomalous dimensions of the singlet twist-four operators. In our paper \cite{Aprile:2017bgs}, we used the resolved mixing for the singlet channel presented here in more detail to make a closed-form resummed prediction for the double discontinuity of the correlator $\langle \mathcal{O}_2 \mathcal{O}_2 \mathcal{O}_2 \mathcal{O}_2 \rangle$ at order $1/N^4$. In particular in \cite{Aprile:2017bgs} we already presented and used the result for the anomalous dimensions (\ref{etan0n}) in the singlet case $n=0$. We were then able to use a polylogarithmic ansatz to construct a crossing symmetric amplitude, which was fixed almost uniquely\footnote{For details see the discussion in Section 6 of \cite{Aprile:2017bgs}.} by the double discontinuity. From this predicted amplitude we then extracted a closed-form all-spin formula for the $1/N^4$ correction to the singlet twist-four anomalous dimensions. The resulting formula agrees with the dimensions quoted in \cite{Alday:2017xua}.

\section{Four-point correlators of half-BPS operators in $\NN=4$ SYM }
\label{sec_generalities}

Half-BPS scalar operators in $\NN=4$ SYM transform in the irrep $[0,p,0]$ of the  
$R$-symmetry group $SO(6)\subset SU(2,2|4)$ and have protected dimension $\Delta=p$.
At weak coupling, these operators can be described by the single-trace operators 
\beq
\mathcal{O}^{I}_p=\Nphield\, C_I^{i_1,\ldots i_p}\, \Tr\left( \phield_{i_1}\ldots \phield_{i_p} \right),\qquad I=1,\ldots {\rm dim}[0,p,0]\,
\eeq
where $\phield_{i=1,\ldots,6}$ are elementary fields in the adjoint of $SU(N)$, and the $C_I^{i_1\ldots i_k}$ 
provide a real basis of traceless symmetric tensors for the irrep $[0,p,0]$.  
At strong 't Hooft coupling, $\mathcal{O}^{I}_p$ is dual to a scalar field $\varphi^{I}_p$ of type IIB supergravity compactified on $AdS_5\times S^5$. 
According to the AdS/CFT correspondence, the mass of $\varphi^I_p$ is related to the dimension of $\mathcal{O}^I_p$ through the formula $m^2L^2= p(p-4)$, 
where $L$ is the AdS radius, and the corresponding irrep is obtained from Kaluza-Klein reduction on the five-sphere \cite{Kim:1985ez}.

Here we will be interested in four-point correlation functions. 
A generic four-point correlator will transform as a singlet inside the product  $\otimes_{i=1}^{4}[0,p_i,0]$.
Handling the $SO(6)$ structure can be conveniently done as follows,  
\beq\label{def_halfBPS}
\OO_p = \Nphield \, y^{i_1}\ldots y^{i_p} \Tr\left( \phield_{i_1}\ldots \phield_{i_p} \right),\qquad  \vec{y} \cdot \vec{y} = 0\ ,
\eeq 
where $y^{i}$ is a complex null vector parametrizing the coset space $SU(4)/ S(U(2)\times U(2))$. 
In the context of AdS/CFT  bulk fields $\varphi^I_p$ are parametrized by harmonic variables on a different coset space, $S^5\sim SO(6)/SO(5)$,  
therefore the representation \eqref{def_halfBPS} is not directly available.
Four-point correlators obtained in this representation can be re-expressed in terms of the other 
and reduced to the following general form
\beq\label{diagram_correlator}
\la \OO_{p_1}(\vec{x}_1)\OO_{p_2}(\vec{x}_2)\OO_{p_3}(\vec{x}_3)\OO_{p_4}(\vec{x}_4)\ra
= \sum_{ \{ d_{ij}\} }\left( \prod_{1\le i<j\le 4} \left( g_{ij} \right)^{ d_{ij}} \right) A_{ \{ d_{ij} \}}(u,v)\,,
\eeq
where the propagator $g_{ij}$ and the cross ratios $(u,v)$ are defined as
\beq
g_{ij}= \frac{\vec{y}_{ij}^{\,2}}{r_{ij} },
\qquad
u = \frac{ r_{12} r_{34} }{ r_{13} r_{24}  }\,,\qquad v=\frac{ r_{14} r_{23}  }{ r_{13} r_{24} }\,,\qquad 
r_{ij}=(\vec{x}_i-\vec{x}_j)^{\,2}, \qquad \vec{y}_{ij}^{\,2} = \vec{y}_i \cdot \vec{y}_j\,.
\eeq
The sum runs over all possible partitions $\{d_{ij}\}$ such that 
\beq\label{conditionsdij}
d_{ij}=d_{ji},\qquad d_{ii}=0,\qquad \sum_{j} d_{ij}=p_i\quad \forall\, i=1,\ldots 4.
\eeq
We shall refer to the expression \eqref{diagram_correlator} as the diagrammatic representation of the correlator. 
In the diagram a line connecting two black bullet points $i$ and $j$ will correspond to a superpropagator $g_{ij}$. 
In free field theory the diagrammatic representation of the correlator follows directly from Wick's theorem. 
The functions $A_{\{d_{ij}\} }$ are constant in $(u,v)$ and only depend on the rank of the gauge group. 
A simple case study is $\la\OO_2\OO_2\OO_2\OO_2\ra$ which is fully crossing symmetric:\\
\begin{minipage}{\linewidth}
          \begin{center}
             \begin{tikzpicture}
		\def\x1{2}
		\def\y1{3}
		\def\step{1.5}
		\def\distaone{3.25}
		\def\distatwo{6.5}
		\def\colore{black}

\draw (\x1,\y1+\step) node {\rule{0pt}{2cm}};

		\draw (\x1-3,	\y1+.75) 	node 	{ $\la\OO_2\OO_2\OO_2\OO_2\ra\quad = \quad A_{disc.}\quad$};
		\draw (\x1-3,	.75) 		node	 	{ $\phantom{\la\OO_2\OO_2\OO_2\OO_2\ra\quad = \quad} \ A_{conn.}\quad $};
		
		\draw[thick] (\x1-.5,	\y1-.3) -- 			(\x1-.5,	\y1+\step+.3);
		\draw[thick] (\x1-.5,	\y1-.3) -- 			(\x1-.5+.2,	\y1-.3);
		\draw[thick] (\x1-.5,	\y1+\step+.3) -- 	(\x1-.5+.2,\y1+\step+.3);
		
		\draw[thick,\colore] (\x1-.1,		\y1) --	 (\x1-.1,		\y1+\step);
		\draw[thick,\colore] (\x1+.1,		\y1) -- 	(\x1+.1,		\y1+\step);
		\draw[thick,\colore] (\x1-.1+\step,	\y1) -- 	(\x1-.1+\step,	\y1+\step);
		\draw[thick,\colore] (\x1+.1+\step,	\y1) --	 (\x1+.1+\step,	\y1+\step);
		\draw[thick,\colore] (\x1+\distaone,	\y1+.1) -- 			 (\x1+\distaone+\step,		\y1+.1);
		\draw[thick,\colore] (\x1+\distaone,	\y1-.1) -- 			 (\x1+\distaone+\step,		\y1-.1);
		\draw[thick,\colore] (\x1+\distaone,	\y1+\step+.1) --		 (\x1+\distaone+\step,		\y1+\step+.1);
		\draw[thick,\colore] (\x1+\distaone,	\y1+\step-.1) --		 (\x1+\distaone+\step,		\y1+\step-.1);
		\draw[thick,\colore] (\x1+.1+\distatwo,	\y1) --			 (\x1+.1+\distatwo+\step,	\y1+\step);
		\draw[thick,\colore] (\x1-.1+\distatwo,		\y1) --			 (\x1-.1+\distatwo+\step,	\y1+\step);
		\draw[thick,\colore] (\x1+.1+\distatwo,	\y1+\step) -- 		(\x1+.1+\distatwo+\step,	\y1);
		\draw[thick,\colore] (\x1-.1+\distatwo,		\y1+\step) -- 		(\x1-.1+\distatwo+\step,	\y1);

		\filldraw (\x1,		\y1)				 circle (.15);
		\filldraw (\x1+\step,	\y1) 		 		 circle (.15);
		\filldraw (\x1,		\y1+\step) 		 circle (.15);
		\filldraw (\x1+\step,	\y1+\step) 	 	 circle (.15);
		\draw (\x1+\step,	\y1)	 		 	 node[right=0.6*\step cm,above=0.3*\step cm] {$+$};
		\filldraw (\x1+\distaone,		\y1)				 circle (.15);
		\filldraw (\x1+\distaone+\step,	\y1) 				 circle (.15);
		\filldraw (\x1+\distaone,		\y1+\step) 		 circle (.15);
		\filldraw (\x1+\distaone+\step,	\y1+\step) 	 	 circle (.15);
		\draw (\x1+\distaone+\step,	\y1)  				node[right=0.6*\step cm,above=0.3*\step cm] {$+$};
		\filldraw (\x1+\distatwo,		\y1)				 circle (.15);
		\filldraw (\x1+\distatwo+\step,	\y1) 				 circle (.15);
		\filldraw (\x1+\distatwo,		\y1+\step) 		 circle (.15);
		\filldraw (\x1+\distatwo+\step,	\y1+\step) 		 circle (.15);
		\draw (\x1+\distatwo+\step,	\y1) 				 node[right=0.8*\step cm,above=0.3*\step cm] {$+$};

		\draw[thick] (\x1+\distatwo+\step+.5,		\y1-.3) --			(\x1+\distatwo+\step+.5,	\y1+\step+.3);		
		\draw[thick] (\x1+\distatwo+\step+.5-.2,	\y1-.3) -- 			(\x1+\distatwo+\step+.5,	\y1-.3);
		\draw[thick] (\x1+\distatwo+\step+.5-.2, 	\y1+\step+.3)-- 		(\x1+\distatwo+\step+.5,	\y1+\step+.3);

\draw (\x1+\distatwo+\step+1,\y1) node {\rule{2cm}{0pt}};

		\def\x2{2}
		\def\y2{0}

		\draw[thick] (\x2-.5,	\y2-.3) --			 (\x2-.5,		\y2+\step+.3);
		\draw[thick] (\x2-.5,	\y2-.3) --			 (\x2-.5+.2,	\y2-.3);
		\draw[thick] (\x2-.5,	\y2+\step+.3) -- 	(\x2-.5+.2,		\y2+\step+.3);

		\draw[thick,\colore] (\x2,\y2)  rectangle (\x2+\step,\y2+\step);
		
		\draw[thick,\colore] (\x2+\distaone,	\y2) -- 		(\x2+\distaone+\step,		\y2);
		\draw[thick,\colore] (\x2+\distaone,	\y2+\step) -- 	(\x2+\distaone+\step,		\y2+\step);
		\draw[thick,\colore] (\x2+\distaone,	\y2) --		(\x2+\distaone+\step,		\y2+\step);
		\draw[thick,\colore] (\x2+\distaone,	\y2+\step) -- 	(\x2+\distaone+\step,		\y2);
		
		\draw[thick,\colore] (\x2+\distatwo,		\y2) -- 		(\x2+\distatwo,			\y2+\step);
		\draw[thick,\colore] (\x2+\distatwo+\step,	\y2) -- 		(\x2+\distatwo+\step,		\y2+\step);
		\draw[thick,\colore] (\x2+\distatwo,		\y2) -- 		(\x2+\distatwo+\step,		\y2+\step);
		\draw[thick,\colore] (\x2+\distatwo,		\y2+\step) -- 	(\x2+\distatwo+\step,		\y2);

		\filldraw (\x2,		\y2)			 	 circle (.15);
		\filldraw (\x2+\step,	\y2) 				 circle (.15);
		\filldraw (\x2,		\y2+\step) 		 circle (.15);
		\filldraw (\x2+\step,	\y2+\step) 	 	 circle (.15);
		\draw (\x2+\step,	\y2) 				node[right=0.6*\step cm,above=0.3*\step cm] {$+$};
		
		\filldraw (\x2+\distaone, 		\y2)			 	 circle (.15);
		\filldraw (\x2+\distaone+\step,	\y2) 		 		 circle (.15);
		\filldraw (\x2+\distaone,		\y2+\step) 		 circle (.15);
		\filldraw (\x2+\distaone+\step,	\y2+\step) 	 	 circle (.15);
		\draw (\x2+\distaone+\step,	\y2) 				node[right=0.6*\step cm,above=0.3*\step cm] {$+$};
		\filldraw (\x2+\distatwo,		\y2)			 	circle (.15);
		\filldraw (\x2+\distatwo+\step,	\y2) 		 		circle (.15);
		\filldraw (\x2+\distatwo,		\y2+\step) 		circle (.15);
		\filldraw (\x2+\distatwo+\step,	\y2+\step) 		circle (.15);	

		\draw[thick] (\x2+\distatwo+\step+.5,			\y2-.3) --			 (\x2+\distatwo+\step+.5,	\y2+\step+.3);		
		\draw[thick] (\x2+\distatwo+\step+.5-.2,		\y2-.3) -- 			 (\x2+\distatwo+\step+.5,	\y2-.3);
		\draw[thick] (\x2+\distatwo+\step+.5-.2,		\y2+\step+.3)--		 (\x2+\distatwo+\step+.5,	\y2+\step+.3);

\draw (\x2,\y2-.8) node {\rule{0pt}{.8cm}};

	      \end{tikzpicture}
           \end{center}
 \end{minipage}
An explicit computation gives $A_{disc.} = 4(N^2-1)^2$ and $A_{conn.}=16(N^2-1)$. 
Non trivial  $(u,v)$ dependence arises both at loop level in perturbation theory 
\cite{Eden:1998hh,Eden:2000mv,Eden:2011we,Eden:2012tu,Drummond:2013nda,Chicherin:2015edu,Bourjaily:2015bpz,Bourjaily:2016evz}, 
and at strong 't Hooft coupling from holography~\cite{Liu:1998ty,hep-th/9903196,Arutyunov:1999fb,Arutyunov:2000py,Arutyunov:2002fh,Arutyunov:2003ae,Dolan:2006ec,Berdichevsky:2007xd,Uruchurtu:2008kp,Uruchurtu:2011wh}.

It is sometimes conventional to introduce together with the cross ratios $(u,v)$, the $SU(4)$ invariants $\sigma$ and $\tau$ defined by:
\beq
\frac{ g_{13} g_{24} }{ g_{12} g_{34} }= u\,\sigma,\qquad \frac{ g_{14} g_{23} }{ g_{12} g_{34} } = \frac{u\,\tau}{v}\ .
\eeq
Each diagram can then be expressed as a monomial in $\sigma$ and $\tau$ multiplied by a kinematic prefactor.
Without loss of generality, we can assume the operators at locations $\vec{x}_1,\vec{x}_2,\vec{x}_3,\vec{x}_4$ 
have dimensions $p_1\ge p_2\ge p_3 \ge p_4$, respectively.
Then we pull out an overall prefactor from the correlator
\begin{itemize}
\item[1)] 
	$\Prefp= g_{12}^{d}\, g_{13}^{p_1-d}\, g_{23}^{p_2-d}\, g_{34}^{p_4}\ $ with $\ d=\frac{p_1+p_2-p_3+p_4}{2}>0\ $ if $p_2+p_3 \geq p_1+p_4$, 
\item[2)] \vskip 0.2truecm
	$\Prefp=g_{12}^{p_2}\, g_{13}^{p_3-d}\, g_{14}^{p_4-d}\, g_{34}^{d}\ $ with $\ d=\frac{-p_1+p_2+p_3+p_4}{2}>0\ $ if $p_2+p_3 \leq p_1+p_4$,
\end{itemize}
The prefactors can be displayed diagrammatically as:\\[.4cm]
%
\begin{minipage}{\linewidth}
      \centering
      \begin{minipage}{0.3\linewidth}
          \begin{center}
             \begin{tikzpicture}
		\def\x1{1}
		\def\y1{1}
		\def\colore{black}

		\draw (\x1+1,\y1+3) node {case 1)};
		\draw[\colore,very thick] (\x1,\y1)--(\x1+2,\y1) 		node[left=1cm, below=1pt]		 {\scriptsize $p_2{-}d$} ;
		\draw[\colore,very thick] (\x1,\y1)--(\x1,\y1+2) 		node[below=1cm, left=0pt]		 {\scriptsize $d$} ;
		\draw[\colore,very thick] (\x1,\y1+2)--(\x1+2,\y1)		node[above=1.5cm,left=.4cm] 		 {\scriptsize $p_1{-}d$};  
		\draw[\colore,very thick] (\x1+2,\y1+2)--(\x1+2,\y1)  	node[above=1cm, right=0pt] 		 {\scriptsize $p_4$};
		
		\draw[white,very thick] (\x1,\y1)--(\x1,\y1-1);
	
		\filldraw (\x1,\y1)		 circle (.25);
		\filldraw (\x1+2,\y1) 		circle (.25);
		\filldraw (\x1,\y1+2) 		circle (.25);
		\filldraw (\x1+2,\y1+2) 	circle (.25);
	
		\draw[text=white,text centered] (\x1,\y1)         	node     {\footnotesize $p_2$};
		\draw[text=white,text centered] (\x1+2,\y1)     	node     {\footnotesize $p_3$};
		\draw[text=white, text centered] (\x1,\y1+2)     	node     {\footnotesize $p_1$};
		\draw[text=white,text centered] (\x1+2,\y1+2) 	node     {\footnotesize $p_4$};
		\node at (2,\y1-1-.25) {$d:=\frac{p_1+p_2-p_3+p_4}{2}$};
		\draw (\x1+1,\y1-1.75) node {\rule{0pt}{.4cm}};
	      \end{tikzpicture}
           \end{center}
       \end{minipage}
%
%
      \begin{minipage}{0.4\linewidth}
          \begin{center}
             \begin{tikzpicture}
		\def\x1{1}
		\def\y1{1}
		\def\colore{black}

		\draw (\x1+1,\y1+3) node {case 2)};
		\draw[\colore,very thick] (\x1,\y1)--(\x1,\y1+2) 			node[below=1cm, left=0pt] 		 {\scriptsize $p_{2}$} ;
		\draw[\colore,very thick] (\x1,\y1+2)--(\x1+2,\y1+2) 		node[left=1cm, below=1.3cm] 		 {\scriptsize $p_3{-}d$} ;
		\draw[\colore,very thick] (\x1,\y1+2)--(\x1+2,\y1)			node[above=1.75cm,left=.3cm]		 { \scriptsize $p_4{-}d$};  
		\draw[\colore,very thick] (\x1+2,\y1+2)--(\x1+2,\y1)  		node[above=1cm, right=0pt] 		 {\scriptsize $d$};
		
		\draw[white,very thick] (\x1,\y1)--(\x1,\y1-1); 		
	
		\filldraw (\x1,\y1)		 circle (.25);
		\filldraw (\x1+2,\y1) 		circle (.25);
		\filldraw (\x1,\y1+2) 		circle (.25);
		\filldraw (\x1+2,\y1+2) 	circle (.25);
	
		\draw[text=white,text centered] (\x1,\y1)         	node     {\footnotesize $p_2$};
		\draw[text=white,text centered] (\x1+2,\y1)     	node     {\footnotesize $p_3$};
		\draw[text=white, text centered] (\x1,\y1+2)     	node     {\footnotesize $p_1$};
		\draw[text=white,text centered] (\x1+2,\y1+2) 	node     {\footnotesize $p_4$};
		\node at (2,\y1-1-.25) 						    {$d:=\frac{-p_1+p_2+p_3+p_4}{2}$};
		\draw (\x1+1,\y1-1.75) node {\rule{0pt}{.4cm}};
	      
	      \end{tikzpicture}
           \end{center}
         \end{minipage}
  \end{minipage}
In both cases, the values of $d$ is uniquely fixed by solving the conditions \eqref{conditionsdij}. These two cases cover the entire range of non-vanishing possibilities for ordered $p_1,p_2,p_3,p_4$.
With the appropriate prefactor, we can rewrite
\beq\label{pre}
\la \OO_{p_1}(\vec{x}_1)\OO_{p_2}(\vec{x}_2)\OO_{p_3}(\vec{x}_3)\OO_{p_4}(\vec{x}_4)\ra=\Prefp\, \GG_{p_1p_2p_3p_4}(u,v,\sigma,\tau),
\eeq
where $\GG$ is now polynomial in $\sigma$ and $\tau$.  

The ``partial non-renormalization" theorem~\cite{Eden:2000bk} 
puts further constraints on the form of $\GG_{p_1p_2p_3p_4}$. In particular, $\GG_{p_1p_2p_3p_4}$ admits the splitting
\beq\label{SC1}
\GG_{p_1p_2p_3p_4} = \FFp + \Intri(u,v,\sigma,\tau)\,\HHp(g)\ , 
\eeq
where $\Intri(u,v,\sigma,\tau)$ is the following degree two polynomial,\\
\beq\label{Intriuv}
\Intri(u,v,\sigma,\tau)= v+ \sigma^2 u v + \tau^2 u + \sigma v (v-1-u) + \tau (1-u-v)+\sigma\tau u (u-1-v)\ ,
\eeq\\ 
and the key point is that $\FFp$ is independent of the coupling constant $g_{YM}$ whilst all the  non-trivial dependence on $g_{YM}$ appears in $\HHp$. 


\subsection{Large-N correlation functions at strong 't Hooft coupling}

The AdS/CFT correspondence predicts, in the regime of strong 't Hooft coupling corresponding 
to classical supergravity, the leading large-$N$ behaviour of the correlation functions 
$\la \mathcal{O}_{p_1}^{I_1} \mathcal{O}_{p_2}^{I_2} \mathcal{O}_{p_3}^{I_3} \mathcal{O}_{p_4}^{I_4} \ra$.
We briefly sketch how the computation goes, and we make some important remarks. 

The action for the collection of KK modes $\{\, \mathpzc{f}_{\,k}\}_{k\ge 1}$ on AdS$_5\times$S$^5$ can be written as, 
\beq\label{sugraIIB}
{S}_{\rm sugra}= \frac{N^2}{8\pi^2 L^3} \int d\Omega\, \left( \mathcal{L}_{(2)}+ \mathcal{L}_{(3)}+\mathcal{L}_{(4)} +\ldots \right)
\eeq
where $d\Omega$ is the measure on AdS$_5$, and $L$ its radius. We shall denote by $z$ the bulk coordinate, and by $\vec{\xx}$ the boundary coordinates. 
The index $n$ on $\mathcal{L}_{(n)}$ indicates the number of fields, in particular $\mathcal{L}_{(3)}$ and $\mathcal{L}_{(4)}$ 
contain cubic and quartic interactions among KK modes. These include graviton and gauge fields. Self interactions 
and interactions among different $SO(6)$ representations are mediated by the potential. 
The action is explicitly known up to fourth order \cite{Arutyunov:1999fb}. 
The radius of AdS$_5$ can be set to one because we will not consider curvature corrections.\footnote{ 
Curvature and loop corrections to the supergravity effective action have been discussed in \cite{Metsaev:1987ju} }

To start with, let us focus on a specific mode $\overline{\mathpzc{f}}(z,\vec{\xx}\,)$, among the many in the KK tower.
In the saddle point approximation, valid at large $N$,  the field $\overline{\mathpzc{f}}(z,\vec{\xx}\,)$ propagates 
in the bulk according to its equation of motion, $(\nabla^2-m^2)\, \overline{\mathpzc{f}}= J[\{\mathpzc{f}_{\,k}\}]$, where the source term
$J$ depends on the totality of the fields coupling to $\overline{\mathpzc{f}}$.
The general solution for $\overline{\mathpzc{f}}(z,\vec{\xx}\,)$ can be written in terms of the bulk Green function 
$\mathbb{G}_{bb}$, and the bulk-to-boundary propagator $\mathbb{G}_{b\partial}$, as follows
\bea
\overline{\mathpzc{f}}(z,\vec{\xx}\,)&=&  
							\mathpzc{f}^{0}(z,\vec{\xx}\,)+\int  dz d^4\vec{\xx}\,'\ \mathbb{G}_{bb}(z,\vec{\xx}\, ; z',\vec{\xx}\,'\,) \, J[\{\mathpzc{f}_{\,k}(z',\vec{\xx}\,'\,)\}]\ ,\\
\overline{\mathpzc{f}}^{0}(z,\vec{\xx}\,)&=&  
							\int d^4\vec{\xx}\,' \ \mathbb{G}_{b\partial}(z,\vec{\xx}\,;\vec{\xx}\,')\ \mathpzc{S}(\vec{\xx}\,')\ .
\eea
where $\mathpzc{f}^{0}$ solves the homogeneous equation of motion with boundary conditions $\mathpzc{S}(\vec{\xx}\,')$. 
According to the AdS/CFT correspondence, $\mathpzc{S}(\vec{\xx}\,')$ is identified with the boundary source 
which couples to the operator dual to $\overline{\mathpzc{f}}(z,\vec{\xx}\,)$.
The perturbative expansion around the homogeneous solutions $\{\mathpzc{f}^0_{\,k}(z',\vec{\xx}\,')\}$ defines the corresponding series expansion for $J$, i.e. 
$J=J_{(2)}+J_{(3)}+\ldots$ where the label indicates again the number of boundary fields $\mathpzc{S}_k(\vec{\xx}\,')$ at each order. Finally,
evaluating the action on-shell can be interpreted diagrammatically as summing over tree-level Witten diagrams.
The result is the following generating functional for the boundary sources:
\bea
&&
S_{\rm sugra}= 
\int d^4{\vec{\xx}}_1d^4{\vec{\xx}}_2\ \mathpzc{S}_{k_1} (\vec{\xx}_1) \mathpzc{S}_{k_2}(\vec{\xx}_2)\,  \mathpzc{D}^{(2)}_{k_1k_2}(\vec{\xx}_1,\vec{\xx}_2) + \nonumber \\
&&
\rule{1.4cm}{0pt}
\  \int d^4{\vec{\xx}}_1d^4{\vec{\xx}}_2 d^4{\vec{\xx}}_3 \  \mathpzc{S}_{k_1} (\vec{\xx}_1) \mathpzc{S}_{k_2}(\vec{\xx}_2) \mathpzc{S}_{k_3}(\vec{\xx}_3) \mathpzc{D}^{(3)}_{k_1k_2k_3}(\vec{\xx}_1,\vec{\xx}_2,\vec{\xx}_3) + \nonumber\\
&&
\rule{1.4cm}{0pt}
\ \ \int d^4{\vec{\xx}}_1d^4{\vec{\xx}}_2 d^4{\vec{\xx}}_3d^4{\vec{\xx}}_4\  \mathpzc{S}_{k_1} (\vec{\xx}_1)\mathpzc{S}_{k_2}(\vec{\xx}_2)\mathpzc{S}_{k_3} (\vec{\xx}_3)\mathpzc{S}_{k_4}(\vec{\xx}_4)  \mathpzc{D}^{(4)}_{k_1k_2k_3k_4}(\vec{\xx}_1,\vec{\xx}_2,\vec{\xx}_3,\vec{\xx}_4)+\ldots \notag\\
\label{sugra_schematic}
\eea
Here the functions $\mathpzc{D}^{(i=2,3,4)}(\{ \vec{\xx}_i\})$ are proportional to $N^2$ according to \eqref{sugraIIB}. 
Correlators of $n$ operators can then be computed by taking $n$ functional derivatives w.r.t to the dual sources. 
In particular, 
\beq\label{sugrapathintegral}
\la \mathcal{O}^{I_1}_{p_1}(\vec{x}_1) \mathcal{O}^{I_2}_{p_2}(\vec{x}_2) \mathcal{O}^{I_3}_{p_3}(\vec{x}_3) \mathcal{O}^{I_4}_{p_4}(\vec{x}_4) \ra
= \prod_{n=1}^4\frac{\delta}{\delta\mathpzc{S}_{ n}(\vec{x}_n)}\, e^{- S_{\rm sugra}}\Big|_{\mathpzc{S}_{n}=0}\ .
\eeq
Two-point and three-point functions obtained from AdS supergravity manifestly
agree with CFT expectations \cite{Lee:1998bxa, DHoker:1999jke}. 
In the supergravity conventions all two-point functions are normalized to $N^2$, 
however it is always possible to redefine the sources so as to match the normalization given in \eqref{def_halfBPS}. 

Four-point correlators are more interesting and require quite involved manipulations.
Explicit Witten diagram computations have been carried out in the cases of 
$\la\OO_i\OO_i\OO_i\OO_i\ra$ for $i=2,3,4$ \cite{Arutyunov:2000py,Arutyunov:2002fh,Arutyunov:2003ae}, 
$\la \OO_2\OO_2\OO_q\OO_q\ra$ \cite{Berdichevsky:2007xd,Uruchurtu:2008kp}  
and $\la\OO_{k+2}\OO_{k+2}\OO_{q-k}\OO_{q+k}\ra$ \cite{Uruchurtu:2011wh} for arbitrary $q$ and $k$.
Despite complications, the end result is neat and the generalization to arbitrary dimensions 
has been conjectured in terms of a simple Mellin amplitude \cite{Rastelli:2016nze}, which we review in detail in the next section. 
Indications about the correctness of this conjecture also come from explicit supergravity computations \cite{Arutyunov:2017dti}.

It is clear from the form of $S_{\rm sugra}$ in \eqref{sugra_schematic} that upon taking functional derivatives w.r.t the sources, 
a four-point correlator  will get a leading contribution from disconnected two-point functions $\mathpzc{D}^{(2)}_{k_1k_2}\mathpzc{D}^{(2)}_{k_3k_4}$ (when it exists)
plus the subleading contribution $\mathpzc{D}^{(4)}_{k_1k_2k_3k_4}$, which is $1/N^2$ suppressed. 
The latter will contain a dynamical term with $\log(u)$ singularity, 
but it will also contain a subset of the corresponding free field connected correlator.
Therefore, it will be useful to consider the splitting 
\beq
\GG^{\rm sugra}_{p_1p_2p_3p_4}=\GG_{p_1p_2p_3p_4}^{\rm free} + \GG_{p_1p_2p_3p_4}^{\rm dyna }
\eeq
where $\GG_{p_1p_2p_3p_4}^{\rm free}$ can be computed and studied independently from supergravity. 
This free theory contribution will play an important role for the consistency of the AdS/CFT correspondence,
and will be discussed in Section\,\ref{twist2cancellation}, in the context of the superconformal OPE.
The form of $\GG_{p_1p_2p_3p_4}^{\rm dyna }$ is uniquely given by 
\beq
\GG_{p_1p_2p_3p_4}^{\rm dyna }=\Intri(u,v,\sigma,\tau)\, \mathcal{H}^{\rm dyna}_{p_1p_2p_3p_4}\ ,
\eeq
consistent with partial non-renormalisation~\eqref{SC1}.

\subsection{Tree-Level Supergravity from  Mellin space} 

In \cite{Rastelli:2016nze} the authors conjectured a formula for the function 
$\Hdyna$ at leading order in the classical supergravity approximation. 
With the convention that $p_1\ge p_2\ge p_3\ge p_4$,
\bea
&& \Hdyna_{p_1p_2p_3p_4}=\normap\ u^{d_{12}} v^{d_{23}}\, 
		\frac{(r_{13} r_{24})^{p_2+2} }{r_{13}^{\Sigma-p_1-p_3} r_{14}^{\Sigma-p_1-p_4} r_{34}^{\Sigma-p_3-p_4} }\, \oint \dm \ \MM(s,t,\sigma,\tau)\,\Gamma_{p_1p_2p_3p_4}
		\rule{1cm}{0pt}
		\label{Rastelli1}
\eea
where $\dm= dsdt\prod_{i<j} r_{ij}^{-c_{ij}}$ is the measure in Mellin space and the $c_{ij}$ are given by
\begin{align}
c_{12}&=\frac{p_1+p_2-s}{2} & c_{14}&=\frac{p_1+p_4-t}{2}& c_{24}&=\frac{s+t+4-p_1-p_3}{2}\ ,\\[0.2cm]
c_{34}&=\frac{p_3+p_4-s}{2} & c_{23}&=\frac{p_2+p_3-t}{2}& c_{13}&=\frac{s+t+4-p_2-p_4}{2}\ .
\end{align}
The other quantities are
\beq\label{sigmadef}
\Sigma=\frac{p_1+p_2+p_3+p_4}{2},\qquad \Gamma_{p_1p_2p_3p_4}=\prod_{i<j}\Gamma[c_{ij}]\ .
\eeq
Let us notice that $\sum_{j} c_{ij}=p_i+2$, thus $\Hdyna$ has weight zero under rescalings $r_{ij}\rightarrow\lambda r_{ij}$ and therefore is only function of the cross ratios. 
The Mellin amplitude is
\beq\label{MellinA}
\MM(s,t,\sigma,\tau)=
				\sum_{i+j+k\,=\,d_{34}-2}\ \frac{ a_{ijk}\, \sigma^i\tau^j }{ \rule{0pt}{.4cm} (s-\tilde{s}+2k)(t-\tilde{t}+2j)(\muu - \tilde{\muu}+2i)}\ ,
\eeq
with 
$\mu\equiv 2\Sigma-4 -s-t\,$, 
\bea
\tilde{s}&=&p_3+p_4-2,\\
\tilde{t}&=&{\rm min}\{ p_1+p_4,p_2+p_3\}-2, \label{tm}\\
\tilde{\mu}&=&{\rm min}\{ p_1+p_3,p_2+p_4\}-2 \label{um},
\eea
and
\beq
a_{ijk}= 
		8\frac{(L-2)!}{i!j!k!}\left[ \left( 1+ |\Sigma-p_2-p_4|\right)_i \left( 1+ |\Sigma-p_2-p_3|\right)_j \left( 1+|\Sigma-p_3-p_4|\right)_k \right]^{-1}\  .
\eeq
Finally, $\normap\sim 1/N^2$ is an undetermined normalization. 

The integration contour in Mellin space is taken to lie between the left- and right-moving poles of the Mellin integrand.  
The right-moving poles are defined in the $s$ and $t$ variables and can be found both in the Gamma functions and in the rationals of the Mellin amplitude. 
The left-moving poles are given by expressions involving the combination $s+t$.

Formula \eqref{Rastelli1} is very remarkable, and gives access to four-point correlators of any quadruplet of half-BPS operators. 
It has been obtained as the solution of a bootstrap problem which does not rely on the AdS/CFT correspondence. 
Inputs from the knowledge of tree-level Witten diagram in supergravity have been cleverly encoded in the ansatz for $\MM(s,t,\sigma,\tau)$.
However, there are other consistency checks based on the presence/absence of operators in the spectrum of $\NN=4$ SYM at strong coupling 
that $\Hdyna_{p_1p_2p_3p_4}$ must satisfy. These were not directly used in the bootstrap problem, 
and have to do with $\GG_{p_1p_2p_3p_4}^{\rm free}$. We will see, in the context of the superconformal OPE,
that all these consistency checks are indeed passed and we will use them to determine the overall normalisation, $\normap$ for the cases of interest.

In the remainder of this section we outline a simple algorithm which converts $\Hdyna_{p_1p_2p_3p_4}$ into a sum of 
$\Dbar_{\delta_1\delta_2\delta_3\delta_4}$ functions, with $\delta_i$ depending on the charges $p_j$. 
This rewriting will be advantageous when 
we will look at the OPE decomposition of $\Hdyna_{p_1p_2p_3p_4}$. 
In fact, any $\Dbar_{\delta_1\delta_2\delta_3\delta_4}$ with integer $\sigma\equiv(\delta_1+\delta_2-\delta_3-\delta_4)/2\ge 0$ can be written very explicitly as 
\beq\label{DbaruY}
\Dbar_{\delta_1\delta_2\delta_3\delta_4}= u^{-\sigma}\, \Dbar_{\delta_1\delta_2\delta_3\delta_4}^{\rm\, sing} +  \Dbar_{\delta_1\delta_2\delta_3\delta_4}^{\rm\, analytic}+\,   \log(u)\,\Dbar_{\delta_1\delta_2\delta_3\delta_4}^{\rm\, log}\ , 
\eeq
where
\bea\label{Dsingularseries}
\Dbar_{\delta_1\delta_2\delta_3\delta_4}^{\rm\,sing}&=& 
									\sum_{n=0}^{\sigma-1} \frac{(-u)^{n}}{n!}\ 
									\Gamma[\sigma-n]\ \Lambda^{\delta_3\delta_4}_{\delta_1-\sigma\delta_2-\sigma}(n)\ {F}^{\,\delta_2-\sigma+n | \delta_3+n}_{\, \delta_3+\delta_4+2n}(1-v)\ ,\\
 \label{Dlogseries}
\Dbar_{\delta_1\delta_2\delta_3\delta_4}^{\rm\, log}&=& 
									(-)^{\sigma+1}  \sum_{n=0}^\infty \frac{u^n}{n!(\sigma+n)!}\  
									\Lambda^{\delta_1\delta_2}_{\delta_3+\sigma\delta_4+\sigma}(n)\  {F}^{\, \delta_2+n |\delta_3+\sigma+n}_{\, \delta_1+\delta_2+2n}(1-v)\ ,
\eea
with\footnote{ $\Dbar_{\delta_1\delta_2\delta_3\delta_4}^{\rm\, sing}=0$ when $\sigma=0$.}
\begin{align}
F^{a|b}_c(x)\equiv\,_2 F_1[a,b;c](x),
& &
\Lambda^{\delta_1\delta_2}_{\delta_3\delta_4}(n)\equiv \frac{ \Gamma[\delta_1+n]\Gamma[\delta_2+n]\Gamma[\delta_3+n]\Gamma[\delta_4+n] }{\Gamma[\delta_1+\delta_2+2n]}\ .
\end{align}
The expression for $\Dbar_{\delta_1\delta_2\delta_3\delta_4}^{\rm\, analytic}$ will not be relevant for our discussion, and can be found in Appendix \ref{app-Dfns}.
In Appendix \ref{app-Dfns} we also explain how to relate $\Dbar_{\delta_1\delta_2\delta_3\delta_4}$ to $\Dbar_{1111}$ by the action of certain differential operators. 
Since $\Dbar_{1111}$ has a simple representation in terms of polylogarithms \cite{Usyukina:1992jd},
\beq\label{one-loop-box}
\Dbar_{1111}=-\log(u)\,\frac{\Li_1(\x)-\Li_1(\xb)}{\x-\xb}+2\, \frac{\Li_2(\x)-\Li_2(\xb)}{\x-\xb}\ ,
\eeq
defining $\Dbar_{\delta_1\delta_2\delta_3\delta_4}$ from $\Dbar_{1111}$ provides a resummation of the series expansions in \eqref{DbaruY}.

The conversion algorithm is based on the following observation:
For the Mellin integral attached to a generic monomial $\sigma^i\tau^j$, i.e.
\beq\label{step1}
\frac{(r_{13} r_{24})^{p_2+2} }{r_{13}^{\Sigma-p_1-p_3} r_{14}^{\Sigma-p_1-p_4} r_{34}^{\Sigma-p_3-p_4} }\,  
\oint \dm\  \frac{ \Gamma_{p_1p_2p_3p_4} }{(s-\tilde{s}+2k)(t-\tilde{t}+2j)(\mu - \tilde{\mu}+2i)}\ ,
\eeq
it is possible to identify $s-\tilde{s}$, $t-\tilde{t}$ and $\mu-\tilde{\mu}$ with three out of the six $c_{ij}$ appearing in $\Gamma_{p_1p_2p_3p_4}$. 
Therefore, we can rewrite the integrand as a sum of products of six $\Gamma$ functions with arguments shifted compared to $\Gamma_{p_1p_2p_3p_4}$. 
The precise form of this sum depends on the specific values of $i,j$. For concreteness, let us give a simple example, and assume
that $p_1+p_4\le p_2+p_3$ and $p_1+p_3<p_2+p_4$. From \eqref{tm} and \eqref{um} we find
\begin{align}
s-\tilde{s}= -2 c_{34} +2,& & t-\tilde{t}= -2 c_{14}+2,& & \mu-\tilde{\mu} =-2 c_{13} +2,
\end{align}
therefore
\bea\label{gamma}
&&
\frac{ \Gamma_{p_1p_2p_3p_4} }{(s-\tilde{s}+2k)(t-\tilde{t}+2j)(\mu-\tilde{\mu} +2i)}=\nn\\[.3cm]
&&
\rule{3cm}{0pt} \frac{1}{8}\frac{\Gamma[c_{34}] }{ c_{34} -(k+1) } \frac{\Gamma[ c_{14}] }{ c_{14}- (j+1) } \frac{ \Gamma[c_{13}]}{c_{13}- (i+1) } \Gamma[c_{12}]\Gamma[c_{23}]\Gamma[c_{24}]\ . 
\eea
We now wish to write this as a sum of terms in which  the $c_{ij}$ dependence only appears in the $\Gamma$s. 
To this effect we make use of the identity
\begin{align}
	\frac{\Gamma[c]}{c-k-1} = \sum_{s=1}^{k+1} \Gamma[c-s] \frac{\Gamma[k+1]}{\Gamma[k-s+2]}
\end{align}
to rewrite the first three factors on the right hand side of~\eqref{gamma}. The final expression has the form, 
\beq
\frac{ \Gamma_{p_1p_2p_3p_4} }{(s-\tilde{s}+2k)(t-\tilde{t}+2j)(\hat{u} - \tilde{u}+2i)}=\sum_{\{s_{34},s_{14},s_{13}\}} k_{ \{s_{34},s_{14},s_{13}\}} \Gamma^{shift}_{p_1p_2p_3p_4}[s_{34},s_{14},s_{13}]\,,
\eeq
where the shifts $\{s_{34},s_{14},s_{13}\}$ are integers, and we defined 
\begin{align}
	k_{ \{s_{34},s_{14},s_{13}\}}=\frac{\Gamma[k+1]}{\Gamma[k-s_{34}+2]}\frac{\Gamma[j+1]}{\Gamma[j-s_{14}+2]}\frac{\Gamma[i+1]}{\Gamma[i-s_{13}+2]}
\end{align}
 and
\beq
\Gamma^{shift}_{p_1p_2p_3p_4}[s_{34},s_{14},s_{13}]=\Gamma[c_{34}-s_{34}]\Gamma[ c_{14}-s_{14}] \Gamma[c_{13}-s_{13}] \Gamma[c_{12}]\Gamma[c_{23}]\Gamma[c_{24}]\ .
\eeq
Recalling the definition of $\Dbar_{\delta_1\delta_2\delta_3\delta_4}$ in Mellin space, 
\beq
\Dbar_{\delta_1\delta_2\delta_3\delta_4}= \frac{(r_{13} r_{24})^{\delta_2}}{ r_{13}^{\Sigma'-\delta_1-\delta_3}r_{14}^{\Sigma'-\delta_1-\delta_4} r_{34}^{\Sigma'-\delta_3-\delta_4} }\oint \dm'\ \prod_{i<j} \Gamma[c'_{ij}]\,,\qquad
\sum_{j} c'_{ij}=\delta_i\,,
\eeq
it is now evident that \eqref{step1} can be written as a sum of $\Dbar_{\delta_1\delta_2\delta_3\delta_4}$ functions in which
\beq
\begin{array}{ccl}
\delta_1&=& p_1+2-s_{14}-s_{13},\\
\delta_2&=&p_2+2,\\
\delta_3&=&p_3+2-s_{13}-s_{34},\\
\delta_4&=&p_4+2-s_{14}-s_{34}.
\end{array}
\eeq
In this case, $\Sigma'-\delta_i-\delta_j=\Sigma-p_i-p_j+s_{ij}$, for $(i,j)=\{(1,3),(1,4),(3,4)\}$, 
and the dependence on the shifts cancel against $\dm'/\dm.$ 
In the most generic case, 
identities among $\Dbar$ functions (see Appendix \ref{app-Dfns}) might be needed in order to obtain an expression in which $\sigma\ge0$.

Once implemented, the algorithm generates a $\Dbar_{\delta_1\delta_2\delta_3\delta_4}$ representation of $\Hdyna_{p_1p_2p_3p_4}$ for arbitrary values of $p_1\ge p_2\ge p_3\ge p_4$. 
A four-point correlator with a generic configuration of charges can then be obtained upon acting with permutation symmetries of $\GG_{p_1p_2p_3p_4}^{\rm dyna }$ and  the prefactor~\eqref{pre}.
We list some examples relative to $\Hdyna_{pp qq}$ with $p\le q$,
\begin{align}
\label{22nnexample}
\HH_{22\n\n}^{\rm dyna}=\ u^\n\, \ &
							\, \Dbar_{\n,\n+2,2,2}\,,\\[.3cm]
\HH_{33\n\n}^{\rm dyna}=\ u^\n \Big[&
\,\sigma \Dbar_{\n-1,\n+2,2,3}  +\tau  \Dbar_{\n-1,\n+2,3,2} \nn \\[0cm]
&  +\frac{1}{\n-2}  \Dbar_{\n,\n+2,2,2} + \left( \frac{1}{\n-2} +\sigma+\tau\right)  \Dbar_{\n,\n+2,3,3}\Big]\,,
\label{33nnexample}	\\[.3cm]
\HH_{44\n\n}^{\rm dyna}=\ u^\n \ \bigg[& 
2\sigma\tau \left( \Dbar_{\n-2, \n+2, 3, 3} + \Dbar_{ \n-1, \n+2, 3, 4} + \Dbar_{ \n-1,  \n+2, 4, 3}+ \Dbar_{\n,  \n+2, 4, 4} \right)\nn\\
&+\sigma^2 \left( \Dbar_{ \n-2, \n+2, 2, 4} + \Dbar_{\n-1, \n+2, 3, 4} + \frac{1}{2}\Dbar_{\n, \n+2, 4, 4}\right)\nn\\					
&+\tau^2 \left( \Dbar_{ \n-2,  \n+2, 4, 2} + \Dbar_{ \n-1,  \n+2, 4, 3} + \frac{1}{2} \Dbar_{\n,  \n+2, 4, 4}\right)\nn\\
&+\frac{2}{\n-3} \sigma \left( \Dbar_{ \n-1, \n+2, 2, 3} + \Dbar_{\n-1, \n+2, 3, 4} + \Dbar_{\n, \n+2, 3, 3} + \Dbar_{\n, \n+2, 4, 4}\right)\nn\\
&+\frac{2}{\n-3} \tau \left( \Dbar_{ \n-1, \n+2, 3,2} + \Dbar_{\n-1, \n+2, 4, 3} + \Dbar_{\n, \n+2, 3, 3} + \Dbar_{\n, \n+2, 4, 4}\right)\nn\\
&+\frac{2}{(\n-2)(\n-3)}\, \left( \Dbar_{\n, \n+2, 2, 2} +  \Dbar_{\n,\n+2, 3, 3} + \frac{1}{2} \Dbar_{\n, \n+2, 4, 4}\right) \bigg]\ .
\label{44nnexample}
\end{align}
The overall $u^\n$ has a simple interpretation, 
and in fact it will imply that only long multiplets with twist $\ge 2\n$ contribute to the leading $\log(u)$ singularity. 
The set of correlators of the form $\la\OO_p\OO_p\OO_\n\OO_\n\ra$, with $\n\ge p$, 
is perhaps the simplest generalization of the conjecture \cite{Dolan:2006ec} for $\n=p$. 
In particular, for fixed value of $p$, the combination of $\Dbar_{\delta_1\delta_2\delta_3\delta_4}$ 
functions attached to each monomial $\sigma^i\tau^j$ changes according to a simple pattern in $\n$.  
This pattern is already visible at $p=5$, 
\begin{align}
\HH_{55\n\n}^{\rm dyna}=u^\n\ \bigg[&
\sigma^3 \sum_{k=0}^3 \frac{1}{k!} \Dbar_{\n-3+k,\n+2,2+k,5}+ \tau^3 \sum_{k=0}^3 \frac{1}{k!} \Dbar_{\n-3+k,\n+2,5,2+k}\nn\\
&
+3\sigma^2\tau \sum_{k=0}^2 \frac{1}{k!} \left( \Dbar_{ \n-3+k,  \n+2, 3+k, 4} + \Dbar_{ \n-2+k,  \n+2, 3+k, 5}  \right)\nn\\
&+3\tau^2\sigma \sum_{k=0}^2 \frac{1}{k!} \left( \Dbar_{ \n-3+k,  \n+2, 4,3+k} +  \Dbar_{ \n-2+k,  \n+2, 5,3+k}  \right)\nn\\
& +\frac{3\sigma^2}{\n-4}  \sum_{k=0}^2 \frac{1}{k!} \left( \Dbar_{\n-2+k, \n+2, 2+k, 4} + \Dbar_{\n-2+k,\n+2,3+k,5}\right)\nn\\
& +\frac{3\tau^2}{\n-4}   \sum_{k=0}^2 \frac{1}{k!} \left( \Dbar_{\n-2+k, \n+2, 4,2+k} + \Dbar_{\n-2+k,\n+2,5,3+k}\right)\nn\\
&+\frac{6\sigma\tau}{\n-4} \Big(  \Dbar_{\n, \n+2, 4, 4}+\Dbar_{\n,\n+2,5,5}+\Dbar_{\n-1,\n+2,4,5}+\Dbar_{\n-1,\n+2,5,4}\nn\\
& \rule{2cm}{0pt}\Dbar_{\n-2,\n+2,3,3}+\Dbar_{\n-2,\n+2,4,4}+ \Dbar_{\n-1,\n+2,3,4}+ \Dbar_{\n-1,\n+2,4,3}\Big)\nn \\
&+\frac{6\sigma } {(\n-4)(\n-3)}  \sum_{k=0}^2 \frac{1}{k!} \left( \Dbar_{\n-1,\n+2,2+k,3+k}+\Dbar_{\n,\n+2,3+k,3+k} \right)\nn\\
&+\frac{6\tau } {(\n-4)(\n-3)}  \sum_{k=0}^2 \frac{1}{k!} \left( \Dbar_{\n-1,\n+2,3+k,2+k}+\Dbar_{\n,\n+2,3+k,3+k} \right)\nn\\
&+\frac{6}{(\n-4)(\n-3)(\n-2)} \sum_{k=0}^3 \frac{1}{k!} \Dbar_{\n,\n+2,2+k,2+k}\bigg]
\label{55nnexample}
\end{align}
Finally, let us notice again the simplicity of the Mellin amplitude \eqref{MellinA} compared to the $\Dbar_{\delta_1\delta_2\delta_3\delta_4}$ representation of the correlator.

\section{$\mathcal{N}=4$ superconformal OPE }
\label{SC_OPE}

As will be explained further in section~\ref{sec4},  we need to perform a superconformal block decomposition of the leading 
and subleading in $1/N^2$ correlators $\langle \cO_{p_1} \cO_{p_2} \cO_{p_3} \cO_{p_4} \rangle$. There has been a great deal of work on superconformal blocks in $\cN=4$ SYM both from the pioneering work of Dolan and Osborn~\cite{Dolan:2000ut,Dolan:2001tt,Dolan:2002zh,Dolan:2004iy} and more recently~\cite{Bissi:2015qoa} as well as supergroup theoretic approaches~\cite{Heslop:2002hp,Doobary:2015gia}.  In this section we review the methods we use in this paper.

The OPE of two half-BPS operators is 
\beq
\OO_{p_1}(x_1)\OO_{p_2}(x_2) \sim 
							\sum_{\OOL}\, g_{12}^{\frac{p_1+p_2-\Delta}{2}}\ C_{p_1p_2 \OOL }\  \mathscr{L}^{(l)}(x_{12},\partial_{x_2})*\OOL^{(l)}(x_2)\ .
\eeq
The sum runs over all primary operators
$\OOL^{(l)}$ of dimension $\Delta$, spin $l$, which belong to the $SU(4)$ representations,
\beq\label{SU4OPE}
[0,p_1,0]\otimes[0,p_2,0] =
					\sum_{k_1=0}^{p_1}\sum_{k_2=0}^{p_1-k_1}[\, k_1,p_{2}-p_1+2k_2,k_1\,] \qquad (p_1\le p_2)
\eeq
Descendants are obtained by the action of the derivative operator $\mathscr{L}_{}^{(l)}(x_{12},\partial_{x_2})$ on the primaries. 
A manifest $\NN=4$ formulation of the OPE can be obtained by reorganizing the sum over operators into supermultiplets. Therefore,
inserting the OPE of $\OO_{p_1}(x_1)\OO_{p_2}(x_2)$ and $\OO_{p_3}(x_3)\OO_{p_4}(x_4)$ into the four-point correlator we obtain the representation
\bea
\la \OO_{p_1}\OO_{p_2}\OO_{p_3}\OO_{p_4}\ra&=& \Pref^{\rm (OPE)}_{\,\{p_i\}}
\sum_{\{\tau,\,l,\,\mathfrak{R}\}} A^{\{p_i\}}_{\,\mathfrak{R}}(\tw |l)\  \SCW_{\,\mathfrak{R}}^{\{p_i\}}(\tw |l)
\label{SCOPE1}
\eea
where $\tw=(\Delta-l)/2$ and $\SCW_{\,\mathfrak{R}}^{\{p_i\}}(\tw |l)$ are superconformal blocks described below.
Here the sum over representations runs over those which belong to $([0,p_1,0]\otimes[0,p_2,0])\cap([0,p_3,0]\otimes[0,p_4,0])$.
The coefficients $A^{\{p_i\}}_{\,\mathfrak{R}}(\tw |l)$ depend explicitly on the charges and are related to the OPE coefficients by
\beq
A^{\{p_i\}}_{\,\mathfrak{R}}(\tw |l)\ = \sum_{\OOL\in\,\mathfrak{R}} C_{p_1p_2 \OOL } C_{p_3p_4 \OOL}\ ,
\label{SCOPE2}
\eeq
where the sum is over all operators with spin $l$, leading order  dimension $\Delta$ and $SU(4)$ representation $\mathfrak{R}$. 
The prefactor $ \Pref^{\rm (OPE)}$ depends on the ordering of the charges. 
The block decomposition is invariant under swapping points 1 and 2, points 3 and 4 and swapping the pairs of points 1,2 and 3,4. 
Using this symmetry we can clearly always ensure that $p_2\geq p_1$, $p_4 \geq p_3$ and $p_2-p_1 \leq p_4-p_3$. 
Assuming such an ordering, then the following diagram exists in the free theory,\\ 
\begin{minipage}{\linewidth}
          \begin{center}
            \begin{tikzpicture}
		\def\x1{1}
		\def\y1{1}
		\def\colore{black}

		\draw (\x1+1,\y1+2.5) node {\rule{0pt}{.5cm}};
		
		\draw[\colore,very thick] (\x1,\y1)--(\x1+2,\y1+2) 		node[left=1cm, below=1.2cm] {\scriptsize p$_2-$d} ;
		\draw[\colore,very thick] (\x1,\y1)--(\x1,\y1+2) 			node[below=1cm, left=0pt] {\scriptsize d} ;
		\draw[\colore,very thick] (\x1,\y1+2)--(\x1+2,\y1+2)		node[below=.2cm,left=.5cm] { \scriptsize p$_1-$d};  
		\draw[\colore,very thick] (\x1+2,\y1+2)--(\x1+2,\y1)  		node[above=1cm, right=0pt] {\scriptsize p$_3$};
		
		\draw[white,very thick] (\x1,\y1)--(\x1,\y1-1);
	
		\filldraw (\x1,\y1)		 circle (.25);
		\filldraw (\x1+2,\y1) 		circle (.25);
		\filldraw (\x1,\y1+2) 		circle (.25);
		\filldraw (\x1+2,\y1+2) 	circle (.25);
	
		\draw[text=white,text centered] (\x1,\y1)         	node     {\footnotesize p$_2$};
		\draw[text=white,text centered] (\x1+2,\y1)     	node     {\footnotesize p$_3$};
		\draw[text=white, text centered] (\x1,\y1+2)     	node	    {\footnotesize p$_1$};
		\draw[text=white,text centered] (\x1+2,\y1+2) 	node     {\footnotesize p$_4$};
		\node at (7,2) {$d:=\frac{p_1+p_2+p_3-p_4}{2}$};
		\draw (\x1+1,\y1+2.5) node {\rule{0pt}{.5cm}};
		
	      \end{tikzpicture}
           \end{center}
       \end{minipage}
We then take the prefactor as represented by this diagram
\beq\label{OPEprefactor}
\Pref^{\rm (OPE)}_{\,\{p_i\}}=
g_{12}^{d} g_{14}^{p_1-d}g_{24}^{p_2-d}g_{34}^{p_3} \qquad \text{with} \quad p_2 \geq p_1,\ p_4 \geq p_3, \ p_2{-}p_1 \leq p_4{-}p_3\ . 
\eeq
Comparing this with the prefactor (and corresponding diagram) taken out of the supergravity correlator~\eqref{pre}, we see that up to the appropriate permutation of points, the prefactors are the same.

Finally the superconformal blocks themselves,  $\SCW_{\,\mathfrak{R}}^{\{p_i\}}(\tw |l)$,  can be derived using a variety of approaches and were first derived in~\cite{Dolan:2001tt,Dolan:2006ec}. Here we explain them in  a compact fashion in terms of representations of $GL(2|2)$, following~\cite{Heslop:2002hp,Doobary:2015gia} as we now review.

Instead of the cross ratios $u,v,\sigma,\tau$, it will be  useful to use the variables $x_1, x_2$ and $y_1, y_2$: 
\beq
u= \x\xb,\quad v=(1-\x)(1-\xb),\qquad {\sigma}=\frac{1}{\y\yb}, \quad {\tau}= \left(1-\frac{1}{\y}\right)\left(1-\frac{1}{\yb}\right)\ .
\eeq
In terms of these, the degree two polynomial \eqref{SC1}, singled out from the ``partial non-renormalization" theorem,  becomes fully factorized:
\bea
\Intri(u,v,\sigma,\tau) &=& v+ \sigma^2 u v + \tau^2 u + \sigma v (v-1-u) + \tau (1-u-v)+\sigma\tau u (u-1-v)\nn \\[.2cm]
				&=&	\frac{ \left( \x-\y \right) \left( \x-\yb \right) \left( \xb - \y \right) \left( \xb - \yb \right) }{ (\y\yb)^2 }
\eea
Note that $x_{i=1,2}$ and $y_{i=1,2}$ are not to be confused with the space-time variables and internal harmonic variables that were introduced in previous sections. 
The above variables are conformally invariant.  

\subsection{$GL(2|2)$ superconformal partial wave}

Conformal blocks and $SU(4)$ harmonics are commonly introduced in the literature as \cite{Dolan:2000ut,Dolan:2004iy}
\bea
\label{Confblock}
\CB^{\,\tw|l}&=&(-)^l\ 
				 \frac{\x^{\tw+l+1}\, \xb^{\tw}\ \bF_{\tw+l }(\x)\bF_{\tw-1}(\xb)- \xb^{\tw+l+1}\, \x^{\tw}\ \bF_{\tw-1}(\x)\bF_{\tw+l }(\xb) }{\x-\xb}\\[.2cm]
 \label{SU4harm}
Y_{nm}&=&	
			 - \frac{ \bP_{n+1}(\y) \bP_m(\yb)-  \bP_m(\y) \bP_{n+1}(\yb)  }{\y-\yb}
\eea
where ${\bf F}_{\tw}$ is related to$\,\,_2F_1[a,b;c]$ hypergeometrics
and $\bP_n$ is related to Jacobi polynomials ${\rm JP}^{(a|b)}_c$ through the definitions 
\begin{align}
\bF_{\tw}(x)=
			\,_2F_1\left[\tw- \frac{p_{12} }{2},\tw+\frac{p_{34}}{2};2\tw\right](x),  & &
\bP_{n}(y)=
			\, {y}\, {\rm JP}^{(p_1-d_{12}|p_2-d_{12})}_n\left(\frac{2}{y}-1 \right)\ .
\end{align}
In particular, $Y_{nm}$ with $m\leq n$ is a polynomial of degree $n$ in $(\sigma,\tw)$, and $\CB^{\tw|l}$ is analytic in $u$ and $(1-v)$, i.e 
\beq\label{blockmellin}
\CB^{\,\tw|l}
		       = \sum_{n\ge0 \rule{0pt}{.25cm}}\sum_{m\ge{\rm max}(0,l-2n) \rule{0pt}{.25cm} } r_{nm}[\tw,l,p_{12},p_{34}]\, u^{\tw+n} (1-v)^m\ .
\eeq
The series expansion of $\CB^{\,\tw|l}$ begins with leading term $u^{\tw}(1-v)^l$ where $\tw=(\Delta-l)/2$ is half the value of the twist.

$\NN=4$ representations and the corresponding superconformal blocks have been studied extensively in the literature
\cite{Andrianopoli:1998ut, Eden:2001ec,Heslop:2001zm,Dolan:2002zh,Heslop:2002hp,Dolan:2004iy,Bissi:2015qoa,Doobary:2015gia}. 
They can be written as specific linear combinations of terms of the form $\CB \times Y$ corresponding to the component fields appearing in the multiplet. 
This way of writing it depends strongly on the type of multiplet and in particular its shortening conditions. 
A more group theoretic approach was taken in \cite{Doobary:2015gia} giving a uniform description of all superconformal blocks 
via a determinantal formula associated to a $GL(2|2)$ Young tableau which  we review now.

In this approach, an operator is defined on analytic superspace by specifying a $GL(2|2)$ representation via a Young tableau, 
$\ula$, together with a charge $\gamma$, $\cO^{\gamma,\ula}$. The allowed Young tableaux have the general shape%
\footnote{Here we specify the row lengths with the notation $2^\mu$ denoting $2,2,\dots,2$, with $\mu$ entries in the list, that is  $\mu$ rows of length 2.} $\ula=[\lambda_1,\lambda_2,2^{\mu_2},1^{\mu_1}]$.

\begin{align}\label{hook}
\begin{tikzpicture}[scale=.5]
\draw (0,10) -- (20,10) -- (20,9) -- (0,9) -- (0,10);
\node at (10,9.5) {$\leftarrow\lambda_1\rightarrow$};
\draw (0,9) -- (16,9) -- (16,8) -- (0,8) -- (0,9);
\node at (8,8.5) {$\leftarrow\lambda_2\rightarrow$};
\draw (0,8) -- (2,8) -- (2,3) -- (0,3) -- (0,8);
\node at (1,5.7) {$\begin{array}{c}\uparrow\\ \mu_2\\ \downarrow \end{array}$};
\draw (0,3) -- (1,3) -- (1,0) -- (0,0) -- (0,3);
\node at (.55,1.5) {$\begin{array}{c}\uparrow\\ \mu_1\\ \downarrow \end{array}$};
\end{tikzpicture}
\end{align}

The translation to standard quantum numbers depends on the precise shape and is summarised by the table below:
%
%
\begin{align}\label{table}
\begin{array}{|c||c|c|c|c|}
\multicolumn{5}{c}{ 
\begin{array}{l}
\text{ Translation between $\cN=4$  superconformal reps and superfields $\cO^{\gamma\ula}$}  \\
\rule{0pt}{0cm}
\end{array}
}\\
\hline
GL(2|2) \text{ rep }\ula     					&    (\Delta{-}l)/2 			& l     			&  \mathfrak{R} 			  	& \text{multiplet type} \\\hline
[0]                                     					&    \gamma/2          			&0    				& [0,\gamma,0]                            	& \text{half BPS}       \\\hline
\left[1^\mu\right]                                 		 	& \gamma/2             			&0				&   [\mu,\gamma{-}2\mu,\mu]		&\text{quarter BPS} \\ 
\left[\lambda, 1^\mu\right]\ (\lambda\geq 2)   	& \gamma/2             			&\lambda{-}2      	&  [\mu,\gamma{-}2\mu{-}2,\mu] 	&  \text{semi-short}  \\\hline
{ [\lambda_1,\lambda_2,2^{\mu_2},1^{\mu_1}]\  (\lambda_2\geq 2)} 
									& \gamma/2{+}\lambda_2{-}2	&\lambda_1{-}\lambda_2&[\mu_1,\gamma{-}2\mu_1{-}2\mu_2{-}4,\mu_1]&\text{long} \\ \hline
\end{array}
\end{align}
Note that in the case of long multiplets the description of a superconformal representation in terms of $\cO^{\gamma \ula}$ is not unique. Indeed if $\mu_2>2$ then the representation is unchanged if we map 
\begin{align}\label{longdegen}
	\lambda_1 \rightarrow \lambda_1 +1, \ 	\lambda_2 \rightarrow \lambda_2 +1, \ \mu_2\rightarrow \mu_2-1,\  \gamma\rightarrow \gamma-2.
\end{align}
The leading term in the long multiplet $\cO^{\gamma\ula}$ can be written schematically in the form $  \partial^{\lambda_1-\lambda_2} \Box^{\lambda_2-2} \phi^\gamma|_\mathfrak{R}$. Then the above degeneracy in the description of long reps is a reflection of the fact that this is the same representation as  $ \partial^{\lambda_1-\lambda_2} \Box^{\lambda_2-1} \phi^{\gamma-2}|_\mathfrak{R}$.

A further point is that Young tableaux only make sense for integer values of the row lengths. However one can analytically continue the long representations to non-integer values. This is possible because all the long $SL(2|2)$ representations have the same dimension. Specifically we formally allow the first two row lengths $\lambda_1, \lambda_2$ to be non-integer, with the difference $\lambda_1-\lambda_2$ remaining integer. This then allows for anomalous dimensions. We even formally allow the case $\lambda_2 \rightarrow 1$ when $\mu_2=0$. This corresponds to a representation approaching the unitary bound. In the limit when $\lambda_2=1$  multiplet shortening occurs. So as representations, a long rep in this limit splits into two short reps\footnote{We here only consider those representations the four-point function detects.}. Specifically
\begin{align}
	\lim_{\lambda_2\rightarrow 1} \cO^{\gamma [\lambda_2+l,\lambda_2,1^{\mu_1}]}= \cO^{\gamma\, [l+1,1^{\mu_1+1}]} \oplus \cO^{\gamma{-}2\, [l+2,1^{\mu_1}]}\ .
\end{align}

The superconfomal block in all cases is given by the following  determinantal formula 
 \begin{align}\label{detform}
\SCW_{\,\mathfrak{R}}^{\{p_i\}} &= 
						\left(\frac{x_1x_2}{y_1y_2}\right)^{\frac12(\gamma-p_4+p_3)} F^{\alpha\beta\gamma\ula} \qquad 
						\gamma = p_4-p_3, p_4-p_2+2,\dots, {\rm min}(p_1+p_2,p_3+p_4)  \notag\\[.2cm]
F^{\alpha\beta\gamma\ula}\ &=\  (-1)^{ p + 1} D^{-1}
									\det \left( \begin{array}{cc}
															F^X_{\underline\lambda}&R\\[.1cm]
															K_{\underline \lambda}   &   F^Y
											\end{array} \right)\ ,
\end{align}
where the matrix has dimension $(p+2)\times (p+2)$ with
\begin{align}\label{def_alphabeta}
p&=\min\{\alpha,\beta\}, \qquad \alpha=\tfrac12(\gamma-p_{1}+p_2), \quad \beta=\tfrac12( \gamma +p_{3}-p_{4}), 
\end{align}
and for given $\alpha,\beta,\gamma$, and Young tableaux $\ula$, the matrix elements are defined as follows
\begin{align}
(F^X_{\underline \lambda})_{ij}&=\Big([x_i^{\lambda_j  - j}
{}_2F_1( \lambda_j + 1 - j+\alpha, 
\lambda_j + 1 - j+\beta;  2 \lambda_j + 2 - 2 j+\gamma; 
x_i)] \Big)_{  \substack{
		1\leq i
		\leq 2\\
		1\leq j\leq p
	}
}
\nonumber\\[.2cm]
(F^Y)_{ij}&=\Big((y_j)^{i - 1}
{}_2F_1(i -\alpha, i -\beta; 
2 i -\gamma; y_j)  \Big)_{\substack{1\leq i\leq p\\1\leq j
		\leq 2}}
\nonumber\\[.2cm]
(K_\ula)_{ij}&=\Big( -\delta_{i;\,j{-}\lambda_j}\Big)_{\substack{1\leq i\leq p\\ 1\leq j  \leq p}}
\nonumber\\[.3cm]
R&=\left(\begin{array}{cc}
\frac1{x_1-y_1}&\frac1{x_1-y_2}\\[.1cm]
\frac1{x_2-y_1}&\frac1{x_2-y_2}
\end{array}
\right) \qquad 
D=\frac{ (x_1-x_2)	(y_1-y_2)}{(x_1-y_1)(x_1-y_2)(x_2-y_1)(x_2-y_2)  }
\end{align}
The square bracket around the components of $F^X$ indicate that only the regular  part should be taken. So if $\lambda_j<j$ one has to subtract off the first few terms in the Taylor expansion of the hypergeometrics.

This formula as written deals with all cases. Note that the determinant yields a sum of terms each of which contain at most two hypergeometrics in $x_i$ (from the first two rows) and two in $y_i$ (from the last two columns). 
When the determinant is  expanded out, the formula yields different forms depending on whether the multiplet is 1/2 BPS, short or long, due to the different nature of the matrix $K_{\ula}$ in each case. 
All cases can be written in terms of a two-variable or four-variable function. In this paper however we will not need the explicit forms in all cases. Instead we only need the superconformal blocks for long operators, which we use to perform the block expansion of the interacting piece of the correlator~$\HH$. For the free correlator we use an alternative approach outlined in the next section which turns out to be very efficient and much less complicated than the one outlined here. It is particularly useful for performing the free theory analysis which contains all the short multiplets. The study of short multiplets is technically  the most challenging from a superblock point of view.

\subsubsection{Explicit form of the long $\mathcal{N}=4$ superconformal blocks.}

The long multiplets all have Young tableaux containing a two by two block. They thus have Young tableaux of the form $$\ula=[\lambda_1,\lambda_2,2^{\lambda'_2-2},1^{\lambda'_1-\lambda'_2}]$$
That is the first and second rows have length $\lambda_1,\lambda_2$ respectively and the first and second columns  have height $\lambda'_1,\lambda'_2$ respectively,  with $\lambda_1,\lambda_2,\lambda'_1,\lambda'_2 \geq 2$.
In this case the determinantal formula factorises yielding
\begin{align}\label{flong}
	F_{\text{long}}^{\alpha\beta\gamma\ula}(x|y)=(-1)^{\lambda'_1+\lambda'_2}(x_1-y_1)(x_1-y_2)(x_2-y_1)(x_2-y_2)&\times\frac{F_{\lambda _1}^{\alpha \beta \gamma }\left(x_1\right) F_{\lambda _2-1}^{\alpha \beta \gamma
		}\left(x_2\right)- x_1\leftrightarrow x_2}{x_1-x_2}\notag\\&\times\frac{G_{\lambda' _1}^{\alpha \beta \gamma }\left(y_1\right)
	G_{\lambda' _2-1}^{\alpha \beta
	\gamma }\left(y_2\right)-y_1
\leftrightarrow
y_2}{y_1-y_2}
\end{align}
where 
\begin{align}
	F^{\alpha\beta\gamma}_\lambda(x)& :=
	x^{\lambda-1}{}_2F_1(\lambda+\alpha,\lambda+\beta;2\lambda+\gamma;x)\notag\\
	G^{\alpha\beta\gamma}_{\lambda'}(y)&:=
	y^{\lambda'-1}{}_2F_1(\lambda'-\alpha,\lambda'-\beta;2\lambda'-\gamma;y)\ .
\end{align}
From table~\eqref{table}, this gives the superblock corresponding to a long multiplet of half twist $t=\gamma/2+\lambda_2-2$,  spin $l=\lambda_1-\lambda_2$ and $SU(4)$ rep $\mathfrak{R}=[\lambda'_1-\lambda'_2,\gamma-2\lambda'_1,\lambda'_1-\lambda'_2]$.

This can be straightforwardly converted into a $\mathcal{B}\times Y$ notation. From the defintion of $\mathcal{B}$ introduced in \eqref{Confblock} we immediately recognize that
\bea
&&
\frac{F_{\lambda _1}^{\alpha \beta \gamma }\left(x_1\right) F_{\lambda _2-1}^{\alpha \beta \gamma
	}\left(x_2\right)- x_1\leftrightarrow x_2}{x_1-x_2} = 
\frac{1}{ (x_1 x_2)^{2+\frac{\gamma}{2} } } \ \mathcal{B}^{t{+}2|l}\ .
\eea
Similarly, from the definition of $\alpha,\beta$ given in \eqref{def_alphabeta}, and an hypergeometric identity\footnote{
The following identity might be useful
\beq\label{helpful_identity}
\,_2F_1\left[\lambda-\alpha,\lambda-\beta,2\lambda\right](y)
					= y^{-\lambda+\beta} \frac{n!}{(n-2\beta+1)_n} {\rm JP}^{(-\alpha-\beta|\alpha-\beta)}_n\left(\frac{2}{y}-1\right),\quad n\equiv\beta-\lambda\ .
\eeq},
we find 
\bea
&&
-\frac{ G_{\lambda' _2-1}^{\alpha \beta \gamma }\left(y_1\right)G_{\lambda' _1}^{\alpha \beta \gamma }\left(y_2\right)  -y_1 \leftrightarrow y_2}{y_1-y_2} = 
(y_1 y_2)^{ \frac{\gamma}{2} +\frac{p_{34}}2 -2} \frac{(n+1)!m!}{ \rule{0pt}{.45cm} (n{+}2{+}p_{43} )_{n+1} (m{+}1{+}p_{43} )_m } Y_{nm}\qquad
\eea
where we used the definition of $Y_{nm}$ in \eqref{SU4harm} and the identification,
$$m=p_{34}/2+\gamma/2-\lambda'_1,\qquad n=p_{34}/2+\gamma/2-\lambda'_2
$$
The $SU(4)$ representation is then
 $[n-m, 2m+p_{43} ,n-m]$. It is also convenient to define the normalized $SU(4)$ harmonics as follows, 
\be
\Upsilon_{nm}=  \frac{(n+1)!m!}{ \rule{0pt}{.45cm} (n{+}2{+}p_{43} )_{n+1} (m{+}1{+}p_{43} )_m } Y_{nm}\ .
\ee

Including the prefactors from the definition of  $\SCW$, 
and relabelling $\SCW \rightarrow \LL$ to highlight that this is a long operator,  the long superblock then becomes
\bea \label{longmultiplet}
\LL^{\{p_i\}}_{\,nm}(\tw|l)& =& 
\frac{ (\x-\y)(\x-\yb)(\xb-\y)(\xb-\yb)}{  \rule{0pt}{.45cm}  (\y\yb)^2}\, 
\frac{ \CB^{\,2+\tw|l} }{   \rule{0pt}{.45cm} u^{2+ \frac{ p_{43}}{2}  } } \,
\Upsilon_{nm}\ .
\eea

\subsection{Bosonised superblocks}
\label{bosonisedsuperblocks}

Here we outline a novel approach to performing a SCPW analysis, particularly useful for the free theory in $\cN=4$. It is based on the approach of~\cite{Doobary:2015gia}, outlined above and based on analytic superspace: The key observation there is that  superconformal blocks in {\it{generalised}} analytic superspace with $SU(m,m|2n)$ symmetry exhibit a universal structure, thus one can map the correlation functions into a generalised analytic superspace with $SU(m,m|2n)$ symmetry for any $m,n$, perform the appropriate  superblock expansion, and the block coefficients thus obtained will be the same as the ones you would have obtained had you performed the expansion in the original space. In particular it is convenient to map the problem to the generalised conformal group $SU(m,m)$ with $n=0$. 

As we have seen the free theory 4-point function of any four 1/2 BPS operators is given as a sum of products of powers of the superpropagators  $g_{ij}$~\eqref{diagram_correlator}. Now each term in the free theory contains information about operators $\cO^{\gamma\,\ula}$ for a specific value of $\gamma$, namely
\begin{align}\label{gammrel}
\gamma=d_{13}+d_{14}+d_{23}+d_{24}.
\end{align} 
Note that graphically $\gamma$ is simply the number of propagators going from the pair of points 1,2 to the pair 3,4.  Explicitly,  every term in the free theory can be written as
\begin{align}\label{freeterm}
\prod_{i<j}g_{ij}^{d_{ij}}\ =\ 
\Pref^{\rm (OPE)}_{\,\{p_i\}} \times \left(\frac{g_{13}g_{24}}{g_{12}g_{34}}\right)^{\frac12(\gamma-p_4+p_3)} \times \left(\frac{g_{14}g_{23}}{g_{13}g_{24}}\right)^{d_{23}}
\end{align} 
where $\Pref^{\rm (OPE)}_{\,\{p_i\}} $ is the prefactor of~\eqref{OPEprefactor} (with the ordering of the operators as chosen there). Now the second factor in this equation is precisely the factor appearing in front of the superconformal blocks in~\eqref{detform}. Thus the superblock decomposition reduces to the problem of decomposing the final factor in~\eqref{freeterm} in terms of superconformal blocks $F^{\alpha\beta\gamma\ula}$.
This final factor is simply
\begin{align}\label{freeterm2}
\left(\frac{g_{14}g_{23}}{g_{13}g_{24}}\right)^{d_{23}}	= \left(\frac{(1-y_1)
		(1-y_2)}{(1-x_1)(1-x_2)}\right)^{d_{23}} = \text{sdet}^{-{d_{23}}}(1-Z) \qquad \quad Z\sim \left( \begin{array}{cc|cc}
	x_1&&&\\&x_2&&\\ \hline&&y_1&\\&&&y_2
	\end{array}\right)\,,
\end{align}
where we write the result in terms of a diagonal $GL(2|2)$ matrix $Z$. 

In the conventional approach one would then simply expand this in terms of $SU(2,2|4)$ superconformal blocks, $F^{\alpha\beta\gamma\ula}$ as described in the previous section, to obtain information about operators $\cO^{\gamma \ula}$. However, the universal structure of $SU(m,m|2n)$ blocks alluded to above suggests an alternative approach, namely, using blocks in a theory with $SU(m,m|2n)$ for different values of $m,n$ compared to the $\cN=4$ SYM case. Any value of $m,n$ will give accurate information on the block coefficients, but not necessarily complete information. This happens because now the block decomposition will only yield information on operators $\cO^{\gamma \ula}$ where $\ula$ is a valid non-zero $GL(m|n)$ Young tableau. For example, choosing $m=2,n=0$, means we consider   $GL(2)$ Young tableaux, which will only give information about operators with  maximally two row tableaux.  Whereas choosing $m=3,n=0$ will give information on operators whose Young tableaux have up to three rows etc. On the other hand one could consider $m=0$. Then we are considering $SU(0|n)$ tableau which are just ``transposed" $SU(n)$ tableau, where columns and rows are swapped.%
\footnote{This is simply because generalised symmetrisation of  odd indices for a supergroup corresponds to anti-symmetrisation.}
Thus Young tableaux in the $m=0$ case have maximally $n$ columns.

In $\cN=4$ we have $GL(2|2)$ Young tableau which have the hook structure given in~\eqref{hook}, namely at most two rows have length greater than two and at most two columns have length greater than two. On the other hand expanding the structure in~\eqref{freeterm2} in terms of super Schur polynomials using the Cauchy identity (see~\cite{Doobary:2015gia} for details in this context), one can see that the corresponding Young tableaux have maximal height given by the power $d_{23}$.  Furthermore the corresponding blocks must also then have corresponding Young tableaux of height $d_{23}$ or less. This means that performing the expansion with $m=d_{23},n=0$ will give {\it{complete information}} on all the conformal blocks.

The advantage of using $SU(m,m)$ blocks (with $m=d_{23}$)  instead of $SU(2,2|4)$ blocks,  is that they are much simpler (at least conceptually), and are given by the compact formula
\begin{align}\label{eq:25}
F^{\alpha\beta\gamma\underline\lambda}(\underline x) &=\frac{\det\Big(
	x_i^{\lambda_j+m-j}{}_2F_1(\lambda_j{+}1{-}j{+}\alpha ,\lambda_j{+}1{-}j{+}\beta;2\lambda_j{+}2{-}2j{+}\gamma;x_i)\Big)_{1\leq i,j \leq
		m}}{\det\Big(
	x_i^{m-j}\Big)_{1\leq i,j \leq
		m}}\ .
\end{align}
Note that the denominator here is the famous Vandermonde determinant and can be rewritten $\prod_{i<j}(x_i-x_j)$.

We have converted the superconformal blocks to the $SU(m,m)$ theory, but we also need to convert the free correlator itself. This is straightforward. We simply replace the terms in~\eqref{freeterm} as
\begin{align}\label{freetermqq}
\left(\frac{g_{14}g_{23}}{g_{13}g_{24}}\right)^{d_{23}}= \left(\frac{1}{(1-x_1)\dots (1-x_m)}\right)^{d_{23}} = \text{sdet}^{-{d_{23}}}(1-Z) \qquad \quad Z\sim \left( \begin{array}{ccc}
x_1&&\\&\ddots&\\ &&x_m
\end{array}\right)\,.
\end{align} 
Thus performing a superconformal decomposition of free theory four-point functions of half BPS operators in $\cN=4$ SYM becomes equivalent to simply decomposing objects of this form into blocks of the form~\eqref{eq:25}.

Notice that the new $SU(m,m)$ functions and blocks in general depend on more variables than the original $SU(2,2|4)$ ones. Thus one may suspect that, although the blocks are conceptually much simpler, computationally this approach would be slower. However, since we are only interested in Young tableau of specific shapes, and in particular below the third row, they have at most 2 boxes, then we correspondingly only need to perform a very limited expansion in the variables $x_3, \dots, x_m$. Also notice that it is convenient to multiply both sides by the Vandermonde determinant. Then the blocks themselves are holomorphic. 

Finally, note that the procedure as outlined above gives information on free theory operators $\cO^{\gamma\ula}$ for fixed $\gamma$. This is fine for short muliplets as they are uniquely defined by this description, but as discussed around~\eqref{longdegen}, for long operators  the description is degenerate. Thus to obtain the OPE coefficient related to a  specific long representation one will have to sum over all the $\gamma$'s consistent with that representation.

Let us illustrate this procedure with a few simple examples.

\subsubsection*{Twist two contribution to $\langle \cO_2 \cO_2 \cO_2 \cO_2 \rangle$}

Firstly consider the twist 2 sector in the  $\langle \cO_2\cO_2\cO_2\cO_2 \rangle$ free theory. In the free theory the twist two operators are semi-short (recall they combine with other short operators to become long in the interacting theory).
They correspond to semi-short operators in~\eqref{table} with $\gamma=2,\mu=0$. Now the full free correlator is given below~\eqref{conditionsdij}. But only two of the six terms (the fourth and fifth) have~\eqref{gammrel} $\gamma=d_{13}+d_{14}+d_{23}+d_{24}=2$. Thus to extract all information about twist two operators from the free theory we perform the following expansion
\begin{align}
	A_{conn} (g_{12}g_{23}g_{34}g_{14} + g_{12}g_{24}g_{34}g_{13}) = \Pref \times A_{conn}(1+ \det{}^{-1}(1-Z) ) = \Pref \times \sum_{\ula} A_{2 \ula} F^{112\ula}(\underline x)\ ,
\end{align}
where here 
$$\Pref := \Pref^{\rm (OPE)}_{\,\{p_i\}} \times \left(\frac{g_{13}g_{24}}{g_{12}g_{34}}\right)^{\frac12(\gamma-p_4+p_3)} = g_{12}g_{34}g_{13}g_{24}$$
since $\gamma=2,p_i=2$.
This formula can be understood in terms of an $SU(m,m|2n)$ theory for any values of $m,n$. The values of the CPW coefficients $A_{2 \ula}$ will not depend on the group. Moreover the Cauchy identity implies that the left had side is a sum of Schur polynomials of one row only, and so the case $m=1,n=0$ will capture all the relevant information in this case. In this case there is only one variable $x_1$ and the superconformal blocks involve just a single Hypergeometric. In summary therefore the twist two operator CPW coefficients can be deduced from the  following decomposition
\begin{align}
A_{conn}\left(	1+\frac1{1-x_1}\right) = \sum_{\lambda_1} A_{2 [\lambda_1]} x_1^{\lambda_1}{}_2F_1(\lambda_1+1,\lambda_1+1;2\lambda_1+2;x_1)
\end{align}
which has the well known solution for twist two operators~\cite{Dolan:2001tt}
\begin{align}\label{twist2}
	A_{2[\lambda_1]}= 2 A_{conn} \frac{(\lambda_1)!^2}{(2\lambda_1)!}\ .
\end{align}

\subsubsection*{Higher twist singlet contribution to $\langle \cO_2 \cO_2 \cO_2 \cO_2 \rangle$}

Let us now consider the contribution of higher twist long singlet operators to $\langle \cO_2 \cO_2 \cO_2 \cO_2 \rangle$. The maximal value $\gamma$ can take for this correlator is 4 (see~\eqref{detform}). 
Comparing with~\eqref{table} we see that the only way we can achieve a long singlet multiplet is if $\gamma=4$, $\mu_1=\mu_2=0$. Three of the six terms in the free correlator~\eqref{conditionsdij} have $\gamma=d_{13}+d_{14}+d_{23}+d_{24}=4$ (the first, third and sixth). Thus to extract all information about twist four singlet operators we consider the expansion
\begin{align}
A_{disc} (g_{13}^2g_{24}^2 + g_{14}^2g_{23}^2) +A_{conn} g_{13}g_{23}g_{24}g_{14}  &= \Pref \times \Big(A_{disc}(1+ \det{}^{-2}(1-Z) ) + A_{conn}\det{}^{-1}(1-Z)\Big) \notag\\
&= \Pref \times \sum_{\ula} A_{4 \ula} F^{224\ula}(\underline x)\ ,
\end{align}
where this time
$$\Pref := \Pref^{\rm (OPE)}_{\,\{p_i\}} \times \left(\frac{g_{13}g_{24}}{g_{12}g_{34}}\right)^{\frac12(\gamma-p_4+p_3)} = g_{13}^2g_{24}^2$$
since $\gamma=4,p_i=2$.
Here, since the maximal inverse power of $\det(1-Z)$ is 2, we can recover complete information using the $m=2,n=0$ blocks.\footnote{ In fact for the connected piece one could even use the $m=1,n=0$ blocks as for twist 2 above.} 
Completely explicitly, using~(\ref{eq:25},\ref{freetermqq}), we expand (multiplying both sides by the Vandermonde determinant):
\begin{align}
	&(x_1-x_2)\left(A_{disc}\left(1+ \frac1{(1-x_1)^2(1-x_2)^2} \right) + A_{conn}\frac1{(1-x_1)(1-x_2)}\right)\notag\\&=\sum_{\ula} A_{4 \ula}\det\Big(
	x_i^{\lambda_j+2-j}{}_2F_1(\lambda_j{-}j{+}3 ,\lambda_j{-}j{+}3;2\lambda_j{-}2j{+}6;x_i)\Big)_{1\leq i,j \leq
		2}\ .
\end{align}

From~\eqref{table} we see that the long, twist $2t$, spin $l$,
singlet reps with $\gamma=4$ have CPW coefficients
$$A_{4,\ula=[l+t,t]}$$
in the above expansion.

\subsubsection*{Leading large $N$, higher twist singlet contribution to $\langle \cO_5 \cO_5 \cO_5 \cO_5 \rangle$}

In this paper we will be mostly concerned with the leading in $1/N$ piece of the free theory, and we consider such a case for  higher charge. Consider the leading large $N$ free correlator 
\begin{align}
\langle \cO_5 \cO_5 \cO_5 \cO_5 \rangle = A_{disc} \left(g_{12}^5g_{34}^5 + g_{13}^5g_{24}^5+g_{14}^5g_{23}^5 \right)\ .
\end{align}
These terms correspond to $\gamma=0,10,10$ respectively according to~\eqref{gammrel}. For long singlet reps we need $\gamma=10$ (note that if we considered the full free theory, rather than the leading large $N$ part,  we would have to consider different values of $\gamma=4,6,8,10$ and sum over the results.) Taking out the relevant prefactor from the second two terms we need to perform the following expansion
\begin{align}
 A_{disc}(1+ \det{}^{-5}(1-Z) ) =  \sum_{\ula} A_{10 \ula} F^{55\,10\,\ula}(\underline x)\ .
\end{align}
Here we convert to the $m=5, n=0$,  $SU(5,5)$ theory and so
explicitly, using~(\ref{eq:25},\ref{freetermqq}) we expand (multiplying both sides by the Vandermonde determinant):
\begin{align}
&(x_1-x_2)(x_1-x_3)\dots (x_4-x_5)A_{disc}\left(1+ \frac1{(1-x_1)^5\dots(1-x_5)^5} \right)\notag\\&=\sum_{\ula} A_{10 \, \ula}\det\Big(
x_i^{\lambda_j+5-j}{}_2F_1(\lambda_j{-}j{+}6 ,\lambda_j{-}j{+}6;2\lambda_j{-}2j{+}12;x_i)\Big)_{1\leq i,j \leq
	5}\ .
\end{align}

From the table~\eqref{table}, the long singlet operators of half twist $t$, spin $l$ correspond to the Young tableau $\ula=[l+t-3,t-3,2,2,2]$.
Note that to obtain this data it is enough to expand both sides in positive powers of $x_1,\dots x_5$ up to and including the term $x_1^{l+t+1}x_2^tx_3^4x_4^3x_5^2$, thus only a fairly limited expansion in the variables $x_3,x_4,x_5$ is needed. 

Note that if we were to consider the full $\langle \cO_5\cO_5\cO_5\cO_5\rangle$ free correlator, this will have sectors with different values of $\gamma=0,2,4,..,10$. Thus, since the description of long multiplets in terms of $\cO^{\gamma\ula}$ is not unique (see~\eqref{table} and the discussion below) then the CPW coefficient of a long operator will be given by the sum of all coefficients $A_{\gamma\ula}$ consistent with that rep. For example the OPE coefficient of a long singlet  operator of twist $t$, spin $l$ will be given by the sum of terms
\begin{align}
	A_{10[l+t-3,t-3,2,2,2]}+A_{8[l+t-2,t-2,2,2]}+A_{6[l+t-1,t-1,2]}+A_{4[l+t,t]}\ . 
\end{align}

\subsubsection*{Twist two operators from $\la\cO_p \cO_p \cO_\n \cO_\n\ra$}

Let us now consider the twist two contribution to any correlator of the form $\la\cO_p \cO_p \cO_\n \cO_\n\ra$.
The argument above for the $\la\cO_2\cO_2\cO_2\cO_2\ra $ correlator implies that we must have $\gamma=2$ and  then~\eqref{gammrel} then implies that there are  only two contributing diagrams
\begin{align}
	A g_{12}^{p-2}g_{34}^{n-2}\left(g_{13}g_{24}+g_{14}g_{23}\right)=A \times \Pref \times (1+\det{}^{-1}(1-Z))=\Pref \times \sum_{\ula} A_{2\,\ula} F^{112}(Z) 
\end{align}
Notice that once the prefactor has been divided out the computation is exactly the same as the $\la\cO_2\cO_2\cO_2\cO_2\ra$ case described above with the solution~\eqref{twist2}
\begin{align}
A_{2[\lambda_1]}= 2 A \frac{(\lambda_1)!^2}{(2\lambda_1)!}\ .
\end{align}
Finally, at large $N$ the value of $A$ can be deduced by counting the number of inequivalent planar graphs contributing times the number of colour loops in a double line notation as
\begin{align}\label{PaulA3general}
	A=  N^{p+\n} p^2 \n^2
\end{align}

\section{Free theory and Long-multiplet spectrum}
\label{twist2cancellation}

In this section we obtain an expression for the normalization $\normap$ 
relative to the set of correlators $\la \OO_p \OO_p \OO_\n \OO_\n\ra$.
Let us recall that $\norma_{pp\n\n}$ is automatically obtained from first-principle computations in supergravity.
For example, in the cases $\la\OO_p\OO_p\OO_p\OO_p\ra$ with $p=2,3,4$ and $\la\OO_2\OO_2\OO_\n\OO_\n\ra$ 
for any $\n$. However, it does not follow from the solution of the bootstrap problem \cite{Rastelli:2016nze},
and we will need to determine $\norma_{pp\n\n}$ from an independent analysis. 
The important observation will be the following: The OPE analysis of known supergravity 
four-point correlators \cite{Dolan:2004iy,Arutyunov:2000ku} reveals that in the supergravity certain 
long operators are absent from the spectrum. Therefore, in the decomposition 
\beq
\GG^{\rm sugra}_{p_1p_2p_3p_4}=\GG_{p_1p_2p_3p_4}^{\rm free} + \GG_{p_1p_2p_3p_4}^{\rm dyna }
\eeq
a special cancellation takes place between the sector of $\HH^{\rm\, dyna}$ given by $\Dbar^{\rm\, sing}$ and free theory. 
Building on this observation, we first derive $\norma_{22\n\n}$ and $\norma_{33\n\n}$ 
and we then obtain a formula for $\norma_{pp\n\n}$ which generalizes
the result $\norma_{pppp}$ obtained in \cite{Dolan:2006ec}.


\subsubsection*{Twist-2 long cancellation in $\la\OO_2\OO_2\OO_\n\OO_\n\ra$}
The propagator structure in $\la\OO_2\OO_2\OO_\n\OO_\n\ra$ is easily obtained from the case $\n=2$. 
In fact, the latter is maximally symmetric and contains only two crossing symmetric classes \cite{Arutyunov:2002fh}:
$(2,0,0)$ and $(1,1,0)$, which incidentally can be distinguished in terms of disconnected and connected diagrams. 
When $\n>2$, these two classes breaks into four sub-classes. 
This is shown in the diagrammatic expansion below 
where the extra (red) thick line indicates the additional $\n-2$ propagators $g_{34}$.
\begin{center}
\begin{minipage}{\linewidth}
          \begin{center}
             \begin{tikzpicture}
		\def\x1{2}
		\def\xx1{1.1}
		\def\y1{3}
		\def\step{1.4}
		\def\distaone{3.25}
		\def\distatwo{6.75}
		\def\colore{black}
		\def\coloreagg{red}

\draw (\x1,\y1+\step) node {\rule{0pt}{1.2cm}};

		\draw (\xx1-2.4,\y1+.75) node 		{ $\la\OO_2\OO_2\OO_\n\OO_\n\ra\quad = \quad A_{1}\quad$};
		\draw (\xx1-2.4, 0.9*\step) node	 	{ $\phantom{\la\OO_2\OO_2\OO_n\OO_n\ra\quad = \quad \ A_{3}\quad} $};

		\draw[thick,\colore] (\xx1-.1,		\y1) --	 (\xx1-.1,			\y1+\step);
		\draw[thick,\colore] (\xx1+.1,		\y1) --	 (\xx1+.1,			\y1+\step);
		\draw[thick,\colore] (\xx1-.1+\step,	\y1) -- 	 (\xx1-.1+\step,		\y1+\step);
		\draw[thick,\colore] (\xx1+.1+\step,	\y1) --	 (\xx1+.1+\step,		\y1+\step);
		
		\draw[ultra thick,\coloreagg]     (\xx1+\step,\y1-.1) .. controls (\xx1+\step+.5,\y1+0.2*\step) and (\xx1+\step+.5,\y1+0.8*\step)  .. (\xx1+\step,\y1+\step+.1);
		
		\draw[thick] (\x1+\distaone-1,	\y1-.3) -- 			(\x1+\distaone-1,		\y1+\step+.3) 	node[right=0.3*\step cm,below=0.5*\step cm]  {$A_2$} ;
		\draw[thick] (\x1+\distaone-1,	\y1-.3) --			(\x1+\distaone-1+.2,		\y1-.3);
		\draw[thick] (\x1+\distaone-1,	\y1+\step+.3) -- 	(\x1+\distaone-1+.2,		\y1+\step+.3);

		\draw[thick,\colore] (\x1+\distaone,	\y1+.1) -- 			(\x1+\distaone+\step,		\y1+.1);
		\draw[thick,\colore] (\x1+\distaone,	\y1-.1) -- 			(\x1+\distaone+\step,		\y1-.1);
		\draw[thick,\colore] (\x1+\distaone,	\y1+\step+.1) -- 	(\x1+\distaone+\step,		\y1+\step+.1);
		\draw[thick,\colore] (\x1+\distaone,	\y1+\step-.1) -- 		(\x1+\distaone+\step,		\y1+\step-.1);
		
		\draw[ultra thick,\coloreagg]     (\x1+\distaone+\step,\y1-.1) .. controls (\x1+\distaone+\step+.5,\y1+0.2*\step) and (\x1+\distaone+\step+.5,\y1+0.8*\step)  .. (\x1+\distaone+\step,\y1+\step+.1);
		
		\draw[thick,\colore] (\x1+.1+\distatwo,	\y1) -- 		(\x1+.1+\distatwo+\step,	\y1+\step);
		\draw[thick,\colore] (\x1-.1+\distatwo,		\y1) -- 		(\x1-.1+\distatwo+\step,	\y1+\step);
		\draw[thick,\colore] (\x1+.1+\distatwo,	\y1+\step) -- 	(\x1+.1+\distatwo+\step,	\y1);
		\draw[thick,\colore] (\x1-.1+\distatwo,		\y1+\step) --	(\x1-.1+\distatwo+\step,	\y1);
		
		\draw[ultra thick,\coloreagg]     (\x1+\distatwo+\step,\y1-.1) .. controls (\x1+\distatwo+\step+.5,\y1+0.2*\step) and (\x1+\distatwo+\step+.5,\y1+0.8*\step)  .. (\x1+\distatwo+\step,\y1+\step+.1);

		\filldraw (\xx1,		\y1)			 circle (.15);
		\filldraw (\xx1+\step,	\y1) 			 circle (.15);
		\filldraw (\xx1,		\y1+\step) 	 circle (.15);
		\filldraw (\xx1+\step,	\y1+\step) 	 circle (.15);
		\draw (\xx1+\step,	\y1)	 		 node[right=0.7*\step cm,above=0.3*\step cm] {$+$};
		\filldraw (\x1+\distaone,		\y1)			circle (.15);
		\filldraw (\x1+\distaone+\step,	\y1) 			circle (.15);
		\filldraw (\x1+\distaone,		\y1+\step) 	 circle (.15);
		\filldraw (\x1+\distaone+\step,	\y1+\step) 	 circle (.15);
		\draw (\x1+\distaone+\step,	\y1)  			 node[right=0.9*\step cm,above=0.2*\step cm] {$+\  A_2^{\rm exc.}$};
		\filldraw (\x1+\distatwo,		\y1)			circle (.15);
		\filldraw (\x1+\distatwo+\step,	\y1) 			 circle (.15);
		\filldraw (\x1+\distatwo,		\y1+\step) 	 circle (.15);
		\filldraw (\x1+\distatwo+\step,	\y1+\step) 	circle (.15);
		\draw (\x1+\distatwo+\step,	\y1) 			node[right=0.9*\step cm,above=0.3*\step cm] {$+$};

		\draw[thick] (\x1+\distatwo+\step+.7,		\y1-.3) -- 		(\x1+\distatwo+\step+.7,	\y1+\step+.3);		
		\draw[thick] (\x1+\distatwo+\step+.7-.2,	\y1-.3) -- 		(\x1+\distatwo+\step+.7,	\y1-.3);
		\draw[thick] (\x1+\distatwo+\step+.7-.2,	\y1+\step+.3)-- 	(\x1+\distatwo+\step+.7,	\y1+\step+.3);

\draw (\x1+\distatwo+\step+1,\y1) node {\rule{2cm}{0pt}};

		\def\xx2{1.1}	
		\def\stepsmall{.3}
		\def\x2{2.25}
		\def\y2{0.5}

\draw (\x2,\y2) node {\rule{0pt}{.5cm}};

		\draw[thick] (\xx2-.7,		\y2-.3) --			(\xx2-.7, 		\y2+\step+.3)   node[ right=0.3*\step cm ,below=0.5*\step cm] {$A_3$};
		\draw[thick] (\xx2-.7,		\y2-.3) -- 			(\xx2-.7+.2,	\y2-.3);
		\draw[thick] (\xx2-.7,		\y2+\step+.3) --		(\xx2-.7+.2,	\y2+\step+.3);

		\draw[thick,\colore] (\xx2+\stepsmall,\y2)  rectangle (\xx2+\stepsmall +\step,\y2+\step);
		
		\draw[ultra thick,\coloreagg]     (\xx2+\stepsmall+\step,\y2-.1) .. controls (\xx2+\stepsmall+\step+.5,\y2+0.2*\step) and (\xx2+\stepsmall+\step+.5,\y2+0.8*\step)  .. (\xx2+\stepsmall+\step,\y2+\step+.1);
		\draw[thick,\colore] (\xx2+1.2*\distaone,			\y2) --		 (\xx2+1.2*\distaone,			\y2+\step);
		\draw[thick,\colore] (\xx2+1.2*\distaone+\step,		\y2) --		 (\xx2+1.2*\distaone+\step,	\y2+\step);
		\draw[thick,\colore] (\xx2+1.2*\distaone,			\y2) --		 (\xx2+1.2*\distaone+\step,	\y2+\step);
		\draw[thick,\colore] (\xx2+1.2*\distaone,			\y2+\step) --	 (\xx2+1.2*\distaone+\step,	\y2);
		
		\draw[ultra thick,\coloreagg]     (\xx2+1.2*\distaone+\step,\y2-.1) .. controls (\xx2+1.2*\distaone+\step+.5,\y2+0.2*\step) and (\xx2+1.2*\distaone+\step+.5,\y2+0.8*\step)  .. (\xx2+1.2*\distaone+\step,\y2+\step+.1);

		\draw[thick] (\x2+1.2*\distaone+0.9*\step,		\y2-.3) --			(\x2+1.2*\distaone+0.9*\step,	 \y2+\step+.3);		
		\draw[thick] (\x2+1.2*\distaone+0.9*\step-.2,	\y2-.3) -- 			(\x2+1.2*\distaone+0.9*\step,	 \y2-.3);
		\draw[thick] (\x2+1.2*\distaone+0.9*\step-.2,	\y2+\step+.3)--  	(\x2+1.2*\distaone+0.9*\step,	 \y2+\step+.3);


		\draw[thick,\colore] (\x2+\distatwo+\stepsmall,	\y2) --		 (\x2+\distatwo+\stepsmall+\step,	\y2);
		\draw[thick,\colore] (\x2+\distatwo+\stepsmall,	\y2+\step) -- 	 (\x2+\distatwo+\stepsmall+\step, 	\y2+\step);
		\draw[thick,\colore] (\x2+\distatwo+\stepsmall,	\y2) --		 (\x2+\distatwo+\stepsmall+\step,	\y2+\step);
		\draw[thick,\colore] (\x2+\distatwo+\stepsmall,	\y2+\step) --	 (\x2+\distatwo+\stepsmall+\step,	\y2);	
			
		\draw[ultra thick,\coloreagg]     (\x2+\distatwo+\stepsmall+\step,\y2-.1) .. controls (\x2+\distatwo+\stepsmall+\step+.5,\y2+0.2*\step) and (\x2+\distatwo+\stepsmall+\step+.5,\y2+0.8*\step)  .. (\x2+\distatwo+\stepsmall+\step,\y2+\step+.1);

		\filldraw (\xx2+\stepsmall,			\y2)			 circle (.15);
		\filldraw (\xx2+\stepsmall+\step,		\y2) 			 circle (.15);
		\filldraw (\xx2+\stepsmall,			\y2+\step) 	 circle (.15);
		\filldraw (\xx2+\stepsmall+\step,		\y2+\step) 	 circle (.15);
		\draw (\xx2+\stepsmall+\step,		\y2) 			 node[right=0.9*\step cm,above=0.3*\step cm] {$+\ A_3^{\rm exc.}$};
		\filldraw (\xx2+1.2*\distaone,		\y2)			 circle (.15);
		\filldraw (\xx2+1.2*\distaone+\step,	\y2) 			 circle (.15);
		\filldraw (\xx2+1.2*\distaone,		\y2+\step) 	 circle (.15);
		\filldraw (\xx2+1.2*\distaone+\step,	\y2+\step) 	 circle (.15);
		
		\draw (\xx2+1.2*\distaone+1.5*\step, \y2) 		node[right=0.85*\step cm,above=0.3*\step cm] {$\ +\  \ A_4$};
		\filldraw (\x2+\distatwo+\stepsmall,		\y2)			 circle (.15);
		\filldraw (\x2+\distatwo+\stepsmall+\step,	\y2) 		 	 circle (.15);
		\filldraw (\x2+\distatwo+\stepsmall,		\y2+\step) 	 circle (.15);
		\filldraw (\x2+\distatwo+\stepsmall+\step,	\y2+\step) 	 circle (.15);

\draw (\x2,\y2-.5) node {\rule{0pt}{.5cm}};
		
	      \end{tikzpicture}
           \end{center}
 \end{minipage}
 \label{Figure22nn}
\end{center}
The residual symmetry exchanges $g_{14} \leftrightarrow g_{13}$ and $g_{23}\leftrightarrow g_{24}$. In particular, 
\beq\label{relation22nn}
\begin{array}{ccc}
A_1(u,v)&=&A_1 \left( {u}/{v},{1}/{v}\right),\\[.2cm]
A_4(u,v)&=&A_4 \left( {u}/{v},{1}/{v}\right),
\end{array}
\qquad 
\begin{array}{ccc}
A_2^{\rm exc.}(u,v)&=&A_2\left( {u}/{v},{1}/{v}\right),\\[.2cm]
A_3^{\rm exc.}(u,v)&=&A_3 \left( {u}/{v},{1}/{v}\right).
\end{array}
\eeq
As a result, in free theory, where the $A_{i=1,2,3,4}$ are constants, we shall find $A_2^{\rm exc.}=A_2$ and $A_3^{\rm exc.}=A_3$. 
The remaining coefficients to determine are 
\beq
 A_1^{\rm free}=2q N^{2+\n},\qquad 
 (A_2^{\rm free},\,A_3^{\rm free},\,A_4^{\rm free})=\big(  \, 2\n(\n-2),\, 2\n,\, 2\n(\n-1)\, \big)\frac{A_1^{\rm free}}{N^2\ }\ .
 \eeq
Two exceptions to these formulas are, $A^{\rm free}_2=1$ for $\n=2$, and $A^{\rm free}_2=0$ for $\n=3$.
 The OPE prefactor reduces to $g_{12}^2 g_{34}^{\n}$, which corresponds to the diagram associated with $A_1$. 
The correlator can then be rewritten as,
\begin{align}
\la\OO_2\OO_2\OO_\n\OO_\n\ra^{\rm free}=
						{ g_{12}^2g_{34}^q\, A_1^{\rm free} } \bigg[\ &1+ \nn\\
										& \frac{ 2\n }{N^2}  \left( u\sigma+\frac{u\tau}{v} \right) + \nn \\
										&  \frac{ 2\n }{N^2} \left( (\n-2) \left( u^2\sigma^2+\frac{u^2\tau^2}{v^2} \right) + (\n-1) \frac{u^2\sigma\tau}{v}\right) \bigg]\ .
										\label{free22nn_uv}
										\rule{0.5cm}{0pt}
\end{align}
The dynamical part of the correlation function obtained from tree-level supergravity is \cite{Uruchurtu:2008kp},
\beq
\la\OO_2\OO_2\OO_\n\OO_\n\ra^{\rm dyna}= g_{12}^2 g_{34}^\n\ \norma_{22\n\n}\ \Intri(u,v,\sigma,\tau)\, u^\n\, \Dbar_{\n,\n+2,2,2}\,,
\eeq
which decomposes as follows,
\begin{align}
A_1^{\rm dyna}&= v\,u^q \Dbar_{\n,\n+2,2,2}\,,          &      A_3^{\rm dyna}&= -\left(\frac{1-v}{u} +1\right)\, A_1^{\rm dyna}\,, \\[0.2cm]
A_2^{\rm dyna}&=\frac{v}{u}\, A_1^{\rm dyna}\,,\hspace{0cm}   &      A_4^{\rm dyna}&=-\left(\frac{1+v}{u} -1\right)\, A_1^{\rm dyna}\, \ .
\end{align}
Symmetry properties of $\Dbar_{\n,\n+2,2,2}$, described in the Appendix, imply the relations \eqref{relation22nn}. 
The number of propagator structures equals the number of
$SU(4)$ channel in the correlator. These correspond to the intersection
\beq
([0,2,0]\otimes[0,2,0])\cap ([0,\n,0]\otimes[0,\n,0])=[0,2,0]\otimes[0,2,0]\,,
\eeq 
which splits into the six channels,
\beq\label{020content}
[0,2,0]\otimes[0,2,0]=[0,0,0]\oplus[0,2,0]\oplus[0,4,0]\oplus[2,0,2]\oplus[1,0,1]\oplus[1,2,1]\,,
\eeq
according to \eqref{SU4OPE}. 
In each $SU(4)$ channel we shall find contributions from operators belonging to different $\mathcal{N}=4$ representation. 
For example, in the singlet channel we expect a contribution from the stress energy tensor, which belongs to a short multiplet, 
and a contribution from a twist-$2$ scalar, which belongs to a long multiplet. 
Moreover, long multiplets whose lowest dimension operator belong to $[0,0,0]$ 
have precisely the same $SU(4)$ content of \eqref{020content}, thus will contribute to all six channels.

In free theory, all operators have canonical dimensions and are present in the spectrum.
A proper study of the superconformal OPE is needed in order to recombine all such contributions into supermultiplets \cite{Dolan:2004iy,Doobary:2015gia}.
Once this decomposition is achieved \cite{Rayson:2017mma}, it can be shown that twist-two long contributions cancel between $\GG_{22\n\n}^{\rm free }$ and $\GG_{22\n\n}^{\rm dyna }$
precisely for the supergravity value 
\beq\label{norma22nn}
\norma_{22\n\n}=-\frac{2\n}{(\n-2)!} \, \frac{A_1^{\rm free}}{N^2\ }.
\eeq

We can prove \eqref{norma22nn} using a simpler argument: 
In the $[0,0,0]$ channel of the correlator, the conformal block corresponding to the twist-two 
scalar in the corresponding long multiplet has a series expansion with leading power $u^1(1-v)^0$. As remarked in  \eqref{blockmellin},
conformal blocks corresponding to operators with twist $2t$ and spin $l>0$ are distinguished by the leading power $u^\tw (1-v)^l$. 
Therefore twist-$2$ is the very 
first non trivial contribution at order $1/N^2$, and
the absence of a twist-$2$ long multiplet implies that of the corresponding leading power. 
In terms of propagator structure, the $[0,0,0]$ channel is proportional to,
\beq
\la\OO_2\OO_2\OO_\n\OO_\n\ra\Big|_{[0,0,0]}\sim \ 
					A_1+ \frac{u}{6} \left( \frac{A_3}{v}+A_3^{\rm\,exc.}\right) 
					+ \frac{u^2}{20} \left( \frac{A_2}{v^2} + A_2^{\rm\, exc.} +\frac{1}{3}\frac{A_4}{v} \right)\ .
\eeq
where $A_i=A_i^{\rm free}+A_i^{\rm dyna}$. 
At order $1/N^2$, the twist-two long contribution comes from the second term proportional to $A_3^{\rm\, free}$,
and from $A_1^{\rm\, dyna}\sim \Dbar_{\n,\n+2,2,2}^{}$. The expression for
$\Dbar_{\n,\n+2,2,2}^{\rm\, sing}$ can be obtained from \eqref{Dsingularseries}. 
The limit $v\rightarrow1$ is unambiguous and by equating the two contributions we obtain, 
\beq
2 \n \, \frac{A_1^{\rm free}}{N^2\ } + \norma_{22\n\n}\, \Gamma[\n-1] =0\,,
\eeq
which then leads to the result \eqref{norma22nn}. 
This simpler argument generalizes to $\la\OO_p\OO_p\OO_\n\OO_\n\ra$ for arbitrary $p$ and $\n$. In fact, 
it will always be the case that in the $[0,0,0]$ channel of free theory the first and only 
contribution at order $1/N^2$ comes from a twist-$2$ scalar belonging to the corresponding long multiplet. 
As we now show, minor modifications are needed in the derivation of $\norma_{pp\n\n}$ when $p\ge3$. However, 
taking these into account we will be able to obtain $\norma_{pp\n\n}$ in general.


\subsubsection*{Twist-2 long cancellation in $\la\OO_3\OO_3\OO_\n\OO_\n\ra$}
Similarly to the previous discussion, the propagator structure in $\la\OO_3\OO_3\OO_\n\OO_\n\ra$ follows from that at $\n=3$. 
In this case there are three crossing symmetric classes \cite{Arutyunov:2002fh}: 
$(3,0,0)$ contains three disconnected diagrams; $(2,1,0)$ contains six connected diagrams; and $(1,1,1)$ contains a single connected diagram. 
The symmetry breaking pattern when $\n>3$ splits the three symmetric classes into six sub-classes.
\begin{center}
\begin{minipage}{\linewidth}
          \begin{center}
             \begin{tikzpicture}
		\def\x1{2}
		\def\xx1{1.1}
		\def\y1{3}
		\def\step{1.4}
		\def\distaone{3.25}
		\def\distatwo{6.75}
		\def\colore{black}
		\def\coloreagg{red}

\draw (\x1,\y1+\step) node {\rule{0pt}{.7cm}};

		\draw (\xx1-2.4,		\y1+.75) 			node 		{ $\la\OO_3\OO_3\OO_\n\OO_\n\ra\quad = \quad A_{1}\quad$};

		\draw[thick,\colore] (\xx1-.1,		\y1) -- (\xx1-.1,		\y1+\step);
		\draw[thick,\colore] (\xx1,			\y1) -- (\xx1,		\y1+\step);
		\draw[thick,\colore] (\xx1+.1,		\y1) -- (\xx1+.1,		\y1+\step);
		\draw[thick,\colore] (\xx1-.1+\step,	\y1) -- (\xx1-.1+\step,\y1+\step);
		\draw[thick,\colore] (\xx1+\step,		\y1) -- (\xx1+\step,	\y1+\step);
		\draw[thick,\colore] (\xx1+.1+\step,	\y1) -- (\xx1+.1+\step,\y1+\step);
		
		\draw[ultra thick,\coloreagg]     (\xx1+\step,\y1-.1) .. controls (\xx1+\step+.5,\y1+0.2*\step) and (\xx1+\step+.5,\y1+0.8*\step)  .. (\xx1+\step,\y1+\step+.1);
		
		\draw[thick] (\x1+\distaone-1,\y1-.3) -- 		(\x1+\distaone-1,	\y1+\step+.3) node[right=0.3*\step cm,below=0.5*\step cm]  {$A_2$} ;
		\draw[thick] (\x1+\distaone-1,\y1-.3) -- 		(\x1+\distaone-1+.2,\y1-.3);
		\draw[thick] (\x1+\distaone-1,\y1+\step+.3) -- 	(\x1+\distaone-1+.2,\y1+\step+.3);

		\draw[thick,\colore] (\x1+\distaone,	\y1+.1) -- 		(\x1+\distaone+\step,		\y1+.1);
		\draw[thick,\colore] (\x1+\distaone,	\y1) -- 		(\x1+\distaone+\step,		\y1);		
		\draw[thick,\colore] (\x1+\distaone,	\y1-.1) -- 		(\x1+\distaone+\step,		\y1-.1);
		\draw[thick,\colore] (\x1+\distaone,	\y1+\step+.1) -- (\x1+\distaone+\step,	\y1+\step+.1);
		\draw[thick,\colore] (\x1+\distaone,	\y1+\step) -- 	(\x1+\distaone+\step,		\y1+\step);		
		\draw[thick,\colore] (\x1+\distaone,	\y1+\step-.1) -- (\x1+\distaone+\step,		\y1+\step-.1);
		
		\draw[ultra thick,\coloreagg]     (\x1+\distaone+\step,\y1-.1) .. controls (\x1+\distaone+\step+.5,\y1+0.2*\step) and (\x1+\distaone+\step+.5,\y1+0.8*\step)  .. (\x1+\distaone+\step,\y1+\step+.1);
		\draw[thick,\colore] (\x1+.1+\distatwo,	\y1) -- 		(\x1+.1+\distatwo+\step,	\y1+\step);
		\draw[thick,\colore] (\x1+\distatwo,		\y1) -- 		(\x1+\distatwo+\step,		\y1+\step);
		\draw[thick,\colore] (\x1-.1+\distatwo,		\y1) -- 		(\x1-.1+\distatwo+\step,	\y1+\step);
		\draw[thick,\colore] (\x1+.1+\distatwo,	\y1+\step) -- 	(\x1+.1+\distatwo+\step,	\y1);
		\draw[thick,\colore] (\x1+\distatwo,		\y1+\step) -- 	(\x1+\distatwo+\step,		\y1);		
		\draw[thick,\colore] (\x1-.1+\distatwo,		\y1+\step) -- 	(\x1-.1+\distatwo+\step,	\y1);
		
		\draw[ultra thick,\coloreagg]     (\x1+\distatwo+\step,\y1-.1) .. controls (\x1+\distatwo+\step+.5,\y1+0.2*\step) and (\x1+\distatwo+\step+.5,\y1+0.8*\step)  .. (\x1+\distatwo+\step,\y1+\step+.1);

		\filldraw (\xx1,		\y1)			circle (.15);
		\filldraw (\xx1+\step,	\y1) 		 	circle (.15);
		\filldraw (\xx1,		\y1+\step) 	circle (.15);
		\filldraw (\xx1+\step,	\y1+\step) 	circle (.15);
		\draw (\xx1+\step, 	\y1)	 		node[right=0.6*\step cm,above=0.3*\step cm] {$+$};
		\filldraw (\x1+\distaone,		\y1)				 circle (.15);
		\filldraw (\x1+\distaone+\step,	\y1) 				 circle (.15);
		\filldraw (\x1+\distaone,		\y1+\step) 	          circle (.15);
		\filldraw (\x1+\distaone+\step,	\y1+\step) 	 	 circle (.15);
		\draw (\x1+\distaone+\step,	\y1)  			          node[right=0.9*\step cm,above=0.2*\step cm]  {$+\  A_2^{\rm exc.}$};
		\filldraw (\x1+\distatwo,		\y1)				 circle (.15);
		\filldraw (\x1+\distatwo+\step,	\y1) 				 circle (.15);
		\filldraw (\x1+\distatwo,		\y1+\step) 		 circle (.15);
		\filldraw (\x1+\distatwo+\step,	\y1+\step) 		 circle (.15);
		\draw (\x1+\distatwo+\step,	\y1) 				 node[right=0.8*\step cm,above=0.3*\step cm] {$+$};

		\draw[thick] (\x1+\distatwo+\step+.7,		\y1-.3) --		 (\x1+\distatwo+\step+.7,\y1+\step+.3);		
		\draw[thick] (\x1+\distatwo+\step+.7-.2,	\y1-.3) --		 (\x1+\distatwo+\step+.7,\y1-.3);
		\draw[thick] (\x1+\distatwo+\step+.7-.2,	\y1+\step+.3)--	 (\x1+\distatwo+\step+.7,\y1+\step+.3);

\draw (\x1+\distatwo+\step+1,\y1) node {\rule{2cm}{0pt}};

		\def\xx2{1.1}	
		\def\stepsmall{.225}
		\def\x2{2}
		\def\y2{0+.5}
		\def\distaone{4.1}

		\draw[thick] (\xx2-.7,		\y2-.3) --			(\xx2-.7, 		\y2+\step+.3)   node[ right=0.3*\step cm ,below=0.5*\step cm] {$A_3$};
		\draw[thick] (\xx2-.7,		\y2-.3) -- 			(\xx2-.7+.2,	\y2-.3);
		\draw[thick] (\xx2-.7,		\y2+\step+.3) --		(\xx2-.7+.2,	\y2+\step+.3);

		\draw[thick,\colore] 			(\xx2+\stepsmall-.1,		 \y2)--		 (\xx2+\stepsmall-.1,			\y2+\step);
		\draw[thick,\colore] 			(\xx2+\stepsmall+.1,		 \y2)--		 (\xx2+\stepsmall+.1,			\y2+\step);
		\draw[thick,\colore] 			(\xx2+\stepsmall+\step-.1,	 \y2)--		 (\xx2+\stepsmall+\step-.1,	\y2+\step);
		\draw[thick,\colore] 			(\xx2+\stepsmall+\step+.1, \y2)--	 	 (\xx2+\stepsmall+\step+.1,	\y2+\step);
		\draw[thick,\colore] 			(\xx2+\stepsmall,		 \y2)--	 	 (\xx2+\stepsmall+\step+.1,	\y2);
		\draw[thick,\colore] 			(\xx2+\stepsmall, 		  \y2+\step)--	 (\xx2+\stepsmall+\step+.1,	\y2+\step);

		\draw[ultra thick,\coloreagg]     (\xx2+\stepsmall+\step,\y2-.1) .. controls (\xx2+\stepsmall+\step+.5,\y2+0.2*\step) and (\xx2+\stepsmall+\step+.5,\y2+0.8*\step)  .. (\xx2+\stepsmall+\step,\y2+\step+.1);
		\draw[thick,\colore] 			(\xx2+\distaone-.1,		  \y2)--		 (\xx2+\distaone-.1,			\y2+\step);
		\draw[thick,\colore] 			(\xx2+\distaone+.1,		  \y2)--		 (\xx2+\distaone+.1,			\y2+\step);
		\draw[thick,\colore] 			(\xx2+\distaone+\step-.1,	  \y2)--		 (\xx2+\distaone+\step-.1,		\y2+\step);
		\draw[thick,\colore] 			(\xx2+\distaone+\step+.1, 	  \y2)--	 	 (\xx2+\distaone+\step+.1,		\y2+\step);
		\draw[thick,\colore] 			(\xx2+\distaone,		  \y2) -- 		 (\xx2+\distaone+\step,		\y2+\step);
		\draw[thick,\colore]		        (\xx2+\distaone,			  \y2+\step) --	 (\xx2+\distaone+\step,		\y2);
		
		\draw[ultra thick,\coloreagg]     (\xx2+\distaone+\step,\y2-.1) .. controls (\xx2+\distaone+\step+.5,\y2+0.2*\step) and (\xx2+\distaone+\step+.5,\y2+0.8*\step)  .. (\xx2+\distaone+\step,\y2+\step+.1);

		\draw[thick] (\x2+\distaone+0.9*\step,	\y2-.3) --			(\x2+\distaone+0.9*\step,	\y2+\step+.3);		
		\draw[thick] (\x2+\distaone+0.9*\step-.2,	\y2-.3) -- 			(\x2+\distaone+0.9*\step,	\y2-.3);
		\draw[thick] (\x2+\distaone+0.9*\step-.2,	\y2+\step+.3)-- 		(\x2+\distaone+0.9*\step,	\y2+\step+.3);

		\filldraw (\xx2+\stepsmall,		\y2)			 circle (.15);
		\filldraw (\xx2+\stepsmall+\step,\y2) 		 	 circle (.15);
		\filldraw (\xx2+\stepsmall,		\y2+\step) 	 circle (.15);
		\filldraw (\xx2+\stepsmall+\step,	\y2+\step) 	 circle (.15);
		
		\draw (\xx2+\stepsmall+\step,\y2) node[right=0.9*\step cm,above=0.3*\step cm] {$\ \,+\ A_3^{\rm exc.}$};
		\filldraw (\xx2+\distaone,		\y2)			 circle (.15);
		\filldraw (\xx2+\distaone+\step,\y2) 			 circle (.15);
		\filldraw (\xx2+\distaone,		\y2+\step) 	 circle (.15);
		\filldraw (\xx2+\distaone+\step,	\y2+\step) 	 circle (.15);
		
		\draw (\xx2+\distaone+1.5*\step,\y2) node[right=0.55*\step cm,above=0.3*\step cm] {$+$};

		\def\xx3{1.1}	
		\def\stepsmall{.2}
		\def\x3{2}
		\def\y3{-2.25}

		\draw[thick] (\xx3-.7,		\y3-.3) --			(\xx3-.7, 		\y3+\step+.3)   node[ right=0.3*\step cm ,below=0.5*\step cm] {$A_4$};
		\draw[thick] (\xx3-.7,		\y3-.3) -- 			(\xx3-.7+.2,	\y3-.3);
		\draw[thick] (\xx3-.7,		\y3+\step+.3) --		(\xx3-.7+.2,	\y3+\step+.3);

		\draw[thick,\colore] 			(\xx3+\stepsmall,			\y3)--				 	(\xx3+\stepsmall,		\y3+\step);
		\draw[thick,\colore] 			(\xx3+\stepsmall+\step,		\y3)--		 			 (\xx3+\stepsmall+\step,	\y3+\step);
		\draw[thick,\colore] 			(\xx3+\stepsmall,			\y3+.1)--				 (\xx3+\stepsmall+\step,	\y3+.1);
		\draw[thick,\colore] 			(\xx3+\stepsmall,	 		\y3-.1)--	 			 (\xx3+\stepsmall+\step,	\y3-.1);
		\draw[thick,\colore] 			(\xx3+\stepsmall,			\y3+\step+.1)--			 (\xx3+\stepsmall+\step,	\y3+\step+.1);
		\draw[thick,\colore] 			(\xx3+\stepsmall,	 		\y3+\step-.1)--	 		 (\xx3+\stepsmall+\step,	\y3+\step-.1);

		\draw[ultra thick,\coloreagg]      (\xx3+\stepsmall+\step,\y3-.1) .. controls (\xx3+\stepsmall+\step+.5,\y3+0.2*\step) and (\xx3+\stepsmall+\step+.5,\y3+0.8*\step)  .. (\xx3+\stepsmall+\step,\y3+\step+.1);
		\draw[thick,\colore] (\xx3+\distaone,			\y3) --		 	 (\xx3+\distaone,		\y3+\step);
		\draw[thick,\colore] (\xx3+\distaone+\step,		\y3) -- 			 (\xx3+\distaone+\step,	\y3+\step);
		\draw[thick,\colore] (\xx3+\distaone-.1,		\y3) --			 (\xx3+\distaone+\step-.1,	\y3+\step);
		\draw[thick,\colore] (\xx3+\distaone+.1,		\y3) --		 	 (\xx3+\distaone+\step+.1,	\y3+\step);		
		\draw[thick,\colore] (\xx3+\distaone-.1,		\y3+\step) --		 (\xx3+\distaone+\step-.1,	\y3);
		\draw[thick,\colore] (\xx3+\distaone+.1,		\y3+\step) --		 (\xx3+\distaone+\step+.1,	\y3);

		\draw[ultra thick,\coloreagg]     (\xx3+\distaone+\step,\y3-.1) .. controls (\xx3+\distaone+\step+.5,\y3+0.2*\step) and (\xx3+\distaone+\step+.5,\y3+0.8*\step)  .. (\xx3+\distaone+\step,\y3+\step+.1);

		\draw[thick] 	(\x3+\distaone+0.9*\step,		\y3-.3) --			(\x3+\distaone+0.9*\step,\y3+\step+.3);		
		\draw[thick]	 (\x3+\distaone+0.9*\step-.2,	\y3-.3)	 -- 		(\x3+\distaone+0.9*\step,\y3-.3);
		\draw[thick]	 (\x3+\distaone+0.9*\step-.2,	\y3+\step+.3)	-- 	(\x3+\distaone+0.9*\step,\y3+\step+.3);

		\filldraw 	(\xx3+\stepsmall,		\y3)				 	circle (.15);
		\filldraw 	(\xx3+\stepsmall+\step,	\y3) 		 			circle (.15);
		\filldraw 	(\xx3+\stepsmall,		\y3+\step) 			circle (.15);
		\filldraw 	(\xx3+\stepsmall+\step,	\y3+\step) 	 		circle (.15);
		\draw 	(\xx3+\stepsmall+\step,	\y3) 					node[right=0.95*\step cm,above=0.3*\step cm] {$\ +\ A_4^{\rm exc.}$};
		\filldraw    	 (\xx3+\distaone,		\y3)				 circle (.15);
		\filldraw	 (\xx3+\distaone+\step,	\y3) 		 		 circle (.15);
		\filldraw	 (\xx3+\distaone,		\y3+\step) 		 circle (.15);
		\filldraw	 (\xx3+\distaone+\step,	\y3+\step) 		 circle (.15);
		\draw 	(\xx3+\distaone+1.5*\step,	\y3) 				node[right=0.55*\step cm,above=0.3*\step cm] {$+$};

		\def\xx4{1.1}	
		\def\stepsmall{.2}
		\def\x4{2}
		\def\y4{-5}
		\def\distatwo{8.5}

		\draw[thick] (\xx4-.7,		\y4-.3) --			(\xx4-.7, 		\y4+\step+.3)   node[ right=0.3*\step cm ,below=0.5*\step cm] {$A_5$};
		\draw[thick] (\xx4-.7,		\y4-.3) -- 			(\xx4-.7+.2,	\y4-.3);
		\draw[thick] (\xx4-.7,		\y4+\step+.3) --		(\xx4-.7+.2,	\y4+\step+.3);

		\draw[thick,\colore] 			(\xx4+\stepsmall,			\y4)--				 	 (\xx4+\stepsmall+\step,	\y4+\step);
		\draw[thick,\colore] 			(\xx4+\stepsmall,			\y4+\step)--		 	(\xx4+\stepsmall+\step,	\y4);
		\draw[thick,\colore] 			(\xx4+\stepsmall,			\y4+.1)--				 (\xx4+\stepsmall+\step,	\y4+.1);
		\draw[thick,\colore] 			(\xx4+\stepsmall,	 		\y4-.1)--	 			 (\xx4+\stepsmall+\step,	\y4-.1);
		\draw[thick,\colore] 			(\xx4+\stepsmall,			\y4+\step+.1)--			 (\xx4+\stepsmall+\step,	\y4+\step+.1);
		\draw[thick,\colore] 			(\xx4+\stepsmall,	 		\y4+\step-.1)--	 		 (\xx4+\stepsmall+\step,	\y4+\step-.1);

		\draw[ultra thick,\coloreagg]      (\xx4+\stepsmall+\step,\y4-.1) .. controls (\xx4+\stepsmall+\step+.5,\y4+0.2*\step) and (\xx4+\stepsmall+\step+.5,\y4+0.8*\step)  .. (\xx4+\stepsmall+\step,\y4+\step+.1);
		\draw[thick,\colore] (\xx4+\distaone,			\y4) --		 	 (\xx4+\distaone+\step,		\y4);
		\draw[thick,\colore] (\xx4+\distaone,			\y4+\step) -- 		 (\xx4+\distaone+\step,	\y4+\step);
		\draw[thick,\colore] (\xx4+\distaone-.1,		\y4) --			 (\xx4+\distaone+\step-.1,	\y4+\step);
		\draw[thick,\colore] (\xx4+\distaone+.1,		\y4) --		 	 (\xx4+\distaone+\step+.1,	\y4+\step);		
		\draw[thick,\colore] (\xx4+\distaone-.1,		\y4+\step) --		 (\xx4+\distaone+\step-.1,	\y4);
		\draw[thick,\colore] (\xx4+\distaone+.1,		\y4+\step) --		 (\xx4+\distaone+\step+.1,	\y4);
		
		\draw[ultra thick,\coloreagg]     (\xx4+\distaone+\step,\y4-.1) .. controls (\xx4+\distaone+\step+.5,\y4+0.2*\step) and (\xx4+\distaone+\step+.5,\y4+0.8*\step)  .. (\xx4+\distaone+\step,\y4+\step+.1);
		\draw[thick,\colore] 			(\xx4+\distatwo,			\y4)--				 	(\xx4+\distatwo+\step,	\y4+\step);
		\draw[thick,\colore] 			(\xx4+\distatwo,			\y4+\step)--		 	(\xx4+\distatwo+\step,	\y4);
		\draw[thick,\colore] 			(\xx4+\distatwo,			\y4)--					(\xx4+\distatwo+\step,	\y4);
		\draw[thick,\colore] 			(\xx4+\distatwo,	 		\y4+\step)--	 		(\xx4+\distatwo+\step,	\y4+\step);
		\draw[thick,\colore] 			(\xx4+\distatwo,			\y4)--					(\xx4+\distatwo,			\y4+\step);
		\draw[thick,\colore] 			(\xx4+\distatwo+\step,	 \y4)--	 		 	(\xx4+\distatwo+\step,	\y4+\step);

		\draw[ultra thick,\coloreagg]      (\xx4+\distatwo+\step,\y4-.1) .. controls (\xx4+\distatwo+\step+.5,\y4+0.2*\step) and (\xx4+\distatwo+\step+.5,\y4+0.8*\step)  .. (\xx4+\distatwo+\step,\y4+\step+.1);

		\draw[thick] 	(\x4+\distaone+0.9*\step,		\y4-.3) --			(\x4+\distaone+0.9*\step,\y4+\step+.3);		
		\draw[thick]	 (\x4+\distaone+0.9*\step-.2,	\y4-.3)	 -- 		(\x4+\distaone+0.9*\step,\y4-.3);
		\draw[thick]	 (\x4+\distaone+0.9*\step-.2,	\y4+\step+.3)	-- 	(\x4+\distaone+0.9*\step,\y4+\step+.3);

		\filldraw 	(\xx4+\stepsmall,		\y4)				 	circle (.15);
		\filldraw 	(\xx4+\stepsmall+\step,	\y4) 		 			circle (.15);
		\filldraw 	(\xx4+\stepsmall,		\y4+\step) 			circle (.15);
		\filldraw 	(\xx4+\stepsmall+\step,	\y4+\step) 	 		circle (.15);
		\draw 	(\xx4+\stepsmall+\step,	\y4) 					node[right=0.95*\step cm,above=0.3*\step cm] {$\ +\  A_5^{\rm exc.}$};
		\filldraw    	 (\xx4+\distaone,		\y4)				 circle (.15);
		\filldraw	 (\xx4+\distaone+\step,	\y4) 		 		 circle (.15);
		\filldraw	 (\xx4+\distaone,		\y4+\step) 		 circle (.15);
		\filldraw	 (\xx4+\distaone+\step,	\y4+\step) 		 circle (.15);
		\draw 	(\xx4+\distaone+1.5*\step,	\y4) 				node[right=0.8*\step cm,above=0.3*\step cm] {$\ +\ \ A_6$};
		
		\filldraw    	 (\xx4+\distatwo,		\y4)				 circle (.15);
		\filldraw	 (\xx4+\distatwo+\step,	\y4) 		 		 circle (.15);
		\filldraw	 (\xx4+\distatwo,		\y4+\step) 		 circle (.15);
		\filldraw	 (\xx4+\distatwo+\step,	\y4+\step) 		 circle (.15);
		\draw 	(\xx4+\distatwo+1.5*\step,	\y4) 				node[right=0.6*\step cm,above=0.3*\step cm] {};

\draw (\xx4,\y4-.5) node {\rule{0pt}{.5cm}};

	      \end{tikzpicture}
           \end{center}
 \end{minipage}
 \label{Figure33nn}
\end{center}
In free theory we find $A_{i=2,3,4,5}^{\rm exc.}=A_{i=2,3,4,5}$ and $A_3^{\rm free}=A_4^{\rm free}$, with all the other constants given by
\bea
A_1^{\rm free}=3\n N^{3+\n},\qquad  (A_2^{\rm free},\, A_3^{\rm free},\, A_5^{\rm free},\, A_6^{\rm free})=3\n\left( \n-3,\, 1,\,\n-2, \,2\right)\frac{A_1^{\rm free}}{N^2\ }\ .
\eea
The special cases are $\n=3$, $A^{\rm free}_2=1$ and $\n=4$, $A^{\rm free}_2=0$. 
The OPE prefactor is  $ g_{12}^2 g_{34}^\n$ and we can rewrite the correlator as 
\begin{align}
\label{33nnfreeDO}
\la \OO_3\OO_3\OO_q\OO_q \ra^{\rm free}= g_{12}^2 g_{34}^\n\, A_1^{\rm free} \bigg[\ &1\ + \nn \\
						    & 	\frac{3q}{N^2} \left(  u\sigma +\frac{u\tau}{v}+ u^2\sigma^2+\frac{u^2\tau^2}{v^2} +2 \frac{u^2\sigma\tau}{v} \right) + \nn \\[.1cm]
						    &  \frac{3q}{N^2} \bigg( (q-3) \left(u^3\sigma^3 +\frac{u^3\tau^3}{v^3} \right) \nn\\
						    &\rule{3cm}{0pt}  +(q-2)\left( \frac{u^3\sigma^2\tau}{v} + \frac{u^3\sigma\tau^2}{v^2}  \right)\bigg) \bigg]	\ . 
\end{align}
From the Mellin integral \eqref{Rastelli1} we find
\bea
\la\OO_3\OO_3\OO_\n\OO_\n\ra^{\rm dyna}&=&
							g_{12}^3 g_{34}^\n\ \norma_{33\n\n}\ \Intri(u,v,\sigma,\tau)\, \HH^{\rm dyna}\\[.2cm]
\HH^{\rm dyna}_{33\n\n} &=&	u^\n \Big[\   \sigma \Dbar_{\n-1,\n+2,2,3}  +\tau  \Dbar_{\n-1,\n+2,3,2} \nn \\[0cm]
				     & &	\rule{1cm}{0pt}	 +\frac{1}{\n-2}  \Dbar_{\n,\n+2,2,2} + \left( \frac{1}{\n-2} +\sigma+\tau\right)  \Dbar_{\n,\n+2,3,3}\Big]\ .
\eea
Results for $\n>3$ are novel compared to the supergravity literature. 
In $\la\OO_3\OO_3\OO_\n\OO_\n\ra$ there are ten $SU(4)$ channels corresponding to the intersection
\beq
([0,3,0]\otimes[0,3,0])\cap ([0,\n,0]\otimes[0,\n,0])=[0,3,0]\otimes[0,3,0]\ .
\eeq
These include contributions from long multiplets whose lowest dimension operators 
belong to $[0,0,0]$, $[1,0,1]$ and $[0,2,0]$, respectively. These three channels correspond to a decomposition of $\HH^{\rm dyna}$ of the form
\bea
&&
\HH_{33\n\n}^{\rm dyna}=
\frac{u^\n}{\n-2}\,\left[\Dbar_{\n,\n+2,2,2}+\frac{\n+1}{3}\Dbar_{\n,\n+2,3,3}+\frac{q-2}{6}\left(  \Dbar_{\n-1, \n+2, 3, 2}+ \Dbar_{\n-1, \n+2, 2, 3}\right) \right]\Upsilon_{00}\ \notag \\[.2cm]
&& 
+u^\n\,\left[ \left[ \frac{\Dbar_{ \n-1, \n+2, 2, 3}-\Dbar_{\n-1, \n+2, 3, 2}}{2} \right] \Upsilon_{10}
+  \left(\Dbar_{\n-1, \n+2, 3, 2}+\Dbar_{ \n-1, \n+2, 2, 3} + 2 \Dbar_{\n, \n+2, 3, 3}\right)\Upsilon_{11}\right]\nn\\
\label{33nnSU4}
\eea
%
%
A new feature compared to $\la\OO_2\OO_2\OO_\n\OO_\n\ra$ is the presence of several $\Dbar^{}_{\delta_1\delta_2\delta_3\delta_4}$ for each channel. 
This implies a more intricate recombination analysis of the superconformal OPE \cite{Dolan:2004iy,Doobary:2015gia}.   
Nevertheless, since the very first contribution to the $[0,0,0]$ channel in free theory only comes from a twist-two scalar,
the absence of a twist-two long multiplet in the spectrum can be unambiguously detected from
\begin{align}
\la\OO_3\OO_3\OO_\n\OO_\n\ra\Big|_{[0,0,0]}\sim\ & \ A_1 + \frac{u}{6} \left(\frac{A_3}{v}+A_3^{\rm\, exc.}\right)\nn \\
									&
									   +\frac{u^2}{20} \left( \frac{A_4}{v^2}+A_4^{\rm\,exc.}+\frac{1}{3}\frac{A_6}{v}\right)
									   +\frac{u^3}{50} \left(\frac{A_2}{v^3} +A_2^{\rm\,exc.}+\frac{1}{6} \frac{A_5}{v^2}+\frac{1}{6}A_5^{\rm\,exc.}\right)\,.
\end{align}
where $A_i=A_i^{\rm free}+A_i^{\rm dyna}$.
Following a procedure similar to that outlined for $\la\OO_2\OO_2\OO_n\OO_n\ra$,
we find that the  
the relevant terms are $A_3^{\rm free}$ and
\beq
A_1^{\rm dyna}= \frac{  \norma_{33\n\n}  }{\n-2}\,u^\n v  \left( \Dbar_{\n,\n+2,2,2} + \Dbar_{\n,\n+2,3,3} \right)\, ,
\eeq
where the precise form of $ \Dbar^{\rm sing}_{\n,\n+2,2,2}$ and $\Dbar^{\rm sing}_{\n,\n+2,3,3}$ can be obtained from \eqref{Dsingularseries}. 
Importantly, only  $\Dbar^{\rm sing}_{\n,\n+2,2,2}$
will provide the leading power $u^1$, and the other $\Dbar_{\n,\n+2,3,3}$ can be discarded. The equation to be solved is then 
\beq
3\n\,\frac{A_1^{\rm free}}{N^2} + \norma_{33\n\n} \frac{\Gamma[\n-1]}{\n-2} =0
\eeq
and the solution
\beq
\norma_{33\n\n}=-\frac{3\n}{(\n-3)!}\,\frac{A_1^{\rm free}}{N^2}\ .
\eeq


\subsubsection*{Normalization $\norma_{ppnn}$}

The analysis of the singlet channel in $\la\OO_3\OO_3\OO_\n\OO_\n\ra$ captures the generic features of the four point correlator 
$\la\OO_p\OO_p\OO_\n\OO_\n\ra$. Two comments are in order: 
Firstly, the leading contributions to the scalar channel of the correlator is  
\beq
\la\OO_p\OO_p\OO_\n\OO_\n\ra\Big|_{[0,0,0]}\sim \ A_1 + \frac{u}{6} \left(\frac{A_3}{v}+A_3^{\rm\, exc.}\right)\ +\ O(u^2)\ ,
\eeq
where $A_{3}^{\rm free}$ has been given in \eqref{PaulA3general}. Secondly, 
even though several $\Dbar$ functions will contribute to $A^{\rm dyna}_1$ 
only $\Dbar_{\n,\n+2,2,2}$ is relevant for the twist-two cancellation. 
From the definition of $\Intri(u,v,\sigma,\tau)$ and the Mellin formula \eqref{Rastelli1}
we obtain
\beq
A_1^{\rm dyna}= (p-2)! (\n-1)_{2-p}\, u^\n v\, \Dbar_{\n,\n+2,2,2}+\ldots
\eeq 
where the dots stands for those $\Dbar^{}_{\delta_1\delta_2\delta_3\delta_4}$ which do not contribute to the argument. We have assumed $q\ge p$, and
the non trivial coefficient $(p-2)! (\n-1)_{2-p}$ can be checked explicitly in the examples \eqref{44nnexample}-\eqref{55nnexample}.

It then follows from the twist-two long cancellation that
\beq
p\n\,\frac{A_1^{\rm free}}{N^2} + \norma_{pp\n\n}  (p-2)! \Gamma[\n-1] (\n-1)_{2-p}  =0
\eeq
with solution 
\beq
\norma_{pp\n\n}=-\frac{p}{(p-2)!}\frac{q}{(\n-p)!}\, \frac{A_1^{\rm free}}{N^2}\ .
\eeq


\section{Determining strong coupling data from the correlator}
\label{sec4}

Having described the structure of the free theory and tree-level supergravity results that we need, we now proceed to analyse the OPE. The knowledge of the OPE leads to an exact superconformal block representation of any four-point correlator, including both short and long exchanged representations. If we restrict attention to the contribution of long multiplets, which comes from 
the free theory as well as from $\HH^{\rm dyna}_{p_1 p_2 p_3 p_4}$, we find
\bea\label{Mixing1}
\la \OO_{p_1}\OO_{p_2}\OO_{p_3}\OO_{p_4}\ra^{\rm long}&=& N^{ \Sigma  }\,  \Pref ^{\rm (OPE)}
 												\sum_{\tw,\,l,\,\mathfrak{R}} A^{\{p_i\}}_{\,\mathfrak{R}}(\tw|l)\  \LL^{\{p_i\}}_{\, \mathfrak{R} }(\tw|l),\\ 
A^{\{p_i\}}_{\,\mathfrak{R}}(\tw|l)&=& \sum_{\OOL\in\,\mathfrak{R}} C_{p_1p_2\OOL} C_{p_3p_4\OOL}\ .
\eea
Here the operators have been normalised as in  \eqref{def_halfBPS} with $\Sigma= ({p_1+p_2+p_3+p_4})/2$. 
The explicit expression for $\LL^{\{p_i\}}_{\, \mathfrak{R} }(\tw|l)$ can be read from \eqref{longmultiplet}.
Expanding both the dimensions and OPE coefficients up to leading order in $1/N^2$, 
\beq
\Delta_{\OOL}=\Delta^{(0)}_{\OOL} + \frac{2}{N^2} \eta_{\OOL},\qquad C_{p_1p_2\OOL}=C^{(0)}_{p_1p_2\OOL}+\frac{1}{N^2}C^{(1)}_{p_1p_2\OOL}\,,
\eeq
we obtain the following refinement
\begin{align}
\la \OO_{p_1}\OO_{p_2}\OO_{p_3}\OO_{p_4}\ra^{\rm long}= N^{ \Sigma } \,\Pref^{\rm (OPE)} & 
									\Bigg( \sum_{\tw_0} \sum_{l,\, \mathfrak{R}} \AA^{\{p_i\}}_{\,\mathfrak{R}}(\tw_0|l)\  \LL^{\{p_i\}}_{\, \mathfrak{R} }(\tw_0|l) \nn\\[.2cm]
 &\rule{.3cm}{0pt}
+\frac{1}{N^2}\log(u)\, \sum_{\tw_0}\sum_{l,\,\mathfrak{R}} \symbol^{\{p_i\}}_{\,\mathfrak{R}}(\tw_0|l)\ \LL^{\{p_i\}}_{\, \mathfrak{R} }(\tw_0|l)\, +\,  \ldots \rule{.2cm}{0pt} \Bigg) \nn
\end{align}
where at order $1/N^2$ we omitted analytic terms in $u$, which will not be relevant for our discussion. 
Here $\tw_0=(\Delta^{(0)}_{\OOL}-l)/2$ and we defined
\bea
\label{mixing_equation1}
\AA^{\{p_i\}}_{\,\mathfrak{R}}(\tw|l)&=& \sum_{\OOL\in\,\mathfrak{R}} C^{(0)}_{p_1p_2\OOL} C^{(0)}_{p_3p_4\OOL},\\[.2cm]
\symbol^{\{p_i\}}_{\,\mathfrak{R}}(\tw|l)&=& \sum_{\OOL\in\,\mathfrak{R}} \eta_{\OOL}C^{(0)}_{p_1p_2\OOL} C^{(0)}_{p_3p_4\OOL}\ .
\label{mixing_equation2}
\eea
The data on the l.h.s of these equations will be obtained from the explicit form of the correlators. In particular, disconnected free theory determines $\AA_{\mathfrak{R}}(\tw|l)$, whereas $\symbol_{\mathfrak{R}}(\tw|l)$ is obtained from the leading $\log(u)$ singularity of $\HH^{\rm dyna}$.

A fundamental assumption we will make about the supergravity limit
is that the only operators surviving are in one-to-one correspondence 
with single-trace half-BPS operators $\OO_p$ and multi-trace operators  $\OOL_{\tw,l}$ built from products of the $\OO_p$. 
In the large $N$ expansion three point functions of half-BPS operators are $1/N$ suppressed, as the computation \eqref{sugrapathintegral} shows, and in any case contribute to the protected sector in the OPE.
We expect the double-trace operators to be the only long operators $\OOL_{\tw,l}$ to have non-vanishing three-point functions $C_{p_1 p_2 \OOL}^{(0)}$. Triple-trace and higher multi-trace operators are expected to have their three-point functions suppressed by further powers of $1/N^2$, i.e. they will start contributing to $C_{p_1 p_2 \OOL}^{(1)}$ and higher. 
	
In the first instance we will focus on unprotected operators in the singlet representation of $SU(4)$, 
since these are the operators whose data ultimately determine the loop correction ($O(1/N^4)$) 
to $\langle \mathcal{O}_2 \mathcal{O}_2 \mathcal{O}_2 \mathcal{O}_2 \rangle$~\cite{ Aprile:2017bgs  }. 
The exchanged singlet operators in question have the following description in the free theory:
\begin{align}
	K^{\text{free}}_{t,l,i} = \mathcal{O}_{i+1}\Box^{t-i-1}\partial^l \mathcal{O}_{i+1} + \dots \label{kweak}
\end{align}
where the $SU(4)$ indices are understood to be contracted to produce a singlet, 
and the ellipsis denotes similar terms with the space-time derivatives distributed 
differently between the two constituent operators, $\mathcal{O}_{i+1}$. 
The precise combination will not be important here, but importantly there is a unique combination yielding a conformal primary operator.
The operators given in (\ref{kweak}) have spin $l$ and dimension $2t+l$ (i.e. twist $2t$)  while $i=1\dots t{-}1$ labels the $(t-1)$ different operators which have  the same spin and dimension.  As soon as the coupling is turned on, these $(t-1)$ operators will mix and develop anomalous dimensions. 

At strong  coupling with large $N$, the operators again take their free theory dimensions, 
with anomalous dimensions developing at order $1/N^2$. Since the operators (\ref{kweak}) 
are protected at infinite $N$ they all remain present in the spectrum even though they reside in long multiplets. 
It no longer makes sense to write the operators explicitly as~\eqref{kweak}, but the number of operators is the same.
Thus we denote by $K_{t,l,i}$, with $i=1,\dots,t{-}1$,  the corresponding operators at strong coupling. 
They are operators which have well-defined anomalous dimensions at  $O(1/N^2)$. 
This automatically means their two-point functions  are orthogonal at $O(N^0)$ and we can also normalise them,  so we have
 \begin{align}
 	\langle K_{t,l,i} K_{t,l,i'}\rangle = \delta_{ii'}\ .
 \end{align}
Since we only consider them at leading order in $1/N^2$, we will also drop the superscript from the three-point functions $C_{p_1 p_2 K_{t,l,i}}^{(0)}$ and just write $C_{p_1 p_2 K_{t,l,i}}$ instead. 


We wish to obtain the anomalous dimensions $\eta_{t,l,i}$ of the operators $K_{t,l,i}$ as well as their large $N$ three-point functions $C_{p p K_{t,l,i}}$. 
First note that at leading order in the large $N$ limit the OPE of $\mathcal{O}_{p}\mathcal{O}_{p}$ contains the operators $K_{t,l,i}$ for all $t\geq p$. 
Thus for fixed $t$, the four-point correlators $\langle \mathcal{O}_{p} \mathcal{O}_{p} \mathcal{O}_{q} \mathcal{O}_{q} \rangle$ with $p \leq q$
contain information about operators $K_{t,l,i}$ for all $q \leq t$.   
Noting the $p\leftrightarrow q$ symmetry we deduce that there are $t(t-1)/2$ such independent correlators. 
We can then organize the information $\AA^{\{p_i\}}_{\,\mathfrak{R}}(\tw|l)$ coming from each correlator in the free theory at leading order into the following symmetric matrix,
\bea\label{matrixsymbol}
\namedue\Big|_{[0,0,0]}&=&\left( \begin{array}{ccccc} 
\AA^{2222} & \AA^{2233}& \ldots&    \AA^{22tt} \\
  & \AA^{3333} & \ldots &   \AA^{33tt} \\
 &  & \ldots &    \ldots \\
 & & &  \AA^{tttt}
 \end{array} \right) \ .\\
\nn
\eea
In fact, from the form of the large $N$ free theory correlators one can see immediately that the above matrix $\hat{\mathcal{A}}$ is actually diagonal.
Likewise we can organise the information $\symbol^{\{p_i\}}_{\,\mathfrak{R}}(\tw|l)$ coming from the $\log u$ term at order $1/N^2$ in each correlator into another symmetric matrix,
\bea
\label{matrixsymbol2}
\nameuno\Big|_{[0,0,0]}&=&\left( \begin{array}{ccccc} 
\symbol^{2222} & \symbol^{2233} & \ldots & \symbol^{22tt}  \\
  & \symbol^{3333} &   \ldots & \symbol^{33tt} \\
 &  &   \ldots&\ldots \\
 & & & \symbol^{tttt}
 \end{array} \right)\ . 
\rule{0pt}{1cm} \nn
\eea
Both in $\nameuno$ and $\namedue$ we have just given the independent entries in the upper triangular part explicitly.

Consider now the $(t{-}1)$ independent operators $K_{t,l,i}$. They are associated to
$(t-1)^2$ three-point functions $C_{p p K_{t,l,i}}$ where $i=1,\dots t{-}1$ and $p=2,\dots,t $,
and $(t{-}1)$ anomalous dimensions $\eta_{t,l,i} $. 
In total therefore we have $t(t{-}1)$ unknowns that need to be determined. 
Thus the matrices (\ref{matrixsymbol}) and (\ref{matrixsymbol2}) contain the precise amount of data needed!
The reason for this precise matching of degrees of freedom is that the operators 
$K_{t,l,i}$ are  (in one-to-one correspondence with) bilinears of half-BPS single-trace operators.
The matching is thus a remarkable feature of large 't Hooft coupling and large $N$ only, as in general there will be many other types of operators contributing.

Let us now examine the equations \eqref{mixing_equation1}-\eqref{mixing_equation2} in detail, beginning with low twist cases.
To simplify notation a little, we redefine $C_{pp K_{t,l,i}}$ in favor of $c_{pi}$ taking out a universal factor which we find is always present,  
\begin{align}
	(C_{pp K_{t,l,i}})^2 = \frac{(l+t+1)!^2}{(2l+2t+2)!} c_{pi}^2\,, \qquad p=2,\ldots,t, \quad i=1,\ldots,t-1\ .
\end{align}
At fixed twist we expect $c_{pi}$ to depend non trivially on $l$.

\subsection{Twist 4}

Here there is only one operator contributing and it only appears in the simplest correlator $\langle \mathcal{O}_2 \mathcal{O}_2 \mathcal{O}_2 \mathcal{O}_2 \rangle$.
Extracting the relevant superblock coefficient we obtain at leading order (from the disconnected free correlator)
\bea
(C_{22 K_{t,l,1}})^2=\AA^{2222} & \Rightarrow &   c_{21}^2 = \frac{4}{3} (l+1)(l+6)\,, \\
\eta_1(C_{22 K_{t,l,1}})^2=\symbol^{2222} &\Rightarrow &  c_{21}^2 \eta_1 = -64\,.
\eea
This clearly yields
\begin{align}
\eta_1 = -\frac{48}{(l+1) (l+6)},\qquad c_{21} =  \sqrt{\frac{4(l+1) (l+6)}{3}}\ .
\label{twist4sol}
\end{align}
This result has been known for a long time~\cite{Dolan:2001tt}. Note the symmetry $l \rightarrow - 7 - l$.

\subsection{Twist 6}

The situation becomes more interesting when we move to twist 6. Here there are two operators contributing, $K_{3,l,1}$ and $K_{3,l,2}$. The free theory results give:
\bea
c_{21}^2 + c_{22}^2 &=& \frac{2}{5}(l+1)(l+8)\,, \notag \\[.2cm]
c_{31}^2 + c_{32}^2 &=& \frac{9}{40}(l+1)(l+2)(l+7)(l+8)\,, \notag \\[.2cm]
c_{21} c_{31} + c_{22} c_{32} &=& 0\,.\label{twist6}
\eea
It is interesting at this point to compare briefly with the free gauge theory at large $N$. The relevant correlator (disconnected free correlator) is {\em exactly the same} as the one we are discussing here  at strong coupling.   However, despite this one should not be tempted to assume the leading large $N$ three-point functions are also the same at strong and weak coupling.  In the free theory at large $N$ we recall that  the two operators are explicitly given as  $K_{3,l,1} =\mathcal{O}_{2}\partial^l\Box\mathcal{O}_{2}+\dots$ and $K_{3,l,2} =\mathcal{O}_{3}\partial^l\mathcal{O}_{3}+\dots$. Although in general other operators contribute at weak coupling (single trace etc.), at large $N$ only these two contribute (the OPE can easily be performed explicitly via Wick contractions to verify this). Further the three point functions $c^{\text{weak}}_{22}$ and $c^\text{weak}_{31}$ are supressed at this order and thus the solution of the above equations reads simply:
\begin{align}
	c^{\text{weak}}_{22}=c^\text{weak}_{31}=0,\  (c^{\text{weak}}_{21})^2={\frac{2}{5}(l+1)(l+8)},\  (c^{\text{weak}}_{32})^2={\frac{9}{40}(l+1)(l+2)(l+7)(l+8)}\ ,
\end{align}
and the three-point functions $c^{\text{weak}}_{pi}$ are diagonal.

The strong coupling  interpretation of the equations turns out to be very different however, even though it arises from the same free disconnected correlator. The dynamical parts of the correlators give
\bea
c_{21}^2 \eta_1 + c_{22}^2 \eta_2 &=& -96\,, \notag \\[.2cm]
c_{31}^2 \eta_1 + c_{32}^2 \eta_2 &=& -54(l^2 + 9l +44)\,, \notag \\[.2cm]
c_{21} c_{31} \eta_1 + c_{22} c_{32} \eta_2 &=& 432\,, \,
\eea
and in particular the last equation means that here the three-point $c_{pi}$ functions cannot be diagonal. Instead we straightforwardly solve the above equations and obtain the solution
\begin{align}
\begin{split}
\eta_1 = -\frac{240}{(l+1) (l+2)},&\qquad\eta_2= -\frac{240}{(l+7) (l+8)},\\
c_{21} = {-}\sqrt{\frac{2(l+1) (l+2) (l+8)}{5 (2 l+9)}},&\qquad c_{22}=-\sqrt{\frac{2(l+1) (l+7) (l+8)}{5 (2 l+9)}},\\
c_{31} = \sqrt{\frac{9(l+1) (l+2) (l+7)^2 (l+8)}{40 (2 l+9)}},& \qquad c_{32}=-\sqrt{\frac{9 (l+1) (l+2)^2 (l+7) (l+8)}{40 (2 l+9)}}.
\end{split}\label{twist6sol}
\end{align}

\subsection{General twist} 

The first task in attempting to understand the general structure is to generalise the equations we obtain from the correlators via the superconformal block expansion. 
At leading order the situation is simpler, since off-diagonal correlators $\langle \mathcal{O}_{p} \mathcal{O}_{p} \mathcal{O}_{q} \mathcal{O}_{q} \rangle$ with $p\neq q$ are suppressed and therefore the matrix $\namedue$ is diagonal. We have computed a number of explicit examples and spot the pattern%
\footnote{In more detail,  we first computed the cases with  $p=t$ up to 6 and spotted a pattern for these which we then confirmed at $p=7$. Next we considered cases for fixed $p$ general $t$, some of which were already available~\cite{Dolan:2001tt,Doobary:2015gia}. We spotted a pattern for these up to a numerical $p$ dependent coefficient using results up to $p=5$. This final numerical factor we can then fix as a function of $p$ uniquely by comparison with the $p=t$ case.}  that leads to the following general formula,
\begin{align}
&\!\!\AA^{pppp}\Big|_{[0,0,0]}=\notag\\[.2cm]
& \!\! \frac{24 (l+1) (t-2)! (t!)^2 (l+2 t+2) (l+t-1)! ((l+t+1)!)^2 (p+t)! (l+p+t+1)!}{(p+1) (p-2)! ((p-1)!)^3 (2 t)! (t+2)! (l+t+3)! (2 l+2 t+2)! (t-p)! (l-p+t+1)!}.\label{ppt}
\end{align}
Let us notice that $\AA^{pppp}$ has completly factorized form. 
For fixed twist, we can define the matrix of three-point function coefficients 
\beq
C(t| l)=
\left(
\begin{array}{cccc}
C_{22 K_{t,l,1}} & C_{22 K_{t,l,2}}  & \ldots  & C_{22 K_{t,l,t-1} }\\
C_{33 K_{t,l,1}} & C_{33 K_{t,l,2}} & \ldots & \\
\ldots & & &\\
C_{t t K_{t,l,1}} & & & 
\end{array}\right)
\eeq
and rewrite the equations \eqref{mixing_equation1} in matrix form, 
\beq
\tilde{c}\, \tilde{c}^T=\text{Id}_{t-1},\qquad  C=    \widehat{\AA}\,^{\frac{1}{2} }\cdot \tilde{c}(t| l)
\eeq
where the orthonormality property of the matrix $\tilde{c}$ is manifest. 
Equations  \eqref{mixing_equation2} become
\beq\label{rastellimatrix}
\tilde{c}\cdot \text{diag}\left(\eta_1,\ldots,\eta_{t-1}\right)\cdot \tilde{c}\,^T =   \widehat{\AA}^{-\frac{1}{2} }\cdot \nameuno \cdot \widehat{\AA}^{-\frac{1}{2} }
\eeq
The columns of $\tilde{c}(t|l)$, are then eigenvectors of the matrix $ \widehat{\AA}^{-\frac{1}{2} }\cdot \nameuno \cdot \widehat{\AA}^{-\frac{1}{2} } $ and
the anomalous dimensions are the corresponding eigenvalues. 
Notice from the structure of eq. (\ref{rastellimatrix}) (recalling that $\hat{\mathcal{A}}$ is diagonal) the remarkable property that $\det(\widehat{\symbol})$ will factorise. 
From the explicit expressions for $\symbol^{ppqq}$ obtained upon decomposing $\HH^{\rm dyna}$ in superconformal blocks this property is completely obscure.
In particular, $\symbol^{ppqq}$ is found to be proportional to a polynomial in $l$ of degree $2(p-2)$, with $p\le q$, which does not admit real roots. 
Their expressions are cumbersome and thus we will not display them explicitly. 

Let us rewrite in this new notation the solution at twists four and six from eqs. (\ref{twist4sol}) and (\ref{twist6sol}). The $\tilde{c}$ matrix in these two cases is
\bea
\tilde{c}(2|l)=1\,, & \qquad & 
 \tilde c(3|l) =	\left( \begin{array}{cc}
							\sqrt{\frac{l+2}{2 l+9}} 	& \sqrt{\frac{l+7}{2 l+9}} \\[.4cm]
							-\sqrt{\frac{l+7}{2 l+9}} 	& \sqrt{\frac{l+2}{2 l+9}} \\
			\end{array}\right)\,, 
\eea

where it can easily be verified that $\tilde{c}(3|l)\, \tilde{c}(3|l)^T=\text{Id}_2$. We also repeat the formulae for the anomalous dimensions for later convenience,
\bea
\eta_{2,l,1} = \left\{ \begin{array}{l} -\frac{48}{(l+1) (l+6)}\end{array}\right\} &\qquad&  
\eta_{3,l,i} =  \left\{\begin{array}{l} -\frac{240}{(l+1) (l+2)}, -\frac{240}{(l+7) (l+8)} \end{array}\right\}\,,
\eea
We now proceed by performing the superblock expansion to find $\nameuno$ up to higher values of $t\le 12$,  and solve for anomalous dimensions and $\tilde{c}(t|l)$.
From the solution at twist eight we obtain
\bea
\tilde c(4|l)&=&	\left(
	\begin{array}{ccc}
	\sqrt{\frac{7(l+2) (l+3)}{6(2 l+9) (2 l+11)}} 		 &  \sqrt{\frac{5(l+3) (l+8)}{3(2 l+9) (2 l+13)}} 	& \sqrt{\frac{7(l+8) (l+9)}{6(2 l+11) (2 l+13)}} \\[.5cm]
	- \sqrt{\frac{2(l+2) (l+8)}{(2 l+9) (2 l+11)}} 		 &  -\sqrt{\frac{35}{(2 l+9) (2 l+13)}} 			&  \sqrt{\frac{2(l+3) (l+9)}{(2 l+11) (2 l+13)}} \\[.5cm]
	 \sqrt{\frac{5(l+8) (l+9)}{6(2 l+9) (2 l+11)}}		 & - \sqrt{\frac{7(l+2) (l+9)}{3(2 l+9) (2 l+13)}} 	& \sqrt{\frac{5(l+2) (l+3)}{6(2 l+11) (2 l+13)}} \\
	\end{array}
	\right), 
\eea
and 
\bea
\eta_{4,l,i}&=&\left\{\begin{array}{l}
	-\frac{720(l+7)}{(l+1) (l+2) (l+3)}, 
	-\frac{720}{(l+3) (l+8)}, 
	-\frac{720(l+4)}{(l+8) (l+9) (l+10)} 
	\end{array}
	\right\} .
\eea

For higher twists the solution becomes quite lengthy so we find it helpful to introduce a more compact notation for the square root factors. We define
\be
(n) = \sqrt{l+n}\,,\qquad [n] = \sqrt{2l + n}\,.
\ee
With this more compact notation the solution at twist ten takes the form,

\bea
\label{twist10ctilde}
\tilde{c}(5|l)&=&\left(
\begin{array}{cccc}
\sqrt{\frac{3}{2}} \frac{(2) (3) (4)}{[9] [11] [13]} 	&   \sqrt{\frac{5}{2}} \frac{(3) (4) (9)}{[9] [13] [15]}       &    \sqrt{\frac{5}{2}} \frac{(4) (9) (10)}{[11] [13] [17]}    &    \sqrt{\frac{3}{2}} \frac{(9) (10) (11)}{[13] [15] [17]} \\[.3cm]
-\sqrt{\frac{27}{8}} \frac{(2) (3) (9)}{[9] [11] [13]} 	&   -\sqrt{\frac{5}{8}}  \frac{(l+18)(3)}{[9] [13] [15]}       &    \sqrt{\frac{5}{8}}  \frac{(l-5)(10)}{[11] [13] [17]}       &     \sqrt{\frac{27}{8}} \frac{(4) (10) (11)}{[13] [15] [17]} \\[.3cm]
\sqrt{\frac{5}{2}} \frac{(2) (9) (10)}{[9] [11] [13]}   &   -\sqrt{\frac{3}{2}} \frac{ (l-3)(10)}{[9] [13] [15]}        &   -\sqrt{\frac{3}{2}}  \frac{(l+16)(3)}{[11] [13] [17]}      &     \sqrt{\frac{5}{2}} \frac{(3) (4) (11)}{[13] [15] [17]} \\[.3cm]
-\sqrt{\frac{5}{8}} \frac{(9) (10) (11)}{[9] [11] [13]} &   \sqrt{\frac{27}{8}} \frac{(2) (10) (11)}{[9] [13] [15]} &   -\sqrt{\frac{27}{8}} \frac{(2) (3) (11)}{[11] [13] [17]}  &     \sqrt{\frac{5}{8}} \frac{(2) (3) (4)}{[13] [15] [17]} \\[.3cm]
\end{array}
\right),
\eea
and
\bea
\eta_{5,l,i}&=& \left\{
	\begin{array}{l}
	-\frac{1680 (l+7) (l+8)}{(l+1) (l+2) (l+3) (l+4)} ,
	-\frac{1680}{(l+3) (l+4)} ,
	-\frac{1680}{(l+9) (l+10)} ,
	-\frac{1680 (l+5) (l+6)}{(l+9) (l+10) (l+11) (l+12)} 
\end{array}\right\}.\nn\\
\eea

We begin to see intriguing structure in the entries of the matrix as well as in the anomalous dimensions. Note the symmetry $l\rightarrow - 2t - 3 - l$ which is an invariance of the set of anomalous dimensions and an invariance up to signs of the $\tilde{c}$ matrix under a flip about the vertical axis. Note also that at twist ten we see for the first time the appearance of polynomials in $l$ (without a square root) in the numerators of the central entries of \eqref{twist10ctilde}. At twist ten these polynomials are all linear, but their degrees increase as we increase the twist further.

Indeed, proceeding  to compute the next few examples one gets a better idea of the structure. The anomalous dimensions reveal a fairly simple structure that is consistent with the formula
\begin{align}
\eta_{t,l,i}^{[0,0,0]}=	-\frac{2 (t-1)_4 (t+l)_4 }{(l+2 i-1)_6 }\ ,
\label{anomdims}
\end{align}
where $(x)_n = x(x+1)\ldots(x+n-1)$ is the Pochhammer symbol.
Note that the anomalous dimensions are all negative for all physical values of $l$.

The $\tilde c(t|l)$ matrix is trickier to understand. Already from the results up to twist ten 
we note a pattern of square roots of linear factors of $l$. In addition we have seen that in the entries 
towards the centre one finds fewer square root factors in the numerator, and polynomials in $l$ without a square root. 
Note that the entries of the matrix always have a finite (but possibly vanishing) limit as $l\rightarrow \infty$. 
In fact, we can deduce the structure of $\tilde c(t|l)$ for a given twist in terms of an ansatz with some undetermined numbers,
\begin{align}
	\tilde c_{pi}^{[0,0,0]}=&\sqrt{\frac{2^{1-t} (2l+4 i+3) \left((l+i+1)_{t-i-p+1}\right){}^{\sigma_1}
			\left((t+l+p+2)_{i-p+1}\right){}^{\sigma_2}}{\left(l+i+\frac{5}{2}\right)_{t-1}}}  \notag\\
		&\times\sum _{k=0}^{\min (i-1,p-2,t-i-1,t-p)} l^k a^{[0,0,0]}_{(p,i,k)}.
		\label{ctildeansatz}
\end{align}
The powers of the Pochhammer factors inside the square root are signs given explicitly by
\be
\sigma_1 = \text{sgn}(t-p-i+1)\,,\qquad  \sigma_2 = \text{sgn}(i-p+1)\,.
\ee
where $p=2,\ldots,t$ and $i=1,\ldots,t-1$. We notice that the square root structure in $\tilde{c}_{pi}$ 
follows from complicated combinatorics, which nevertheless can be captured by the 
two (non-analytic) sign functions $\sigma_1$ and $\sigma_2$.  Around the outer frame of the matrix,
 the unfixed polynomial has degree 0, i.e. it is simply a constant. Its degree increases as we move towards the inside of the matrix. 
One can readily check (\ref{ctildeansatz}) is consistent with the examples given explicitly above and we have tested the structure up to $t=12$.

Given the ansatz (\ref{ctildeansatz}), we have reduced the problem to that of finding the constants $a(p,i,k)$.
Quite surprisingly, enforcing orthonormality of $\tilde c(t|l)$ uniquely fixes the solution\footnote{We have checked this up to twist 48 ($t=24$).}.
In more detail, we first insist that the first row has unit norm, $\sum_i \tilde c_{2i}^2 =1$. 
This is a linear equation in $a(2,i,0)^2$ with a  unique solution. 
In fact, the constraint is a rational function of  $l$ and so this single equation can fix more than one constant. 
Then, orthogonality of the rows $\sum \tilde c_{pi}\tilde c_{qi}=0$ for $p\neq q$ gives a linear system 
in the remaining variables and uniquely fixes them, up to an overall scale which is fixed by the unit norm condition.

We find it remarkable both that there exist such orthonormal matrices with the structure (\ref{ctildeansatz}) and that the matrix is uniquely fixed by orthonormality as a linear system. The fact that the problem is essentially linear means it can be solved quickly and we have complete data up to $t=24$. This enables us to spot patterns and write down general formulae. 
 
We do not have a completely general formula for the full matrix $\tilde c$ but we do have various cases in closed form. In particular the top row of the matrix is given by the formula 
\be
a^{[0,0,0]}_{(2,i,0)} = \frac{2^{t-1}(2i+2)!(t-2)!(2t-2i+2)!}{3(i-1)!(i+1)!(t+2)!(t-i-1)!(t-i+1)!}\,, \qquad i= 1,\ldots,t-1\,.
\ee
This formula completely specifies all the three-point function of the form $C_{\mathcal{O}_{2} \mathcal{O}_{2} K_{t,l,i}}$ 
which was an essential ingredient in the prediction of the one-loop supergravity correction to 
$\langle \mathcal{O}_2 \mathcal{O}_2 \mathcal{O}_2 \mathcal{O}_2 \rangle$ presented in \cite{Aprile:2017bgs}.

\subsection{Generalisation from $[0,0,0]$ to $[n,0,n]$ representations}

Having given the general structure of the solution to the mixing problem for singlet double-trace operators, 
we may now proceed to analysing more general $SU(4)$ representations. 
Specifically we can investigate operators in the series of representations $[n,0,n]$ which also arise in the OPE 
of correlation functions of the form $\langle \mathcal{O}_p \mathcal{O}_p \mathcal{O}_q \mathcal{O}_q \rangle$. 
For each channel of the form $[n,0,n]$ the structure of this problem is analogous to that of singlet channel. 
In particular, at twist $2t$ a basis of double trace operators in the $[n,0,n]$ representation will have the schematic form 
\be
\{ \mathcal{O}_{2+n} \Box^{t-n-2} \partial^l \mathcal{O}_{2+n}, \mathcal{O}_{3+n} \Box^{t-n-3} \partial^l \mathcal{O}_{3+n}, \ldots , \mathcal{O}_{t} \Box^{0} \partial^l \mathcal{O}_{t}\}\,.
\ee
and we expect $(t-1-n)$ superconformal primary operators. 
As for the singlet double trace operators in \eqref{kweak}, 
the precise form of these primary operators is a specific linear combination of the element of the basis, with derivatives acting on the two constituent operators.
These operators again have integer classical dimensions for infinite $N$ and receive anomalous dimensions at order $1/N^2$.

The analysis of the $[n,0,n]$ channel for fixed $n$ follows a very similar logic to that presented in the singlet case.
Once again we conclude that the series of correlators 
$\langle \mathcal{O}_p \mathcal{O}_p \mathcal{O}_q \mathcal{O}_q \rangle$ for $n+2 \leq p \leq q \leq t$ 
provides the right amount of information needed in order to solve for  
anomalous dimensions and three-point functions of the exchanged double trace operators. 
From the general form of the long superconformal blocks \eqref{longmultiplet} 
it is straightforward to isolate the appropriate channel,
and organize the data from the superblock expansion into the symmetric matrices $\nameuno\Big|_{[n,0,n]}$ and $\namedue\Big|_{[n,0,n]}$. 

Before presenting our general results we go through some specific examples.

\subsubsection{[1,0,1]}
In this channel the matrices $\nameuno\Big|_{[1,0,1]}$ and $\namedue\Big|_{[1,0,1]}$ have the form

\bea\label{matrixsymbol}
\nameuno\Big|_{[1,0,1]}=\left( \begin{array}{ccccl} 
\symbol^{3333} & \symbol^{3344} & \ldots & \symbol^{33tt}  \\
  & \symbol^{4444} &  \ldots & \symbol^{44tt} \\
 &  & \ldots  &\ldots \\
 & &  & \symbol^{tttt}
 \end{array} \right)\ , \\
\rule{0pt}{1cm} \nn\\
\namedue\Big|_{[1,0,1]}=\left( \begin{array}{ccccl} 
\AA^{3333} & \AA^{3344} &  \ldots & \AA^{33tt}  \\
  & \AA^{4444} &  \ldots & \AA^{44tt} \\
 &  & \ldots  &\ldots \\
 & &  & \AA^{tttt}
 \end{array} \right) \ ,\\
\nn
\eea
where $\namedue$ is diagonal with entries
\bea
\AA^{pppp}\Big|_{[1,0,1]}&=& \frac{15(p-2)(t-1)(t+2) (l+t)(l+t+3) }{(p+2) (t-2) (t+3) (l+t-1) (t+l+4) }\AA^{pppp}\Big|_{[0,0,0]}
\label{ppt[101]}
\eea
We can then introduce the orthonormal matrix $\tilde{c}(t|l)$ and start solving explicitly the mixing problem.  
For illustration, let us look at the first three cases:

At twist six there is only one operator, therefore
\bea
\tilde{c}(3|l)=1\qquad \eta_{3,l,1}=-\frac{144}{(3 + l) (6 + l)}
\eea
At twist eight there are two operators, and we find

\bea
\tilde{c}(4|l)&=&
\left(
\begin{array}{cc}
\sqrt{	\frac{l+2}{ 2l+11}	 } 	& 	\sqrt{  \frac{l+9}{2 l+11} }  \\[.6cm]
- \sqrt{	\frac{l+9}{2l+11}  } 	& 	\sqrt{  \frac{l+2}{2 l+11} }
\end{array}
\right)  
\eea
with anomalous dimensions
\bea
\eta_{4,l,i}&=& \left\{\begin{array}{l} -\frac{560 (8 + l)}{(2 + l) (4 + l) (7 + l)}, -\frac{560 (3 + l)}{(4 + l) (7 + l) (9 + l)} \end{array}\right\}
\eea

At twist ten it is becoming evident that the structure of eigenvectors and anomalous dimension found in the singlet case generalises to $[1,0,1]$ with minor modification. In particular

\bea
\tilde{c}(5|l)=
\left(
\begin{array}{ccc}
	 \sqrt{   \frac{	9(l+2)(l+3) }{8 (2l+11)(2l+13)   }} 		& 	 \sqrt{  \frac{7 (l+3)(l+10)}{ 4(2 l+11)(2l+15) } }     &    \sqrt{ \frac{9(l+10)(l+11)}{8(2l+13)(2l+15)}}\\[.6cm]
-	 \sqrt{   \frac{	2(l+2)(l+10) }{ (2l+11)(2l+13) }} 		& 	-\frac{ 3\sqrt{7} }{ \sqrt{  (2 l+11)( 2l+15)  }}  		  &    \sqrt{ \frac{2(l+3)(l+11)}{(2l+13)(2l+15)}}\\[.6cm]
 	\sqrt{	    \frac{  	7(l+10)(l+11)}{ 8(2l+11)(2l+13)  }}	&      - \sqrt{  \frac{9( l+2)(l+11)}{4 (2l+11)(2l+15)  } }      &     \sqrt{ \frac{7(l+2)(l+3)}{8(2l+13)(2l+15)}}
\end{array}
\right)
\eea
with anomalous dimensions
\bea
\eta_{5,l,i}&=& \left\{\begin{array}{l} -\frac{1440 (9 + l)}{(2 + l) (3 + l) (5 + l)}, -\frac{1440}{(5 + l) (8 + l)}, -\frac{1440 (4 + l)}{(8 + l) (10 + l) (11 + l)} \end{array}\right\}
\eea
The solution of the mixing problem up to $t=12$ can be found straightforwardly and leads to the expression

\beq
\eta_{t,l,i}^{[1,0,1]}=-\frac{2 (t-2)t (t+1)  (t+3 )(t+l-1)(t+l+1)(t+l+2)(t+l+4)  }{(l+2i)_{6} }
\eeq
for the anomalous dimensions, and

\begin{align}
	\tilde c_{pi}^{[1,0,1]}=&\sqrt{\frac{2^{1-t} (2l+4 i+5) \left((l+i+1)_{t-i-p+1}\right){}^{\sigma_1}
			\left((t+l+p+2)_{i-p+2}\right){}^{\sigma_2}}{\left(l+i+\frac{7}{2}\right)_{t-2}}}  \notag\\
		&\times\sum _{k=0}^{\min (i-1,p-3,t-i-2,t-p)} l^k a^{[1,0,1]}_{(p,i,k)}.
		\label{ctildeansatz_[101]}
\end{align}
for the entries of the $\tilde{c}(t|l)$ matrix,
with $\sigma_1 = {\rm sgn}(t-p-i+1)$, and  $\sigma_2 = {\rm sgn}(i-p+2)$, and $p=3,\ldots,t$ and $i=1,\ldots,t-2$.
The orthogonality condition of the matrix again determines completely the value of these $a(p,i,k)$ at any twist.

\subsubsection{From $[2,0,2]$ to $[n,0,n]$}

In this section, we present general formulae for the matrices $\nameuno$ and $\namedue$ given in terms of disconnected free theory data, anomalous dimensions and orthonormal $\tilde{c}(t|l)$ matrices. 

Let us begin from free theory, where we have obtained the following result, 
\bea
&&\AA^{pppp}(\tw|l)\Big|_{[n,0,n]} = \\[.2cm]
&&\rule{1cm}{0pt} \frac{ p^2 }{n! p!(p-1)!}\frac{  (n+2)_{n+3}  }{(p+1+n)!(p-2-n)!}\times \frac{ (t!)^2 }{ (2t)!} \frac{ (l+1)((1 + l + t)!)^2(l+2t+2)}{( 2 l + 2 t+2)!}\times\nn \\[.2cm]
&&\rule{1cm}{0pt} (l+t-p+2)_{p-2-n} (l+t+4+n)_{p-2-n} (l+1+t-n)_n(l+1+t+2)_n  \times\nn\\[.2cm]
&&\rule{1cm}{0pt} (t-p+1)_{p-2-n} (t+3+n)_{p-2-n}(t-n)_n(t+2)_n \nn 
\eea
Introducing the $\tilde{c}(t|l)_{[n,0,n]}$ matrices and computing $\nameuno_{[n,0,n]}$ 
for a large number of twist and several values of $n$ we have been able to fit and test both the anomalous dimensions and the entries of $\tilde{c}(t|l)$ with the following formulae:
For the anomalous dimensions we find,
\be
\eta^{[n,0,n]}_{t,l,i}=-\frac{2 (t-1 - n )t (t+1)  (t+2 + n )(t+l-n)(t+l+1)(t+l+2)(t+l+3+n)  }{(l+2i+n-1)_{6} }
\label{genanomdims}
\ee
 and for the entries of the $\tilde{c}(t|l)$ matrix,
\begin{align}
	\tilde c_{pi}^{[n,0,n]}=&\sqrt{\frac{2^{1-t} (2l+4 i+3+2n) \left((l+i+1)_{t-i-p+1}\right){}^{\sigma_1}
			\left((t+l+p+2)_{i-p+n+1}\right){}^{\sigma_2}}{\left(l+i+n+\frac{5}{2}\right)_{t-n-1}}}  \notag\\
		&\times\sum _{k=0}^{\min (i-1,p-n-2,t-n-i-1,t-p)} l^k a^{[n,0,n]}_{(p,i,k)}.
		\label{genctildeansatz}
\end{align}
The signs are given explicitly by
\be
\sigma_1 = {\rm sgn}(t-i-p+1)\,, \qquad \sigma_2 = {\rm sgn}(i-p+n+1)\,.
\ee
All unspecified coefficients $a(p,i,k)$ are again determined by imposing orthogonality of $\tilde{c}$.

\section{Analysis of the spectrum of anomalous dimensions}

Let us now analyse some general behaviour of the spectrum of anomalous dimensions that we found. Here we follow some of the arguments discussed in \cite{1410.4717}. Let us consider a very large, but finite value of $N$. From eq. (\ref{anomdims}) we find that our expression for the full twist of the operator $K_{t,l,i}$ in the singlet channel is
\be
\Delta^{[0,0,0]}_{t,l,i} - l = 2t - \frac{2}{N^2} \eta^{[0,0,0]}_{t,l,i} + \ldots = 2t - \frac{4}{N^2}\frac{(t-1)_4 (t+l)_4 }{(l+2 i-1)_6 } + \ldots\,.
\ee
We note that the numerator of the anomalous term behaves like $t^8$ for large $t$ and that the coefficient is negative. Keeping the leading terms for large $t$ we find
\be
\Delta^{[0,0,0]}_{t,l,i} - l  = 2t - \frac{4}{N^2}\biggl(\frac{t^8}{(l+2 i-1)_6 }+ O(t^7) \biggr)+\ldots\,.
\ee
As argued in \cite{1410.4717} these two facts imply that for some large classical twist $t$ the correction term will dominate over the classical term. Indeed for $t \sim N^{\frac{2}{7}}$ we find the two terms are of the same order and so the anomalous dimension formula inevitably requires corrections to avoid violating the unitarity bound.

\begin{figure}[!t]
    \centering
                \includegraphics[scale=0.5]{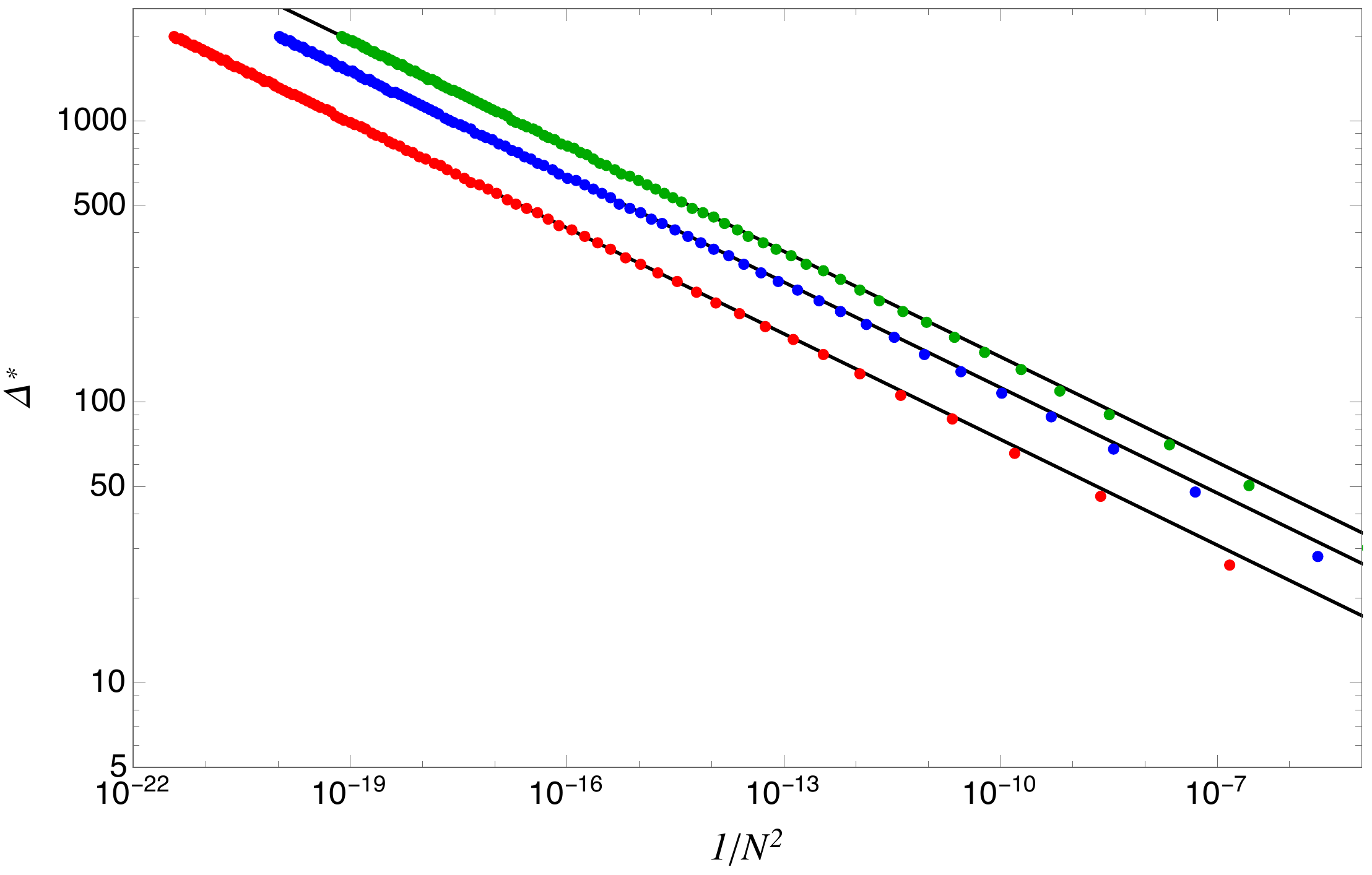}
    \caption{Varying $t$ we show on a log-log plot the value of the dimension $\Delta^\star_l$ at the crossing point $\Delta^{[0,0,0]}_{t+1,l,1} - \Delta^{[0,0,0]}_{t,l,t-1}=0$ as function of ${1}/{N^2}$ for $l=0$ (red), $l=2$ (blue) and $l=4$ (green). The best fit given by the solid black line is $\Delta^\star\approx u_l /N^{1/4}$ with $u_{l=0,2,4}\approx\{4,6.3,8\}$. }
    \label{fig-deltaint}
\end{figure}

In fact we can argue that one needs corrections even before $t$ reaches values of order $N^{\frac{2}{7}}$. Since we have resolved the mixing of the $(t-1)$ operators with the same classical twist $2t$ we may consider the differences in their dimensions. As already observed, the anomalous dimensions are all negative and so as one increases $1/N^2$ away from zero the dimensions decrease. One can see from the formula (\ref{anomdims}) that dimension of the operator with twist $i=1$ decreases fastest and the dimension of the operator with $i=t-1$ decreases slowest. We can then consider the slowest descending operator, $K_{t,l,t-1}$ at level $t$ and the fastest descending one $K_{t+1,l,1}$ at level $(t+1)$. The difference in their dimensions is
\be
\Delta^{[0,0,0]}_{t+1,l,1} - \Delta^{[0,0,0]}_{t,l,t-1} = 2 - \frac{4}{N^2}\frac{t^8}{(l+1)_6} + O(t^7)
\ee
Hence we find for $t \sim N^{\frac{1}{4}}$ that the two operators will become degenerate and then cross over in the values of their dimensions. Such level crossing should not occur at generic points in moduli space, it should only be associated with points of increased symmetry, such as the free theory limit. Thus we conclude that before we reach this point further corrections to the anomalous dimensions become relevant.
A plot of the value of the dimension at the crossing point against $1/N^2$ for $l=0,2,4$ is displayed in Fig. \ref{fig-deltaint}.

It also interesting to consider the properties of the anomalous dimensions as functions of the spin $l$~\cite{Alday:2007mf,1212.3616,1212.4103,Alday:2017gde}.
In particular, the anomalous dimensions are conjectured to be negative, monotonic and convex as a function of $l$, at least for large enough $l$. 

Our results for all anomalous dimensions are negative for any values of $t$ and $l$.
By examining their precise form (\ref{anomdims}) (which are simply rational functions involving linear factors in $l$) one can straightforwardly see that for all values of $i\leq (t+1)/2$ the anomalous dimensions satisfy monotonicity and convexity for all values of $l > 1-2i$. This is simply  because at large $l$, $\eta_{t,l,i} \sim -\frac{2 (t-1) t (t+1) (t+2)}{l^2}$ is monotonic and convex, and for decreasing values of $l$, the first zero or pole is the negative pole at $l = 1-2i$.\footnote{The case $i=(t-1)/2$ for $t$ odd is in fact a special case, but in fact has no zero's and again the first special point is a negative pole at $l=-t-4$. It is thus negative, monotonic and convex for all $l>-t-4$.}

For  $i> (t+1)/2$, as we reduce the value of $l$, the anomalous dimension hits a zero at $l=-t$ before it reaches the  pole at $l = 1-2i$. Thus monotonicity and convexity break down at some point with convexity breaking down first. By considering the equation $\partial^2_l \eta_{t,l,i} =0$ we can study for which value of $l$ convexity breaks down as we reduce $l$. Assuming large $t$ (so we can approximate the resulting large polynomial equation with its highest powers)  the breakdown in convexity occurs at $l\sim 2 \sqrt{2} i+4 i-\sqrt{2} t-3 t$.
This is negative for $i<\frac{3+\sqrt{2}}{2 \left(2+\sqrt{2}\right)}t \sim 0.646 t$ and so the anomalous dimension is still convex, and monotonic for physical $l$ in this range. For operators with $i>0.646t$ on the other hand,  the anomalous dimension ceases to be convex for some finite positive value of $l$. 
The worst offender is the operator with the maximal value of $i=t-1$. This ceases to be convex for $l$ below approximately $ (1+\sqrt{2})t\sim 2.41t$.

\section{Conclusions}

We have presented a detailed analysis of the double trace spectrum of $\mathcal{N}=4$ super Yang-Mills theory in the supergravity limit. We have shown that the known tree-level supergravity results contain all the necessary information to resolve the degeneracy of the double trace operators in the large $N$ limit. Here we have focussed on the correlation functions of the form $\langle \mathcal{O}_p \mathcal{O}_p \mathcal{O}_q \mathcal{O}_q \rangle$ since these are sufficient to resolve the degeneracy of the double-trace operators in the $[n,0,n]$ representations of $SU(4)$. Similar methods can be applied to the more general cases $\langle \mathcal{O}_{p_1} \mathcal{O}_{p_2} \mathcal{O}_{p_3} \mathcal{O}_{p_4} \rangle$ to resolve the mixing for more general representations.

Our results for the leading order OPE coefficients and anomalous dimensions are surprisingly simple, even given the very compact Mellin space form of the tree-level supergravity correlators given in \cite{Rastelli:2016nze}. The fact that the anomalous dimensions admit such a simple formula as (\ref{genanomdims}) is remarkable. Even more remarkable perhaps is the universal structure we find in the orthogonal $\tilde{c}$ matrices. The fact that orthogonal matrices $\tilde{c}$ of the form (\ref{ctildeansatz_[101]}) exist at all is surprising. We should point out that modifications of the structure of the square root factors in (\ref{ctildeansatz_[101]}) typically lead to no orthogonal solution at all. Indeed the structure of the $\tilde{c}$-matrices in the $[n,0,n]$ case was first guessed based on this structure before being explicitly identified by analysing the relevant channels of the OPE. It would be very interesting to understand whether the structure (\ref{ctildeansatz_[101]}) arises due to some as yet unidentified simplicity which could suggest more about the higher order $1/N$ corrections to the quantities we have derived in this work.

The results we have presented here for the singlet channel have already been used in \cite{Aprile:2017bgs} to contruct a prediction for the one-loop correction to the $\langle \mathcal{O}_2 \mathcal{O}_2 \mathcal{O}_2 \mathcal{O}_2 \rangle$ correlator. Certainly similar analyses could be carried out to make one-loop predictions for more general correlators. This would rely on resolving the mixing for more general representations than we have examined here. 

Finally, while we have focussed on $\mathcal{N}=4$ super Yang-Mills theory here, the phenomenon of large $N$ degeneracy and the need for resolving mixing is presumably common to many holographic theories. Essentially the phenomenon arises because of the presence of a compact factor (here an $S^5$) in the gravity background which leads to the presence of a Kaluza-Klein tower of modes related to the massless gravity modes. For fixed twist and spin one will then typically have many double-trace operators one can consider and these will generically mix. It would be interesting to consider both other models and the generic structure of large $N$ CFTs further.

\section*{Acknowledgements}
FA would like to thank Gleb Arutyunov, Jorge Russo, Kostas Skenderis, Arkady Tseytlin, and Konstantin Zarembo for discussions on related topics. JMD and HP are supported by the ERC Grant 648630.
PH acknowledges support from an STFC Consolidated Grant ST/L000407/1 and also National Science Foundation under Grant No. NSF PHY-1125915. FA acknowledges support from STFC through Consolidated Grant ST/L000296/1.

\appendix
\section{$\overline{D}$-functions}
\label{app-Dfns}
The analytic part of a $\Dbar$-function is given by
\begin{align}
	\Dbar_{\delta_1\delta_2\delta_3\delta_4}^{\rm analytic}= 
	(-)^\sigma \sum_{n,m\ge 0} \frac{u^n }{n!(\sigma+n)!} \Lambda^{\delta_1\delta_2}_{\delta_3+\sigma\delta_4+\sigma}(n) 
				\frac{ (\delta_2+n)_m (\delta_3+\sigma+n)_m }{ (\delta_1+\delta_2+2n)_m }\,\mathfrak{f}_{nm} \frac{(1-v)^m}{m!\ }
\end{align}
where 
\bea
\mathfrak{f}_{nm}
		&= &\Big[ +\psi(n+1)+\psi(\sigma+1+n) + 2 \psi(\delta_1+\delta_2+2n+m) \rule{0pt}{.6cm} \nn\\
& & \rule{.5cm}{0cm} -\psi(\delta_4+\sigma+n)-\psi(\delta_1+n) -\psi(\delta_3+\sigma+n+m) -\psi(\delta_2+n+m)\Big]\,  \nn
\eea
and we recall the definition
\beq
\Lambda^{\delta_1\delta_2}_{\delta_3\delta_4}(n)\equiv \frac{ \Gamma[\delta_1+n]\Gamma[\delta_2+n]\Gamma[\delta_3+n]\Gamma[\delta_4+n] }{\Gamma[\delta_1+\delta_2+2n]}\ .
\eeq
In general, the full $\Dbar$-functions can be recursively generated by the action of differential operators on the four-dimensional scalar one-loop box integral $\Phi^{(1)}(u,v)$, for which there is an explicit expression in terms of polylogarithms, see equation~(\ref{one-loop-box}).
Starting with $\Dbar_{1111}(u,v):=\Phi^{(1)}(u,v)$, when $\delta_i$, and $\Sigma=(\delta_1+\delta_2+\delta_3+\delta_4)/2$ are integers one can generate any $\Dbar_{\delta_1\delta_2\delta_3\delta_4}$ from the following recursion relations~\cite{Arutyunov:2002fh}:
\def\du {\partial_u}
\def\dv {\partial_v}
\begin{align}
\Dbar_{\delta_1+1,\delta_2+1,\delta_3,\delta_4}&=-\du \Dbar_{\delta_1\delta_2\delta_3\delta_4},\nn\\
\Dbar_{\delta_1,\delta_2,\delta_3+1,\delta_4+1}&=(\delta_3+\delta_4-\Sigma-u\du)\Dbar_{\delta_1\delta_2\delta_3\delta_4},\nn\\
\Dbar_{\delta_1,\delta_2+1,\delta_3+1,\delta_4}&=-\dv\Dbar_{\delta_1\delta_2\delta_3\delta_4},\nn\\
\Dbar_{\delta_1+1,\delta_2,\delta_3,\delta_4+1}&=(\delta_1+\delta_4-\Sigma-v\dv)\Dbar_{\delta_1\delta_2\delta_3\delta_4},\nn\\
\Dbar_{\delta_1,\delta_2+1,\delta_3,\delta_4+1}&=(\delta_2+u\du+v\dv)\Dbar_{\delta_1\delta_2\delta_3\delta_4},\nn\\
\Dbar_{\delta_1+1,\delta_2,\delta_3+1,\delta_4}&=(\Sigma-\delta_4+u\du+v\dv)\Dbar_{\delta_1\delta_2\delta_3\delta_4},
\end{align}
The $\Dbar$-functions obey many transformation identities (stemming from the permutation symmetries of the one-loop box integral), one of which is the permutation property
\begin{align}
	\Dbar_{\delta_1\delta_2\delta_3\delta_4}(u,v) = v^{\delta_1+\delta_4-\Sigma}\Dbar_{\delta_2\delta_1\delta_4\delta3}(u,v) = u^{\delta_3+\delta_4-\Sigma}\Dbar_{\delta_4\delta_3\delta_2\delta_1}(u,v),
\end{align}
which can be used to convert a $\Dbar$-function with negative $\sigma$ into one with $\sigma\geq0$, as required for the decomposition shown in equation~(\ref{DbaruY}).\\
In some cases it is useful to use the reflection identity
\begin{align}
	\Dbar_{\delta_1\delta_2\delta_3\delta_4}(u,v) = \Dbar_{\Sigma-\delta_3,\Sigma-\delta_4,\Sigma-\delta_1,\Sigma-\delta_2}(u,v)
\end{align}
to bring a $\Dbar$-function into a more convenient form.

Finally, under crossing transformations of the cross-ratios $(u,v)$ the $\Dbar$-functions behave as
\begin{align}
	\Dbar_{\delta_1\delta_2\delta_3\delta_4}(u,v)&=\Dbar_{\delta_3\delta_2\delta_1\delta_4}(v,u),\nn\\
												 &=u^{-\delta_2}\Dbar_{\delta_4\delta_2\delta_3\delta_1}\left(\frac{1}{u},\frac{v}{u}\right),\nn\\
												 &=v^{\delta_4-\Sigma}\Dbar_{\delta_2\delta_1\delta_3\delta_4}\left(\frac{u}{v},\frac{1}{v}\right).
\end{align}


\end{document}